%% file: FCCE.tex
\documentclass[twoside]{cernyrep}

\usepackage{lipsum} 
\usepackage{mathptmx}
\usepackage{afterpage}
\usepackage{graphicx}
\usepackage[table]{xcolor}
\usepackage{csquotes}
\usepackage{multirow}
\usepackage{tikz}
\usepackage{pgf-pie}
\usepackage[normalem]{ulem}
\usepackage{datetime2}
\usepackage{array}
\usepackage{soul}
\usepackage{enumerate}
\usepackage[referable]{threeparttablex}
\usepackage{footnote}
\usepackage{comment}
\usepackage{siunitx}
\usepackage[bookmarks, bookmarksdepth=3, colorlinks=true, pdftex, linkcolor=black, citecolor=black, urlcolor=black]{hyperref}

\usepackage{lscape}

\newcolumntype{L}[1]{>{\raggedright\let\newline\\\arraybackslash\hspace{0pt}}m{#1}}
\newcolumntype{P}[1]{>{\raggedright\let\newline\\\arraybackslash\hspace{0pt}}p{#1}}
\newcolumntype{C}[1]{>{\centering\let\newline\\\arraybackslash\hspace{0pt}}m{#1}}

\usepackage[acronym,toc,section=section]{glossaries}
\usepackage{glossary-superragged}

\makeglossaries

\numberwithin{table}{section}
\numberwithin{figure}{section}

\usepackage[hang,flushmargin]{footmisc}
\fancypagestyle{plain}{}
\newcommand{\epem}{$\mathrm{e^+e^-}$}

\DeclareSIUnit\barn{b}
\DeclareSIUnit\micron{$\mu$m}
\newcommand{\ttbar}{t$\mathrm{\bar t}$}

\usepackage{etoolbox}
\AtBeginEnvironment{thebibliography}{\interlinepenalty=10000}
\newcommand{\murl}[1]{\hfil\penalty 100 \hfilneg \hbox{\url{#1}}}

\newlength{\oddmarginwidth}
\newlength{\evenmarginwidth}
\fancyhfoffset[LO,RE]{\oddmarginwidth}
\fancyhfoffset[LE,RO]{\evenmarginwidth}
\hypersetup{
     colorlinks = true,
     linkcolor = black,
     anchorcolor = blue, 
     citecolor = blue, 
     filecolor = blue, 
     urlcolor = blue 
     }

\begin{document}

\title{Comparative evaluation of future collider options}
\pagenumbering{roman}
\setcounter{page}{1}

\thispagestyle{empty}
\setlength{\unitlength}{1mm}
\begin{picture}(0.001,0.001)
\put(-8,8){CERN Yellow Reports: Monographs}
\put(110,8){CERN-2025-011}

\put(-8,-40){\huge \bfseries Comparative evaluation of future collider options} 
\put(-8,-50){\Large \bfseries Future Colliders Comparative Evaluation Working Group}


\put(3,-90){\Large Corresponding authors:}
\put(8,-98){\Large G.~Arduini, CERN}
\put(8,-106){\Large N.~Mounet, CERN}

\put(53,-250){\includegraphics{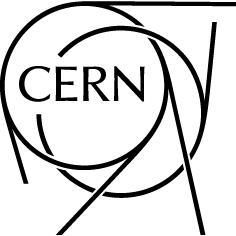}}

\end{picture}
\newpage

\thispagestyle{empty}
\mbox{}
\vfill

\begin{flushleft}
CERN Yellow Reports: Monographs\\
Published by CERN, CH-1211 Geneva 23, Switzerland\\[3mm]

\begin{tabular}{@{}l@{~}l}
 ISBN & 978-92-9083-718-3 (paperback) \\
 ISBN & 978-92-9083-719-0 (PDF) \\
 ISSN & 2519-8068 (Print)\\ 
 ISSN & 2519-8076 (Online)\\ 
 DOI & \url{https://doi.org/10.23731/CYRM-2025-0011}\\
\end{tabular}\\[5mm]

Copyright \copyright{} CERN, 2025\\[1mm]
\raisebox{-1mm}{\includegraphics[height=12pt]{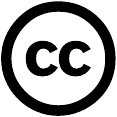}}
Creative Commons Attribution 4.0\\[5mm]

This volume should be cited as:\\[1mm]
Comparative evaluation of future collider options,\\Future Colliders Comparative Evaluation Working Group \\
CERN Yellow Reports: Monographs, CERN-2025-011 (CERN, Geneva, 2025)\\ 
\url{https://doi.org/10.23731/CYRM-2025-0011}.\\[3mm]

Accepted in Dec.\ 2025 by the \href{http://library.cern/about_us/editorial_board}{CERN Reports Editorial Board} 
(contact \href{mailto:Carlos.Lourenco@cern.ch}{Carlos.Lourenco@cern.ch}).\\[1mm]
Published by the CERN Scientific Information Service (contact \href{mailto:Jens.Vigen@cern.ch}{Jens.Vigen@cern.ch}).\\[1mm]
Indexed in the \href{https://cds.cern.ch/collection/CERN\%20Yellow\%20Reports?ln=en}{CERN Document Server} and in \href{https://inspirehep.net/}{INSPIRE}.\\[1mm]
Published Open Access to permit its wide dissemination, as knowledge transfer is an integral part of the mission of CERN.
\end{flushleft}



\author{\textbf{Future Colliders Comparative Evaluation Working Group:} G.~Arduini$^{a,\ast}$, M.~Benedikt$^a$, F.~Gianotti$^a$, K.~Jakobs$^{b}$, M.~Lamont$^a$, R.~Losito$^a$, M.~Meddahi$^a$, J.~Mnich$^a$, N.~Mounet$^{a,\ast\ast}$, D.~Schulte$^a$, F.~Sonnemann$^a$, S.~Stapnes$^a$, F.~Zimmermann$^a$}

\institute{\vspace{5mm}
$^a$CERN, Geneva, Switzerland \\
$^b$Universit\"at Freiburg, Germany\\\vspace{10pt}
$^\ast$ gianluigi.arduini@cern.ch \\
$^{\ast\ast}$ nicolas.mounet@cern.ch\\
\vspace{10pt}
{\begin{center} December 19\textsuperscript{th}, 2025 \end{center}}
}

\begin{abstract}

In anticipation of the completion of the High-Luminosity Large Hadron Collider (HL-LHC) programme by the end of 2041, CERN is preparing to launch a new major facility in the mid-2040s. According to the 2020 update of the European Strategy for Particle Physics (ESPP), the highest-priority next collider is an electron–positron Higgs factory, followed in the longer term by a hadron–hadron collider at the highest achievable energy. 

The CERN directorate established a Future Colliders Comparative Evaluation working group in June 2023. This group brings together project leaders and domain experts to conduct a consistent evaluation of the Future Circular Collider (FCC) and alternative scenarios based on shared assumptions and standardized criteria.

This report presents a comparative evaluation of proposed future collider projects submitted as input for the Update of the European Strategy for Particle Physics.  These proposals are compared considering main performance parameters, environmental impact and sustainability, technical maturity, cost of construction and operation, required human resources, and realistic implementation timelines.

An overview of the international collider projects within a similar timeframe, notably the CEPC in China and the ILC in Japan is also presented, as well as a short review of the status and prospects of new accelerator techniques.

\end{abstract}


\graphicspath{ {./logos/} }


\maketitle
\tableofcontents

\newpage
\pagenumbering{arabic}

\null
\thispagestyle{empty}
\addtocounter{page}{-1}
\newpage

\input{include/01-ExecutiveSummary/ExecutiveSummary}

\newpage
\input{include/02-Introduction/introduction.tex}

\newpage
\input{include/02b-CriteriaandMetrics/Criteria_and_Metrics}

\newpage
\input{include/03-OptionsatCERN/OptionsatCERN}

\newpage
\input{include/04-Beyond2050/Beyond2050}

\newpage
\input{include/05-OptionsoutsideCERN/OptionsoutsideCERN}

\newpage
\input{include/06-ReuseLHCTunnel/ReuseLHCtunnel}

\newpage
\input{include/07-NewAcceleratorTechnology/NewAcceleratorTechnology}


\clearpage
\newpage

\section*{Acknowledgements}
\addcontentsline{toc}{section}{Acknowledgements}

We warmly thank Dave Newbold, Mike Seidel for the organisation of the European Accelerator \acrshort{rd} Platform Review for the \acrfull{ldg}, as well as its chair, Norbert Holtkamp, and the Review Panel (Mei Bai, Frederick Bordry, Nuria Catalan-Lasheras, Barbara Dalena, Massimo Ferrario, Andreas Jankowiak, Robert Rimmer, Herman ten Kate, Peter Williams). The review has provided extremely valuable input for this comparative evaluation.

We sincerely thank the following colleagues for discussions and input: Erik Adli, Bernhard Auchmann, Nicolas Bellegarde, Caterina Bloise, Oliver Br\"uning, Liam Bromiley, Xavier Buffat, Simone Campana, John Farmer, Gerardo Ganis, Jie Gao, Frank Gerigk, Edda Gschwendtner, Philippe Lebrun, Benno List, Jenny List, Ed Mactavish, Peter McIntosh, Shinichiro Michizono, Tatsuya Nakada, Mauro Nonis, John Osborne, Vittorio Parma, Thomas Roser, Nikolaos Sapountzoglou, John Seeman, Thomas Schörner, Markus Schulz, Maxim Titov, Ezio Todesco, Marlene Turner, Anders Unnervik, Jim Virdee, Tim Watson, and Akira Yamamoto.

We are also very grateful to Jens Vigen and Carlos  Louren\c{c}o for their careful proofreading and for providing editorial corrections.

\newpage

\appendix
\addcontentsline{toc}{chapter}{Appendices}
\input{include/08-Annex/A-Criteria}
\newpage
\input{include/08-Annex/B-AdditionalTables}

\clearpage

\input{include/08-Annex/ZZ-AcronymsandGlossary}

\clearpage
\newpage
\fancyhead[LE]{\rule[-1ex]{0pt}{1ex} References}
\begingroup
\begin{flushleft}
\interlinepenalty=10000
\addcontentsline{toc}{chapter}{References}
\bibliography{bibliography.bib}{}
\bibliographystyle{utphys_custom}
\end{flushleft}
\endgroup

\end{document}

%% file: include/01-ExecutiveSummary/ExecutiveSummary.tex
\section*{Executive summary}
\label{sec:ExecutiveSummary}
\addcontentsline{toc}{section}{Executive summary}

\setcounter{figure}{0}
\renewcommand{\thefigure}{ES.\arabic{figure}}

The 2020 update of the \acrfull{espp} identified the construction of an electron–positron Higgs factory, designed to study the Higgs boson with high precision, as the highest priority next collider. In the longer term, Europe should aim at a hadron collider at the highest achievable energy.
In response to these strategic objectives, \acrshort{cern} has consolidated its future accelerator programme around several key initiatives:
\begin{itemize}
    \item The \acrfull{fcc} integrated programme, which has published a detailed feasibility study.
    \item The \acrfull{clic}, which has progressed in the development of its radio-frequency (\acrshort{rf}) technology, now deployed in various applications, and has published a readiness report.
    \item The \acrfull{mc} study, supported by international collaborations including the \acrfull{imcc} and the \acrshort{eu}-funded MuCol Consortium, which is investigating the potential of this novel approach.
    \item Participation in a wide-ranging European accelerator \acrshort{rd} programme, targeting technologies such as \acrfull{hfm}, \acrfull{pwfa} (via the \acrlong{awake}---\acrshort{awake}), Superconducting radio-frequency (\acrshort{srf}) systems, and to a lesser extent \acrfullpl{erl}. The accelerator \acrshort{rd} programme has recently undergone a mid-term international review.

\end{itemize}

In addition, the \acrfull{ilc} remains under discussion in Japan, although progress toward the implementation phase is limited. A proposal of a \acrfull{lcf} at \acrshort{cern}, building on the \acrshort{ilc} design, has been recently developed.

\acrshort{cern} aims to be in a position to begin operating a new collider, by the second half of the 2040s, to follow timely after the completion of the \acrfull{hllhc} programme, planned to be completed by the end of 2041. To meet this timeline, key decisions and initial resource commitments must occur between 2028 and 2030.

Given the differing levels of maturity across various proposals, this report aims to present a comparative evaluation of collider projects that could realistically be operational by 2045--2050. Projects that are unlikely to meet that timeline are discussed separately, along with advanced concepts and options to reuse the \acrshort{lhc} tunnel. The information presented here is based on the content of the proposals submitted to the \acrfull{esppu} 2026.

The following main criteria have been considered for the comparative evaluation of the proposals (\acrshort{clic}, \acrshort{fccee} and \acrshort{lcf}) for \textit{first phase} \epem colliders at \acrshort{cern} and are described in Section~\ref{sec:criteria_metrics}:
\begin{itemize}
    \item main parameters and performance;
    \item environmental aspects;
    \item technical readiness;
    \item accelerator and experiments construction and installation: material costs;
    \item accelerator construction and installation: human resources;
    \item project timeline;
    \item operation: material and personnel costs.
\end{itemize}

Although the proposed \epem  colliders will not use existing \acrshort{cern} accelerators as injectors, their implementation relies heavily on \acrshort{cern}’s expertise, institutional framework and in particular on its existing infrastructure 
providing access to key resources such as electrical power, cryogenics, technical workshops, water, communication and transport networks and making it a very well adapted host site.

\begin{figure}[h!]
    \centering
    \includegraphics[width=0.9\linewidth]{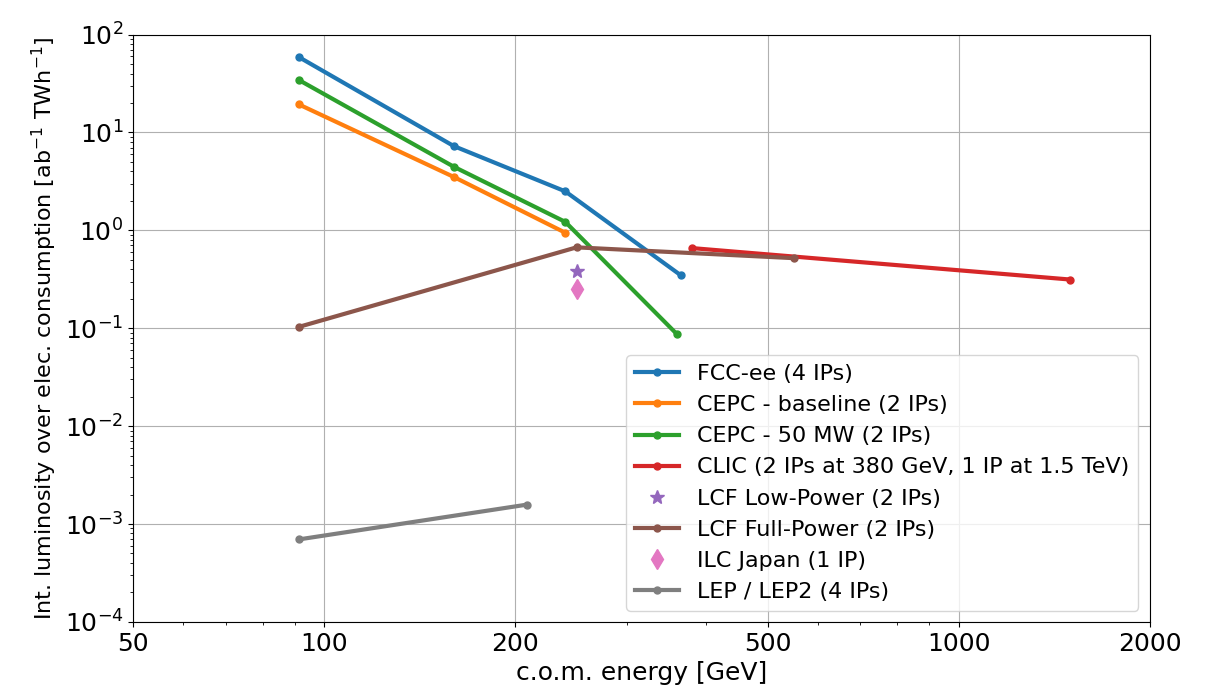}
    \caption{Integrated luminosity over all experiments per year of nominal operation, per unit of electricity consumption for future electron-positron colliders (excluding off-line computing)---see Tables~\ref{tab:Parameters_CLIC_FCC_LCF},~\ref{tab:Param_CLIC_LCF_upgrades} and~\ref{tab:Parameters_ILC_CEPC_rescaled}; the performance has been rescaled to the \acrshort{fccee} operational year for \acrshort{cepc} and to \acrshort{lcf} 250 \acrfull{lp} for the \acrshort{ilc} (see Table~\ref{tab:operational_year}). For \acrshortpl{lc} the total luminosity (including that below 99\% of the nominal \acrshort{com}\ energy) is considered. \acrshort{lep} and \acrshort{lep2} data were respectively taken for the years 1993 and 2000~\cite{bib:Assmann_LEP2,bib:electricity_LEP_1993,bib:electricity_LEP2_2000}. For the \acrshort{ilc}, a single \acrshort{ip} is considered but with two experiments (in ``push-pull'' mode, see Section~\ref{sec:ILC_intro}). The information for \acrshort{lep3} has not been added at this stage.
    }
    \label{fig:int_lumi_per_TWh}
\end{figure}

\acrshort{cern} has a consolidated experience with circular colliders and has successfully operated the \acrfull{lep}.  Multi-ampere \epem ~beams have been demonstrated at the \acrfull{pepii} and \acrfull{kekb}. The prospected luminosity targets for \acrshort{fccee} are challenging, particularly for the operation at the Z-pole, but they will benefit of the experience gained at \acrshort{daphne} with crab-waist operation, the ongoing performance ramp-up of \acrshort{skekb}, which will demonstrate nearly all the required accelerator physics techniques for \acrshort{fccee}, as will the future electron ring for the \acrfull{eic} at \acrfull{bnl}. Differently from \acrfullpl{lc}, energy loss due to synchrotron radiation limits the maximum energy realistically achievable by circular colliders to the \ttbar ~threshold. Circular colliders can naturally accommodate a larger number of \acrfullpl{ip} and their larger repetition rate (determined by the revolution frequency in the \SI{}{kHz} range) allow them to provide significantly higher integrated luminosities per unit of energy consumption than \acrshort{lc}s up to energies of \SIrange{300}{350}{GeV} (see Fig.~\ref{fig:int_lumi_per_TWh}). It is worth noting the significant improvement expected in the integrated luminosity per unit of electricity consumption for future circular colliders as compared to their predecessors, \acrshort{lep} and \acrshort{lep2}. In addition, \acrshort{fccee} offers the capability of varying the collision energy between \SIrange[range-units=single,range-phrase={ and }]{90}{240}{\GeV} seamlessly and at unparalleled luminosity. This flexibility is reduced after the upgrade for operation at the highest energy and operation at lower energy will be possible, albeit at lower luminosity. 

In their baseline design \acrshortpl{lc} can deliver polarized beams (both beams for \acrshort{lcf} and only the electron beam for \acrshort{clic}). The luminosity performance of \acrshort{lc}s relies on high positron production rates, the capability to preserve the emittance of the beams delivered by the damping rings all through the main linac and the \acrfull{bds} and to minimize the chromatic aberrations resulting from the extremely small optics waist at the \acrshort{ip}. This implies achieving an excellent control of the wakefields in the accelerator and of the optics at the final focus to maintain in collision beams with nanometer beam sizes at the \acrshort{ip} in a reproducible fashion, demanding advanced feed-forward systems to minimize the effects of ground motion and temperature variations. The control of optics aberrations, the tuning algorithms and the stabilization against ground motion and temperature variation have been studied at the \acrfull{atf} at \acrshort{kek} in Japan, the \acrfull{fftb} and the \acrfull{facet} at \acrshort{slac} in the \acrshort{us}, though additional~\acrshort{rd} is required. In order to achieve high luminosity \acrshort{lc}s must be operated in the high-beamstrahlung regime resulting in a rather broad luminosity spectrum as a function of the~\acrfull{com} collision energy $\sqrt{s}$. The main parameters and performance of the \epem colliders proposed at \acrshort{cern} and based on the same operational scenario defined in Section~\ref{sec:opyear} are listed in Table~\ref{tab:Parameters_CLIC_FCC_LCF}.

A preliminary analysis of the carbon footprint of the construction of the new collider facilities has been performed, though it is still affected by significant uncertainties given the relatively early phase of the projects and the uncertainty in the evolution of the process of decarbonisation of the energy and commodities market in the next two decades. Nevertheless the analysis has evidenced that the major contributions to \acrfull{ghg} emissions are associated with the construction of the components of the accelerators and detectors and their technical infrastructure, followed by the \acrfull{ce} works. The emissions associated to the generation of the electricity required for operation are lower, thanks to the low carbon intensity of the electricity generated in France, from where the electricity powering \acrshort{cern} facilities is purchased.  The possible re-use of the \acrshort{ce} infrastructure for future upgrades can further benefit the sustainability of the projects. The estimated annual electricity consumption of the collider, its experiments (including their local data centres) and associated injector complex (excluding the other \acrshort{cern} accelerators) amounts to approximately \SI{0.8}{\tera\watt\hour}/y for \acrshort{clic} and \acrshort{lcf} \acrshort{lp} and varies between \SIrange[range-units=single,range-phrase={ and }]{1.2}{1.9}{\tera\watt\hour}/y for \acrshort{fccee} depending on the \acrshort{com} energy. These values should be compared with the present \acrshort{cern} annual electricity consumption of about \SI{1.3}{\tera\watt\hour}/y of which approximately \SI{0.7}{\tera\watt\hour}/y are necessary to operate the \acrshort{lhc} collider and its experiments (injector complex excluded). \acrshort{hllhc} operation will demand additional \SI{0.1}{\tera\watt\hour}/y.

The technical readiness of \acrshort{clic}, \acrshort{fccee} and \acrshort{lcf} are comparable, though \acrshort{lcf} benefits of the developments and industrialization effort conducted for the \SI{1.3}{\GHz} Superconducting radio-frequency (\acrshort{srf}) cavities for the \acrshort{desy} \acrfull{xfel} and \acrshort{slac} \acrfull{lcls2} projects and therefore has an overall higher \acrfull{trl} as compared to the other two proposals. Though significant progress has been made in the stabilization algorithms and in the start-to-end simulations, \acrshort{ip} spot size control and pulse-to-pulse beam position stability remain the main challenges for \acrshort{lc}s.

The cost estimates for the three projects (see Tables~\ref{tab:CLIC_cost}, ~\ref{tab:FCCee_cost} and~\ref{tab:LCF_cost}) have a different level of uncertainty. \acrshort{fccee} has generally a more accurate cost estimate (at class 3 for civil engineering, accelerators and associated technical infrastructure) thanks to the higher level of detail achieved, in particular for the civil engineering and the territorial implementation as the placement scenario has been fully defined. 
In addition, this estimate has been extensively reviewed by \acrshort{cern} governing bodies. 
Civil engineering is the major contributor to the higher cost of \acrshort{fccee}. 
The cost of \acrshort{clic} baseline is the lowest but any considered future upgrade requires additional \acrshort{ce} works.

The personnel required for the construction of the projects (not including contractor personnel and the personnel required for \acrlong{hw}---\acrshort{hw}---and beam commissioning) ranges between \SIrange[range-units=single,range-phrase={ and }]{10000}{15000}{\acrfull{ftey}}. The most labour-intensive activities related to the accelerator installation, as well as \acrshort{hw} and beam commissioning, will only be able to occur towards the end and after completion of the \acrshort{hllhc} programme.

Following the recommendations of the \acrshort{esppu} 2020, a detailed feasibility study for the implementation of \acrshort{fcc} in two stages has been successfully completed and a detailed timeline for the construction established. The \acrshort{clic} project and, in  particular the \acrshort{lcf} project, would require a preparation phase with a corresponding investment of resources to reach a level of implementation details comparable to that of the \acrshort{fcc} feasibility study. This phase, if successful, should be followed by an implementation preparation phase requiring Council approval and by a construction phase whose timeline should be coherent, in terms of resources, with the \acrshort{hllhc} programme.

An estimate of the resources (material and personnel) required for operation of the collider accelerator complex (including technical infrastructure), and of the technical infrastructure of the experiments has been made (see Table~\ref{tab:operation_CLICFCCeeLCF}). The resources needed for the operation of the rest of the accelerator complex (e.g.\,the \acrshort{ps} and \acrshort{sps} complex), its technical infrastructure and the technical and scientist support for the experimental detectors and computing are excluded. The lowest requirements are for \acrshort{clic} baseline option but in all cases they are comparable to those required for the operation of~\acrshort{lhc} though the expected electricity costs will be higher for \acrshort{fccee}, but for a higher integrated luminosity.

The \acrshort{clic} and \acrshort{lcf} proposals contemplate two experiments sharing the luminosity delivered by the main linac and the same experimental cavern, while four experiments (each located in a different experimental cavern) are served simultaneously in \acrshort{fccee} and therefore they could engage a larger experimental community, estimated at approximately 3500 \acrshort{ftey} (2/3 of which are scientists) per experiment.

After approximately 10 years of operation it is proposed to increase the energy of \acrshort{clic} and \acrshort{lcf} to \SI{1.5}{\TeV} and \SI{550}{\GeV}, respectively, essentially using the same technology. For~\acrshort{clic} significant \acrshort{ce} work will be required to increase the length of the tunnel from \SIrange{12.1}{29.6}{\km} and additional higher-gradient accelerating modules will have to be installed. The length of the \acrfull{bds} also needs to be increased and new magnets installed. No additional~\acrshort{ce} work will be necessary for the \acrshort{lcf} energy upgrade as the length of the collider tunnel has been defined for the \acrshort{com} energy of \SI{550}{\GeV} assuming nominal parameters for the \acrshort{srf} cryomodules. The bulk of the upgrade work will consist in installation of additional cryomodules. The expected cost breakdown of the proposed upgrades for the \acrshort{lc}s is presented in Tables~\ref{tab:clic1500cost} and \ref{tab:LCF550_cost}. 

In the \acrshort{fcc} ``integrated programme'' the \epem collider would be replaced by a proton-proton collider (\acrshort{fcchh}) after approximately 15 years of operation. \acrshort{fcchh} will be installed in the same tunnel of \acrshort{fccee}, sharing  much of its \acrshort{ce} and  \acrfull{ti}, and it would operate at a \acrshort{com} energy of at least \SI{85}{\TeV}, extending the physics discovery potential by an order of magnitude in the~\SI{10}{\TeV} \acrfull{pcm} energy range.
The \acrshort{fcchh} baseline assumes \SI{14}{\tesla} magnets which are close to the state of the art of $\mathrm{Nb_3Sn}$ technology. \acrshort{rd} towards \acrshort{hts} magnets is pursued and could enable higher collision energies. 

As a second phase after \acrshort{fccee}, \acrshort{fcchh} could start operation in the mid of the 2070s. On a technically-limited timescale, \acrshort{fcchh}, as a stand-alone project with \SI{14}{\tesla} $\mathrm{Nb_3Sn}$ dipoles, could start operation in the mid of the 2050s. The availability of resources will drive the real schedule. The cost breakdown of the construction of \acrshort{fcchh} as a second phase after \acrshort{fccee} is shown in~\Table~\ref{tab:fcchh_cost}.

The \acrfull{mc} has also the ambition to approach the \SI{10}{\TeV} \acrshort{pcm}. Two conceptual scenarios with a maximum \acrshort{com} energy of \SI{7.6}{\TeV} have been developed for \acrshort{cern}. The \acrshort{mc} injectors would be installed in  \acrshort{sps} and \acrshort{lhc} tunnels. One or two additional tunnels would need to be built for the installation of the collider serving two \acrshort{ip}s. While progress is being made, the \acrshort{mc} has not yet reached a maturity level that gives sufficient confidence in its feasibility. Significant investments in a staged \acrshort{rd} programme requiring conspicuous demonstration facilities are necessary to reach that level. Namely, a demonstration of 6D cooling is a necessary condition to assess the feasibility and performance of a \acrshort{mc}. The technical design of the demonstrator and its construction demand an amount of resources significantly exceeding the present level of \acrshort{rd} resources. A detailed timeline cannot be defined at present but only sketched with some decision points. Given the uncertainty on the feasibility, the construction and installation costs (see Fig.~\ref{fig:Cost_MC}) are only indicative.  

The possibility to reuse the \acrshort{lhc} tunnel to host an \epem collider was proposed for the first time in 2012 and recently revived in preparation of the \acrshort{esppu} 2026. This proposal envisions a collider operating between \SIrange[range-units=single,range-phrase={ and }]{91.2}{230}{\GeV} with two \acrshort{ip}s. The main parameters, listed in Table~\ref{tab:tablep3} (column \acrshort{lep3} - 2025), are based on scaling from earlier conceptual designs and assume the use of \acrshort{fccee} \acrshort{srf} systems. The collider would require a booster installed in the same tunnel to provide top-up injection and operate at a maximum intensity corresponding to a synchrotron radiation power loss of \SI{50}{\MW} per beam as for \acrshort{fccee}. A detailed design of the lattice, magnet system specifications as well as integration of the collider and booster and the associated accelerator and technical systems in the \acrshort{lhc} tunnel (whose diameter is limited to \SI{3.8}{\m} in the arcs and \SI{4.4}{\m} in the \acrlong{lss}---\acrshort{lss}) will need to be produced. Extensive studies would be necessary to bring the level of maturity of this proposal to a level at least comparable to that of the other \epem collider projects above considered. The electricity consumption prospected by the proponents is comparable to that of \acrshort{fccee} at the lowest energy and we can expect that the resources required to operate \acrshort{lep3} will be similar to those necessary for the \acrshort{lhc} operation, though with a prospected lower performance by an order of magnitude as compared to \acrshort{fccee} in terms of integrated luminosity per year of nominal operation. As a result, the reduced initial \acrfull{capex} might be, at least partly, offset by higher operation costs in order to achieve similar integrated luminosities as \acrshort{fccee} over the full programme. In addition it will not be possible to run \acrshort{lep3} at  the \ttbar ~threshold.

Among the initiatives considering the re-use of the \acrshort{lhc}, the \acrfull{lhec} would collide a \SI{50}{\GeV} electron beam, accelerated by a new 3-turn high-current~\acrshort{erl}, with a proton or ion beam from \acrshort{hllhc} producing collisions at a \acrshort{com} energy of \SI{1.18}{\TeV} in the \acrshort{lhc} \acrshort{ip}2, where the \acrshort{alice} experiment is located. The \acrshort{erl} would require the construction of a new tunnel with a length corresponding to 1/3 the circumference of the \acrshort{lhc}. The main parameters of this machine are listed in Table~\ref{tab:LHeC}. The annual electricity consumption of the \acrshort{lhec} would be approximately \SI{1.1}{\tera\watt\hour}/y (excluding the electricity consumption of the \acrshort{lhc} injectors complex). High-current multi-turn energy recovery still needs to be demonstrated for beam powers exceeding \SI{1}{\mega\watt} (well below the nominal \acrshort{lhec} beam power of \SI{2.5}{\giga\watt}). The approval of \acrshort{lhec} could only occur after such a demonstration is successful. 
The completion of the construction of \acrshort{lhec} can only be realistically envisaged after the end of the \acrshort{hllhc} programme. 
The resources required for its operation would be comparable to those necessary for the present operation of \acrshort{lhc} and its injectors implying a delay of at least a decade in the construction of a next-generation flagship collider.

The parameters of the \acrfull{helhc}, a hadron collider in the \acrshort{lhc} tunnel, using \acrshort{hfm} magnets studied as part of the \acrshort{fcc} \acrfull{cds}, are also recalled in Table~\ref{tab:tabhelhc}. Such a collider would be limited to a maximum \acrshort{com} energy of \SI{33}{\TeV}, well below the goal of a \SI{10}{\TeV} \acrfull{pcm}, even if equipped with \SI{20}{\tesla} magnets, and it could start operation only at the beginning of the 2070s.

A short review of the status and prospects of new accelerator techniques is presented in Section~\ref{sec:NewAccTech}. Although significant progress has been made towards the acceleration of higher quality electron beams, several technological and accelerator/plasma physics challenges remain. The level of maturity of the wakefield-based accelerators does not allow to consider them as realistic alternatives to the above proposals, at least on the timescale considered.

An overview of the collider projects proposed for other regions is also available in Section~\ref{sec:outsideCERN}.

\setcounter{figure}{0}
\renewcommand{\thefigure}{\arabic{figure}}
\numberwithin{figure}{section}

%% file: include/02-Introduction/introduction.tex
\section{Introduction and context} 
\label{sec:Introduction}

The 2020 update of the European Strategy for Particle Physics (\acrshort{esppu} 2020) has identified the priorities for the field in Europe, within the global context~\cite{bib:EPPSU}: 
\begin{itemize}
    \item \textit{An electron-positron Higgs factory is the highest-priority next collider. For the longer term, the European particle physics community has the ambition to operate a proton-proton collider at the highest achievable energy. }
    \item \textit{Europe, together with its international partners, should investigate the technical and financial feasibility of a future hadron collider at \acrshort{cern} with a centre-of-mass energy of at least~\SI{100}{\TeV} and with an electron-positron Higgs and electroweak factory as a possible first stage. Such a feasibility study of the colliders and related infrastructure should be established as a global endeavour and be completed on the timescale of the next Strategy update. }
    \item  \textit{Innovative accelerator technology underpins the physics reach of high-energy and high-intensity colliders. It is also a powerful driver for many accelerator-based fields of science and industry. The technologies under consideration include high-field magnets, high-temperature superconductors, plasma wakefield acceleration and other high-gradient accelerating structures, bright muon beams, energy recovery linacs. The European particle physics community must intensify accelerator \acrshort{rd} and sustain it with adequate resources. A roadmap should prioritise the technology, taking into account synergies with international partners and other communities such as photon and neutron sources, fusion energy and industry. Deliverables for this decade should be defined in a timely fashion and coordinated among \acrshort{cern} and national laboratories and institutes.} 
\end{itemize}

Given these recommendations, \acrshort{cern} consolidated its future accelerator programme in 2020 to address the priorities of the \acrshort{esppu} 2020 outlined above by establishing or extending the following initiatives:
\begin{itemize}
    \item \textbf{\acrfull{fcc} feasibility study}: the principal aim was the delivery of a feasibility study report by the end of March 2025. A successful mid-term review took place during Q4 2023, and the final report was recently delivered on time.
    \item \textbf{\acrfull{clic}}: following extensive \acrshort{rd}, the key X-band radio-frequency (\acrshort{rf}) technology has been deployed world-wide in a number of projects as well as for societal applications.  Studies continued  with the aim of delivering a Project Readiness Report which has been submitted to the \acrshort{esppu} 2026.
    \item  \textbf{\acrfull{mc}}: an International Muon Collider Study was launched with a dedicated funding line. An international collaboration has been set-up. It includes the MuCol Consortium established under a grant agreement within the~\acrfull{horizon} and a strong~\acrshort{us} component.
    \item \textbf{Accelerator \acrfull{rd}}: An European Accelerator  \acrshort{rd} roadmap has been outlined~\cite{bib:AccRDRoadmap} and its execution is underway and looking to secure the requisite technology for the next generation of machines. At \acrshort{cern} it is realized through four main lines:
    \begin{itemize}
        \item \acrfull{hfm} \acrshort{rd}: both \acrfull{lts} and \acrfull{hts} magnet development;
        \item \acrshort{rf} technology \acrshort{rd}: both \acrfull{nc} X-band and \acrfull{srf};
        \item Proton-driven \acrfull{pwfa} with the~\acrfull{awake} study;
        \item \acrfull{clear}.
    \end{itemize}
\end{itemize}

There is limited progress in Japan moving to a pre-lab phase for the \acrfull{ilc}. A targeted~\acrshort{rd} phase is ongoing and a proposal for an \acrfull{ilc} implementation at~\acrshort{cern} has been submitted to the \acrshort{esppu} 2026 under the appellation \acrfull{lcf}.

The \acrfull{hllhc} baseline schedule foresees operation until the end 2041. \acrshort{cern} is positioning itself to be able to start operating an~\epem Higgs factory in the second half of the forties. This implies down-selection, a decision, and an initial commitment of resources in 2028, with subsequent project preparation and execution in the following decade.

In the global context it is important to note that:
\begin{itemize}
    \item in its report, the US 2023 \acrfull{p5} has endorsed \textit{an off-shore Higgs factory, located in either Europe or Japan, to advance studies of the Higgs boson following the \acrshort{hllhc}  while maintaining a healthy on-shore particle physics program}, adding that \textit{The \acrshort{us} should actively engage in design studies to establish the technical feasibility and cost envelope of Higgs factory designs. We recommend that a targeted collider panel review the options after feasibility studies converge. At that point, it is recommended that the \acrshort{us} commit funds commensurate with its involvement in the \acrshort{lhc} and \acrshort{hllhc}}. The panel recommended \textit{dedicated \acrshort{rd} to explore a suite of promising future projects. One of the most ambitious is a future collider concept: a~\SI{10}{\TeV} \acrfull{pcm} collider to search for direct evidence and quantum imprints of new physics at unprecedented energies.} and recommended a \textit{targeted collider \acrshort{rd} to establish the feasibility of a~\SI{10}{\TeV} \acrshort{pcm} muon collider. A key milestone on this path is to design a muon collider demonstrator facility}~\cite{bib:P5report}.
    \item In China, the \acrfull{cepc} community continues a robust \acrshort{rd} programme and development of site options and has published a \acrfull{cdr} at the end of 2023~\cite{bib:CEPC_TDR_accelerator}. 
    \item In Japan, despite the reluctance to move into the pre-lab phase, the \acrfull{mext} is actively supporting continued \acrshort{ilc} \acrshort{rd}.
\end{itemize}

The maturity of the above-mentioned proposals differs and as input to  the incoming \acrshort{esppu} a comparative evaluation is presented. 

\textit{First phase future collider proposals} that could become operational at \acrshort{cern} at the latest in 2045--2050 are compared. Other equivalent options proposed elsewhere in the world in the same time frame are also presented. Proposals that are presently not anticipated to match realistically the above time frame (including possible major upgrades or developments) are discussed separately.

The latest version of the proposals for re-use of the \acrshort{lhc} tunnel have been summarized as reference for completeness. A short summary of some advanced collider concepts, for which extensive studies (e.g.\ a \acrshort{cdr}) do not exist, has also been provided.

\subsection{General considerations about a next flagship project at \acrshort{cern}}

\acrshort{cern}’s current accelerator complex, centred on the \acrshort{lhc}, supports a diverse and world leading scientific programme spanning high-energy proton–proton collisions, heavy-ion physics, and a broad array of fixed-target experiments. These include studies of antimatter, nuclear physics with radioactive ions and neutrons, and a variety of precision measurements enabled by unique facilities such as \acrshort{isolde}, \acrshort{ntof}, \acrshort{ad}--\acrshort{elena}, and the East and North Areas.

Looking ahead, the \acrfull{hllhc} upgrade will be installed during \acrlong{ls}~3 (\acrshort{ls}3, 2026--2030). By the end of \acrshort{hllhc} operation in 2041, the total proton-proton integrated luminosity delivered to \acrshort{atlas} and \acrshort{cms} at or above \SI{13}{\TeV} since the beginning of the \acrshort{lhc} is expected to reach \SI{3}{\per\atto\barn}. \acrshort{ls}3 will also see significant investment in consolidating the broader accelerator complex, ensuring its continued effectiveness in the following decades.

The financial and personnel demands of \acrshort{hllhc} operations mean that the commissioning of any major new facility at \acrshort{cern} cannot realistically begin before the end of \acrshort{hllhc} exploitation. Moreover, a transitional period will be required to redeploy resources, complete installation and commissioning activities, and ramp up new infrastructures. Thus, a future collider at \acrshort{cern} cannot realistically start operation before the mid-2040s, regardless of the precise nature of the project and its technical readiness. 

The approval within the current strategic period of a large-scale post-\acrshort{lhc} collider project aiming for operation in the mid-2040s would significantly constrain \acrshort{cern}’s ability to pursue other long-term options. It is therefore critical to balance ambition with realism in assessing near- and long-term pathways.

The \acrshort{fcc} integrated programme foresees a sequential implementation of \acrshort{fccee} followed by \acrshort{fcchh}, with operation potentially spanning from the mid 2040s to the end of the century. Should circumstances evolve, a stand-alone \acrshort{fcchh} collider with operation starting at the end of the 2050s remains a technically viable and potentially competitive option. 

The \acrshort{fcc} feasibility study has included detailed territorial implementation studies and established structured, multi-level engagement with the Host State authorities and local communities. These developments represent a substantial investment in building a sustainable, long-term foundation for the project. In this context, parallel developments for alternative options in the local region should not undermine this process.  Any siting analysis for such alternatives should be limited to conceptual studies making use of existing information, as Host State authorities and local services and communities would not be able to support implementation studies for two big infrastructures.  

While the \acrlong{mc} represents an exciting long-term possibility with excellent physics potential, it remains at an early stage of development. The required \acrshort{rd} is substantial and includes demonstration of key feasibility elements such as 6D cooling. Given its technical immaturity and high resource demands, it cannot be considered as a realistic option for implementation in the strategic timeframe discussed here. The \acrlong{mc} should therefore be pursued through a sustained, staged \acrshort{rd} effort, independent of the decision on the next major collider.

Alternative solutions, such as \acrshort{lep3} or \acrshort{lhec}, would entail major infrastructure investments and operating costs and therefore delay the implementation of a next-generation collider by at least one to two decades. For this reason, they cannot be considered as “bridges”, but rather as alternative options to other proposed colliders. 

A linear collider facility at \acrshort{cern} would preclude the construction of \acrshort{fcchh} due to infrastructure and other constraints in the local, highly-populated region.

%% file: include/02b-CriteriaandMetrics/Criteria_and_Metrics.tex
\section{Criteria and metrics for the comparison} \label{sec:criteria_metrics}

The following main criteria have been considered for the comparative evaluation, in particular for the projects at \acrshort{cern}, when available:
\begin{itemize}
    \item Main parameters and performance
    \item Environmental aspects
    \item Technical readiness
    \item Accelerator and experiments construction and installation: material costs
    \item Accelerator construction and installation: human resources
    \item Project timeline
    \item Operation: material and personnel costs.
\end{itemize}

Consideration has been given also to the following subjects:
\begin{itemize}
    \item Upgrade potential
    \item Size of the experimental community that could be served
    \item Accelerator technology production capabilities and potential buy-in from European institutes.
\end{itemize}

\subsection{Main parameters and performance}\label{sec:MainParametersandPerformance_metrics}
A short description of the colliders and of the associated injector chain with their main components and technologies is given. Re-use of existing infrastructure/accelerators with the required upgrades vs. construction of new ones is addressed.
Peak luminosity as well as yearly integrated nominal performance (as well as expected integrated electricity consumption) are presented based on reference yearly operational schedules (see Section~\ref{sec:opyear}). When different values are assumed by the proponents, corresponding tables are also presented separately in the relevant Appendix.

In order to achieve high luminosity linear colliders must be operated in the high-beamstrahlung regime resulting in a rather broad luminosity spectrum as a function of the~\acrshort{com}\ collision energy $\sqrt{s}$. For these machines both the instantaneous luminosity above 99\% of $\sqrt{s}$ and the total instantaneous luminosity integrated over the full energy spectrum are presented.

The estimates for the electricity consumption include the following infrastructure:
\begin{itemize}
    \item collider;
    \item injectors;
    \item transfer lines;
    \item experiments (which normally represent a small fraction of the overall consumption---a few percent) and their local data centres;
    \item general services and controls;
    \item off-line data-analysis centres;
\end{itemize}
unless specified differently.

Considerations on the complexity of the accelerator proposal and on its impact on performance ramp-up are also made.

\subsubsection{Operational year}\label{sec:opyear}
The breakdown of the operational phases for the~\epem colliders and the hadron colliders and their duration have been defined in Ref.~\cite{bib:Bordry_proposedfuturecolliders} based on past experience. An extrapolation for the muon colliders is also proposed here, though with a larger uncertainty due to the maturity of the related study.
The fraction of electricity consumption (expressed as a percentage of the consumption for nominal operation with beam) required for each operational phase, is also given. These values are used to estimate the yearly electricity consumption. 

\paragraph*{\epem colliders}
The main parameters for circular and linear~\epem colliders are presented in Table~\ref{tab:operational_year} and Fig.~\ref{fig:operational_year}. The annual scheduled physics time amounts to 185 days, with an efficiency for data taking of 75\%. Operation occurs at constant luminosity for both the circular (top-up injection) and linear \epem colliders, with no turn-around time between physics fills (the beams in the circular colliders are injected from the injectors at collision energy and no time is lost for acceleration in the collider). 

\paragraph*{Hadron colliders}
The main parameters are presented in Table~\ref{tab:operational_year_FCChhMuC} and Fig.~\ref{fig:operational_year_FCChh}. 
It is assumed that 160 days of high-luminosity physics operation are scheduled, based on~\acrshort{lhc} experience. A longer~\acrfull{yets}, the required powering tests of the~\acrshort{sc} circuits following it, the necessary recovery time after technical stops and the need of special physics runs at lower luminosity (e.g.\ for luminosity calibration) explain the shorter scheduled physics time as compared to~\epem colliders. 
The machine is assumed to be available for 70\% of the scheduled high-luminosity physics time and to deliver luminosity for experiment data taking (also called ``stable beam'' phase) for 35\% of the scheduled high-luminosity physics time.

\paragraph*{Muon colliders}
The main parameters are presented in Table~\ref{tab:operational_year_FCChhMuC} and Fig.~\ref{fig:operational_year_MuC}.
The breakdown of the operational phases for the~\acrshort{mc} is assumed to be similar to that of the hadron colliders because of the large superconducting magnetic system. No special physics runs are contemplated. Machine availability is assumed to be 70\% as for the hadron colliders. Operation at constant luminosity (top-up mode) is tentatively assumed with no turn-around time between physics fills as the collider is operated at constant energy.

\subsection{Environmental aspects}

Accelerators and detectors construction and operation (including on-line and off-line computing), as any other human activity, make use of natural resources and impact the environment.

~\gls{lcagl}~(\acrshort{lca}) (see Glossary p.~\pageref{sec:Glossary}) is a structured, comprehensive and internationally standardised~(\acrshort{iso}~14040/44) method to quantify some of the potential environmental impacts such as emissions and resources consumed. Existing methodologies identify several impact categories which, apart from~\acrfull{ghg} emissions, are not widely reported and are subject to large uncertainties~\cite{clic_ilc_lca_arup}. For that reason the reporting will be limited to~\acrshort{ghg} emissions expressed in \SI{}{\gram}~CO\textsubscript{2} eq.\ (or multiples).  

Host states legislation requires a precise set of parameters to be presented in the environmental impact study and limits to be respected. It is important to underline that here we provide a simplified set of indicators to evaluate the~\acrshort{ghg} emissions in construction and operation, rather than a list of parameters ready to be submitted to the authorities. The numbers also will have different level of uncertainty due to the different level of maturity of the various studies.  Those numbers have therefore to be considered mostly for the purpose of understanding the main drivers of the environmental impact, and eventually whether there is any margin to reduce such impact. 

\subsubsection{\acrshort{ghg} emissions associated with construction and operation}\label{sec:criteria_ghgemissions}

Where available, the following contributions have been considered:
\begin{itemize}
    \item ``Cradle to gate with options'' (\acrshort{lca} modules A1 to A5---i.e.\ raw material supply, transport, manufacture, transport to work site, construction process on the work site---as defined in~\acrshort{en}~15804) for \acrshort{ce} works. The contribution of the modules A1 to A5 represents more than 80\% of the overall \acrshort{ghg} emissions~\cite{clic_ilc_lca_arup,clic_ilc_lca-accel_arup}. Strategies to reduce carbon footprints---such as using alternative cements, enhancing material efficiency, and adopting green energy---are being proposed, broader advancements in cement production and concrete construction are also underway, but are not yet consolidated. The suitability of the materials, their availability in the required quantities and the associated cost are not fully assessed. Therefore, both the \acrshort{ghg} emissions associated to the utilization of presently used materials, based on Portland (CEMI) cement, as well as those considering low-carbon content concrete are presented. This approach is consistent with the roadmap established by the \acrfull{cembureau}~\cite{bib:cembureau}. \acrshort{cembureau}, the leading representative of the cement industry in Europe, aims to reduce the embodied carbon associated with cement production and usage. To achieve this, \acrshort{cembureau} has established a net zero roadmap, targeting carbon neutrality for the European cement industry by 2050. This roadmap aims for a 37\% reduction in \acrshort{ghg} emissions by 2030 and a 78\% reduction by 2040, ultimately reaching carbon neutrality by 2050.
    \item Module A1 to A3 (i.e.\ raw material supply, transport, manufacture) for the main components of the accelerators, their technical infrastructure and the detectors.
    \item The~\acrshort{ghg} emissions due to the generation of the electricity required for the operation of the accelerators (main collider and injectors), experiments (including local data centres), the worldwide computing grid and their general infrastructure and services~(see below).
\end{itemize}

Although the above list is not exhaustive, it provides an indicative assessment of the relative impact of construction and operation of the accelerators. 

The presented estimates are inherently affected by uncertainties related to the carbon factors used in the~\acrshort{lca}.

\paragraph*{Carbon intensity of electricity generation}\label{sec:GHGElectricity}

The evolution of the carbon intensity of electricity generation (expressed in \SI{}{\gram} CO\textsubscript{2} eq.\ per \SI{}{\kilo\watt\hour}) predicted by~\acrfull{rte}, the French transmission system operator for electricity, has been assumed for projects at~\acrshort{cern} for the period spanning 2030 to 2050. Table~\ref{tab:carbonintensityelectricitygenerationfrance} 
lists the projected ranges for the life cycle emissions encompassing both: a) direct emissions related to the combustion of oil, gas and coal and b) upstream emissions related to the extraction and transport of fuels as well as the construction and end-of-life of production and network infrastructure \cite{bib:rte_futurs_2022}.
In its analysis, \acrshort{rte} accounts for the CO\textsubscript{2} eq.\ emissions from cross-border electricity trading, ensuring that imported electricity emissions are proportionally included in France's total electricity generation mix and exported emissions excluded accordingly. The reported values are therefore based purely on the electricity consumption data.

\begin{table}[h!]
    \caption{Life cycle carbon intensity [\SI{}{\gram} CO\textsubscript{2} eq.\ per \SI{}{\kilo\watt\hour}] of electricity consumption. Each lower (upper) value is the average of the minimum (maximum) carbon intensity values for six possible scenarios for the evolution of the French electricity system~\cite{bib:rte_futurs_2022}. The reference years for \acrshort{rte} are 2019 and 2050. A linear interpolation was applied to estimate the values in the intermediate years.}
    \centering
    \footnotesize
    \begin{tabular}{|c|c|c|c|c|c|}
    \hline
        Country | Year  & 2030        & 2035        & 2040        & 2045        & 2050        \\ \hline
        France          & 39--41    & 32--34    & 25--29    & 19--23    & 14--18       \\ \hline                    
    \end{tabular}
    \label{tab:carbonintensityelectricitygenerationfrance}
\end{table}

\subsubsection{Consumption of land}
Data concerning the surface of land required or affected by the construction work are presented.

\subsubsection{Other considerations}
Optimization measures proposed by the various projects to reduce the carbon footprint during construction and operation, reduce electricity consumption and reuse energy, to make responsible use of natural resources and to limit the environmental impact are briefly discussed.

\subsection{Technical readiness and \acrshort{rd} requirements}

Similarly to what has been done by the Snowmass'21 Implementation Task Force~\cite{bib:SnowMass2021}, the aspects related to the technical readiness of the collider proposals are discussed. These include:
\begin{itemize}
    \item \acrfull{trl} of the accelerator component/subsystems representing the highest technical risk~(see Table~\ref{tab:TRL} in Appendix~\ref{sec:TRLtable}).
    \item  Improvement factor to be achieved between the state of the art and the requirements for the most representative and difficult parameter of the considered component/subsystem. This could also be the required unit cost reduction factor assumed in the estimate of the overall cost, or the electricity consumption reduction assumed in the estimate of the overall electricity consumption, or the required increase in the \acrfull{mtbf} to realistically achieve the prospected accelerator complex availability. 
    \item Effort (personnel and material) required to develop and/or validate the technology underlying the above components/subsystems before (series) production can start, considering the scale of the validation (single component to full-scale).
    \item Technically-limited timescale necessary to bring the component or subsystem to be ready for (series) production.
\end{itemize}

\subsection{Construction and installation costs}\label{sec:criteriacost}

 Unless explicitly indicated, the cost estimates for the construction and installation of accelerators and detectors quoted by the proponents of the projects in 2024 prices are provided. They include the cost of the technical components, materials, contracts, services, civil construction and conventional systems and associated implicit labour such as that provided by a company to produce components, unless otherwise explicitly stated.

The cost estimate includes:
\begin{itemize}
    \item construction costs, i.e.\ from project approval to commissioning;
    \item tooling dedicated for production of components; 
    \item reception tests and pre-conditioning of components;
    \item commissioning of technical systems (without beam);
    \item costs related to land, roads, electricity and water connections as well as for administrative processes;
    \item implicit labour (external companies, \acrlong{fsu}---\acrshort{fsu});
    \item total detector costs;
\end{itemize}
while it does not include:
\begin{itemize}
    \item explicit labour provided by the host institution and the collaborating laboratories, this is provided separately and expressed in \acrfull{ftey};
    \item contingency; 
    \item any potential future inflation;
    \item the costs prior to project approval (construction and \acrshort{rd});
    \item off-line computing;
    \item spares, maintenance;
    \item beam commissioning;

\end{itemize}  
unless explicitly stated.

For the projects considered for implementation at \acrshort{cern}, taxes and customs as well as general laboratory infrastructure and services (these being already available) are not included.

We follow the classification of the~\acrfull{aace}~\cite{bib:AACE_18R97} to assign cost uncertainty classes based on their level of definition (or maturity) assessed from the information available (see Table~\ref{tab:AACEcostScheme} in Appendix~\ref{sec:costClasses}).

When costs have not been provided in 2024 prices, the expected relative cost increase considering inflation and in particular the volatility of many commodity prices due to the economic disruptions in the COVID years 2020--2021, has been quoted. This has been derived considering the evolution of the domestic \gls{ppigl}~(\acrshort{ppi}) of the region from where most of the components are going to be built (see Appendix~\ref{sec:PPI}).

Cost estimates are also provided in~\acrfull{chf} using the exchange rates listed in Table~\ref{tab:exchrates} in Appendix~\ref{sec:exchrates}.

Considering that all the projects discussed will likely become an international endeavour, at least partially, \Gls{pppgl}~(\acrshort{ppp}) conversion rates~\cite{bib:OECD_PPPdata} allow to establish a more realistic value of international (e.g.\ in-kind) contributions than currency-exchange rates. 
The exchange rates corrected for the~\acrshort{ppp} used in this report are listed in Table~\ref{tab:ppp_exchrates} in Appendix~\ref{sec:ppp_exchrates}.

The cost breakdown according to the following main domains for the baseline design has been provided, when available:
\begin{itemize}
    \item Main tunnel accelerators
    \item (pre-)Injectors \& transfer lines
    \item \acrfull{ce}
    \item \acrfull{ti}
    \item Experiments.
\end{itemize}

When available the cost of additional upgrades and options are also presented. 

\subsection{Accelerator construction and installation: human resources}
\label{sec:criteriaMetrics_constructionFTE}

An estimate of the personnel (explicit labour) required for the construction and installation of the accelerators and their technical infrastructure, as well as the technical infrastructure of the experiments, is presented. It does not include contractor personnel and the personnel required in the~\acrshort{rd} phase of the project, nor the personnel required for~\acrfull{hw} and beam commissioning. The estimations of the personnel needs by the various projects presented in this document are compared with those obtained by 
a formula correlating explicit labour in~\SI{}{FTEy} to the project \acrfull{capex} (excluding \acrshort{ce} \cite{bib:Seeman_PrivateComm}) developed by the Snowmass'21~\acrfull{itf}~\cite{bib:SnowMass2021}:
\begin{equation}\label{eq:FTEvsCAPEXconstruction}
    Explicit~labour~\mathrm{[FTEy]}=15.7 \times \left(CAPEX~\mathrm{[2010~ MCHF]}\right)^{0.75}
\end{equation}
where \(CAPEX\) is expressed in 2010~M\acrshort{chf}.

\subsection{Project timeline}\label{sec:timelinedef}

The expected project timeline is presented and commented with the main phases and milestones expected for projects of such extent:
\begin{itemize}
    \item Exploratory studies to assess the most critical feasibility issues and define the~\acrshort{rd} programme (this could include the development of critical components and/or systems, or demonstration facilities required to assess the feasibility, in particular if the concept relies on equipment, systems or performance well beyond the state of the art).
    \item Approval of the~\acrshort{rd} programme.
    \item Technical design, construction and validation of the components, systems, and demonstration facilities. 
    \item \acrfull{cdr} providing a conceptual description of the main components of the accelerators and their infrastructure, resulting from the exploratory study and~\acrshort{rd} results above. The~\acrshort{cdr} should allow to produce a first cost estimate (at least Class 5---see Table~\ref{tab:AACEcostScheme}).
    \item Definition of the placement scenario for the collider complex and its infrastructure after assessment of their territorial compatibility: i.e., respecting the territorial requirements and constraints as well as leveraging territorial assets and developing meaningful synergies such as a co-development activities with the relevant regional stakeholders. This implies a broad spectrum of environmental aspects and considerations including: limiting the consumption of land, developing a credible plan for the management of the excavated materials, limiting the consumption of resources such as electricity and water, keeping technical infrastructures as compact as possible, placing the underground structures away from geologically uncertain locations, optimizing access to electricity lines, water supply and treatment, sewage networks and existing major roads and railway lines, avoiding unfavourable topographic and elevation conditions, and some proximity to spoil dumps. An initial review of the implementation scenario with the relevant regional and local stakeholders is included in this phase.
    \item Preliminary implementation. 
    This includes: reserving all land plots potentially affected by the project for the period of more detailed investigations, geological investigations and studies of: territorial constraints and mitigation/compensation measures, road access, railway access, agricultural studies, forest studies, analysis of the environmental initial state, sustainable energy supply concept. The identification of the legal and regulatory conditions under which the project has to be developed, authorized and implemented (e.g.\  plan approval process and environmental evaluation process) will be required during this phase.
    \item Feasibility Report. It includes the comprehensive conceptual design of the collider, the injectors, the experiments, all technical infrastructures that are required to construct and operate them as well as all directly related civil works, the results of the geological investigations, the results of the preliminary territorial implementation studies with the host region/states, and an estimate of the cost (at least Class 4---see Table~\ref{tab:AACEcostScheme}) and of the personnel required.
    \item First version of a \acrfull{tdr} (or pre--\acrshort{tdr}) outlining the detailed technical specifications and design considerations.
    \item Project approval by the relevant governing body or bodies (e.g.\ Council for the \acrshort{cern} projects).
    \item Environmental evaluation and project authorization processes. Depending on the project's host country, different project authorisation processes will apply. In Europe, large-scale development projects are typically subject to a unique authorisation that is obtained as a result of an ``environmental evaluation'' process. 
    \item Main technologies~\acrshort{rd} completion. This refers to the \acrshort{rd} phase on the critical technologies affecting the main parameters of the accelerators and technical infrastructure and/or having an impact on civil engineering, procurement of components requiring large-scale production, acceptance tests and validation or production with long lead times.
    \item Final version of the \acrshort{tdr} including: functional requirements of the project (performance criteria, safety requirements, regulatory compliance, and any other constraints or expectations), component or module specifications and interfaces; material and equipment specifications, quantities, and sourcing information; testing and validation plan; safety and regulatory compliance. The~\acrshort{tdr} can be delivered only after completion of the \acrshort{rd} phase on the critical technologies of the project and it should allow to assess the cost of the project with an uncertainty corresponding to Class 3 to 2, or better (see Table~\ref{tab:AACEcostScheme}).
    \item Installation. It includes the installation of technical infrastructure inside the shafts and in the tunnels as well as the installation of the accelerator components and of the detectors.
    \item Hardware and beam commissioning of the injectors and collider.
\end{itemize}

\subsection{Collider complex operation: resource requirements}\label{sec:criteriaandmetrics_opresources}

The operation costs will be considered only for options proposed for \acrshort{cern}.

The resources required for operation (material and personnel) of the collider accelerator complex (including technical infrastructure), and of the technical infrastructure of the experiments are included. The resources needed for the operation of the rest of the accelerator complex and its technical infrastructure (e.g.\ the \acrshort{ps} and \acrshort{sps} complex) and the technical and scientist support for the experimental detectors and computing are excluded.

For the annual material costs the following items should be considered~\cite{Brunner:2022usy}:

\begin{itemize}
    \item Maintenance and replacement costs estimated with the following methodology:
    \begin{itemize}
       \item 3\% of powering systems (e.g.\ klystrons,~\acrshort{rf} and magnet power converters) or other ``consumable''~\acrshort{capex}, considering the limited lifetime of these parts; 
       \item 1\% of remaining accelerator~\acrshort{hw} (e.g.\ \acrshort{rf} cavities/structures, magnets, vacuum chambers) or other ``fixed installation''~\acrshort{capex}, for replacements;
       \item 5\% of technical infrastructure (e.g.\ cooling, ventilation, electronics and electrical infrastructures, access and safety systems, cryogenics, computer and network infrastructure, control systems, robotics, lifts and transport system)~\acrshort{capex}. It includes the cost of contractors. As an example, the typical ratio between the maintenance cost and the capital cost of~\acrfull{cv} equipment for industries operating for more than 7000 h/year (\acrshort{cern}~\acrshort{cv} installations operate between 7000 and 8000 h/year) is 7\%  but lower (3\%) for the \acrshort{lhc}~\cite{bib:Nonis_CV_maintenance}. The yearly maintenance costs for the \acrshort{lhc} cryogenics represent approximately 3\% of the corresponding initial \acrshort{capex}~\cite{bib:Delikaris_privatecomm}. These include the cost of helium procurement to compensate for the small but inevitable losses and of liquid nitrogen procurement.

    \end{itemize}
    \item Electricity cost, assuming a price of 80~\acrshort{eur}/\SI{}{\mega\watt\hour}.
    This value is subject to a large uncertainty given the rapidly evolving energy market and the long timespan considered here.
\end{itemize}

The personnel (in~\acrshort{fte}) required for the operation of the collider complex and its experiments' technical infrastructure is estimated considering the present personnel required for operating the \acrshort{lhc} by type of equipment. On average the personnel cost matches the material maintenance and replacement costs. The average cost of a staff \acrshort{fte} at \acrshort{cern} is assumed to be 210~k\acrshort{chf}/y. 

These estimates might be pessimistic and represent an upper limit considering the expected impact of modern controls, \acrfull{ai} and robotics by the time of the operation of the future colliders.

%% file: include/03-OptionsatCERN/OptionsatCERN.tex
\section{Future collider options at \acrshort{cern} for operation starting in 2045--2050 }

\subsection{Main parameters and performance}

\paragraph*{\acrshort{clic}}

\begin{figure}[h!]
 \centering
\includegraphics[width=\textwidth]{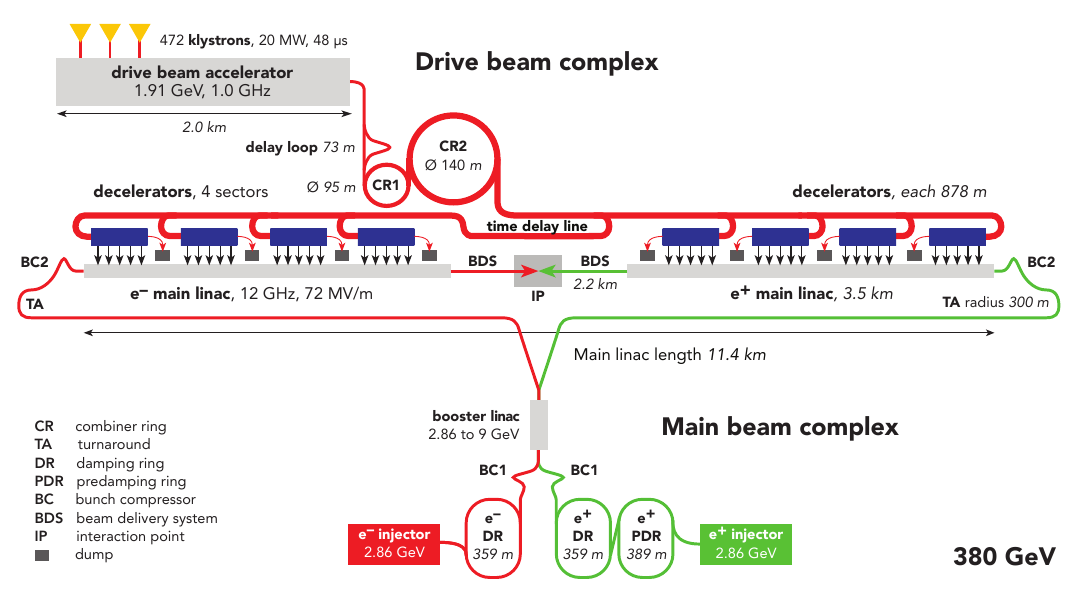}
\caption{Conceptual layout of the \acrshort{clic} collider (\SI{380}{\GeV}).}
\label{fig:CLIC_layout}
\end{figure}

The~\acrfull{clic} is a linear~\epem collider with a proposed initial configuration operating at a \acrshort{com}\ energy of~\SI{380}{\GeV}, with a potential upgrade to~\SI{1.5}{\TeV}~\cite{adli2025compactlineareecollider}. \acrshort{clic} uses a novel two-beam acceleration technique, with~\acrfull{nc} \SI{12}{GHz} accelerating structures operating in the range of~\SI{72}{\mega\volt\per\meter} to~\SI{100}{\mega\volt\per\meter}. A schematic layout of the \acrshort{clic} collider is shown in Fig.~\ref{fig:CLIC_layout}.

The electron beam is produced in a conventional \acrshort{rf} injector, which allows polarisation. The beam emittance is then
reduced in a damping ring.
To create the positron beam, an electron beam is accelerated to \SI{2.86}{\GeV} and sent into a conventional tungsten target generating photons that produce \epem pairs. The positrons are captured and accelerated to \SI{2.86}{\GeV} and their beam emittance is reduced, first in a pre-damping ring and then in a damping ring.
The \acrfull{rtml} accelerates both beams to \SI{9}{\GeV}, compresses their bunch length, and delivers the beams to the main linacs. 
The main linacs accelerate the beams to the collision energy of \SI{190}{GeV}.
The \acrshort{bds} removes transverse tails and off-energy particles with collimators, and compresses the beam to
the small size required at the \acrshort{ip}.
After collision, the beams are transported by the post-collision lines to their respective beam dumps.

The~\acrshort{rf} power for each main linac is provided by a high-current, low-energy~(\SI{1.91}{\GeV}) drive beam that runs parallel to the colliding beam through a sequence of~\acrfull{pets}. The drive beam generates~\acrshort{rf} power in the~\acrshort{pets} and is decelerated; the resulting RF power is then transferred to the accelerating structures using a waveguide network.
The drive beam is generated in a central complex that delivers trains with bunches spaced by~\SI{2.5}{\cm} (i.e. with a repetition frequency of~\SI{12}{\GHz}). This concept reduces the cost compared to powering the structures directly by klystrons, especially for the upgrades to higher energies. The drive-beam and main-beam injectors are located in the \acrshort{cern} Pr\'evessin site.

Ten years of operation are assumed for~\acrshort{clic} in its first, \SI{380}{\GeV} stage~\cite{adli2025compactlineareecollider}). A luminosity ramp-up is foreseen during the first three years, providing successively 10\%, 30\% and 60\% of the design luminosity. The~\acrshort{clic} total programme corresponds to eight years of operation at nominal luminosity. An optional initial stage at \SI{250}{\GeV} is considered. Upgrades at \SIrange[range-units=single,range-phrase={ and }]{550}{1500}{\GeV} are contemplated but require an extension of the tunnel to approximately \SIrange[range-units=single,range-phrase={ and }]{15}{29}{\km}, respectively. Upgrade to even higher energies would require a second drive beam and longer tunnels that might traverse areas with unfavourable geological characteristics.

The main~\acrshort{clic} parameters are listed in Table~\ref{tab:Parameters_CLIC_FCC_LCF}~\cite{adli2025compactlineareecollider}. The repetition frequency of the \acrfull{ml} has been doubled to~\SI{100}{\Hz} as compared to the baseline presented in Ref.~\cite{Brunner:2022usy}. In the updated proposal~\cite{adli2025compactlineareecollider} the \acrshort{clic} \acrshort{ml} distributes the beam pulses between two \acrshort{bds} bringing the beams in collisions at two \acrshort{ip}s where two experiments running in parallel share the luminosity.

Operating the fully installed~\SI{380}{\GeV}~\acrshort{clic} accelerator complex at lower energy $E$ results in lower luminosity, expected to scale roughly as $E^3$~\cite{Brunner:2022usy}, though detailed performance studies have not been performed for operation at the Z-pole with a \SI{380}{\GeV} configuration~\cite{Brunner:2022usy}). 
An installation of just the linac needed for operating at lower energies and an appropriately adapted beam delivery system, would result in higher luminosities though this can only be implemented at the very beginning of the first stage, or during the transition to the second stage. In both cases additional set-up time would be required. At the Z-pole, \SIrange[range-units=single,range-phrase={ and }]{7.5}{135}{\per\femto\barn} can be achieved per year with an unmodified and a modified collider, respectively.

\paragraph*{\acrshort{fccee}}

The~\acrlong{fcc} integrated programme begins with a luminosity-frontier \epem machine (\acrshort{fccee}) spanning \acrshort{com}\ energies from below the Z pole, over the WW pair threshold and ZH production peak, to beyond the top-pair production threshold, later evolving to an energy-frontier hadron collider (\acrshort{fcchh}), as described in Ref.~\cite{
bib:FCCee_ESPP2026}. The presently proposed schedule of the~\acrshort{fcc} programme foresees 14 years of \acrshort{fccee} physics operation 
and 25 years of \acrshort{fcchh} operation, interleaved with a shutdown of 10 years to dismantle the lepton collider and install the hadron collider in the~\acrshort{fcc} tunnel.

\begin{figure}[h!]
 \centering
\includegraphics[width=0.8\textwidth]{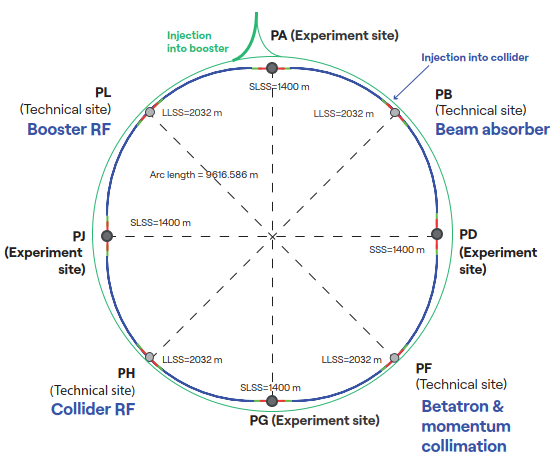}
\caption{Schematic diagram of the \acrshort{fccee} collider.}
\label{fig:FCCee_layout}
\end{figure}

\acrshort{fccee} is a double-ring collider---a schematic view is shown in Fig.~\ref{fig:FCCee_layout}. Its main parameters are listed in Table~\ref{tab:Parameters_CLIC_FCC_LCF}. The two beams collide in  four interaction points. The synchrotron radiation power is restricted to~\SI{50}{\MW} per beam. A full energy booster with an injection energy of~\SI{20}{\GeV}, located in the same tunnel (\SI{5.5}{\m} diameter) as the collider, is used to steadily 
top up the beam currents in the two colliding rings. A \acrfull{he} linac accelerates the electron and positron beams, extracted from a damping ring
at \SI{2.86}{\GeV}, to \SI{20}{\GeV} for injection into the full-energy booster ring. 

From the end of the \acrshort{he} Linac, the electron and positron beams are transferred via new tunnels of \SI{5.5}{\km} total length from surface at \acrshort{cern} Pr\'{e}vessin site directly down to the collider tunnel around \acrshort{fcc} \acrfull{fccpa}. Over the first \SI{4}{\km} both lepton beams can share the same tunnel. Thereafter, the two beams will be split and injected symmetrically around \acrshort{fccpa} at the two ends of the experimental \acrshort{lss}, respectively. 

The injector complex footprint extends over about \SI{1.2}{\km} in length. 
The linacs and damping ring will be installed about \SI{6}{\m} below ground level, using cut and cover techniques. 
Klystron and service buildings will be located on the surface, and connected to the linac tunnels via ducts for waveguides and cables.

The \acrshort{rf} installation in collider and booster is identical for the Z, WW and ZH modes of operation, which is made possible by reverse phase operation at the Z pole. Additional \acrshort{rf} cavities will have to be installed for \ttbar~operation.  
In the collider, during the Z and WW modes, each of the two beams uses half of the installed \acrshort{rf} system, whereas for the ZH and t$\mathrm{\bar t}$ modes with much fewer bunches, both beams go through the full \acrshort{rf} system. 
The switch from beam crossing, in the middle of the 
\acrshort{rf} \acrshort{lss} for Z and WW operation to beam combination and separation at the start and end of this straight for ZH running  is achieved by a combination of magnets and electrostatic separators~\cite{bib:FCCee_ESPP2026}.  
The collision energy can be flexibly changed between these three modes of operation, e.g., year by year, or even faster. 
The klystron galleries are separated from the main tunnel, allowing for installation activities during machine operation.

In total, 14 years of physics operation are foreseen for \acrshort{fccee}. 
Following 1.5 years of beam commissioning, a first operation block contains a four-year run on the Z pole (the first two years of which being at half the design luminosity),  a  two-year run at the WW energy, and a three-year run for ZH Higgs production. 
As mentioned above, the order of the runs in this first block of operation can be modified and even multiple switches between runs can be envisaged. 
Once the first block completed, one year of shutdown will be required for the additional \acrshort{rf} installation, for the second operation block, namely a five-year run around the  $\mathrm{t\bar t}$ threshold.

\paragraph*{\acrfull{lcf}}

\begin{figure}[h!]
\centering
\includegraphics[width=\textwidth]{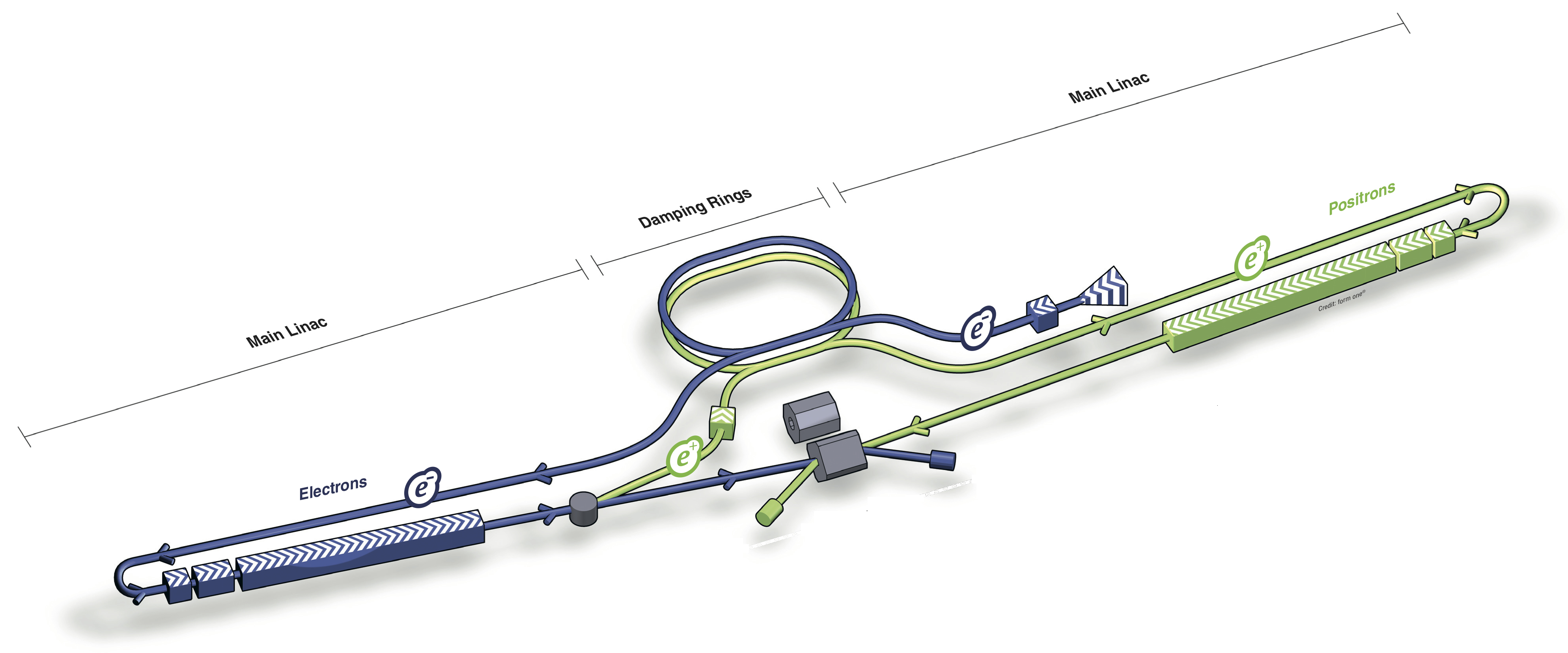}
\caption{Conceptual layout of the \acrshort{lcf}/\acrshort{ilc} collider.}
\label{fig:ILC_layout}
\end{figure}

An additional proposal \cite{LinearCollider:2025lya} for a \epem ~\acrfull{lcf} to be built at \acrshort{cern} has been submitted as input for the update of the \acrshort{espp} 2026. The design of the accelerators is identical to that of the \acrshort{ilc} (see Section~\ref{sec:ILC_intro}), and it is based on bulk-Niobium \acrshort{srf} cavities operated at \SI{1.3}{\GHz} with an accelerating gradient of \SI{31.5}{\mega\volt\per\meter}. The conceptual layout is shown in Fig.~\ref{fig:ILC_layout}. 

The baseline scenario considers a \acrshort{com}\ energy of \SI{250}{\GeV} with the linear collider installed in a tunnel \SI{33.5}{\km} long (to be compared with \SI{20.5}{\km} of the \acrshort{ilc}) offering the possibility to increase the \acrshort{com}\ energy to \SI{550}{\GeV} or higher at later stages and providing operational margin for the baseline \acrshort{com}\ energy by installing some additional cryomodules, if needed. Most of the tunnel would have a circular cross-section of \SI{5.6}{\m} diameter. 
In the baseline configuration---\acrfull{lp} version---the \acrshort{lc} would operate with a repetition rate of \SI{10}{\Hz} (double as compared to that of the \acrshort{ilc}) with the same number of bunches per pulse (1312) serving two experiments installed in two separated \acrshort{ip}s located in the same cavern and sharing the pulses delivered by the collider alternatively. The operation at \SI{10}{\Hz} has implied the increase of the nominal $\mathrm{Q_0}$ of the cavities installed in the cryomodules from $1\times10^{10}$ (\acrshort{ilc} specification) to $2\times10^{10}$ at the same nominal gradient of \SI{31.5}{\mega\volt\per\meter}.

A luminosity ramp-up is foreseen during the first three years, providing successively 10\%, 30\% and 60\% of the design luminosity. An upgrade to higher intensity (doubling the number of bunches per pulse) is also contemplated after five years of operation---\acrfull{fp} version---and it would require a~\acrfull{ls} of one year. The first year after the intensity upgrade would be dedicated to a run at the Z-pole.  The main parameters of the baseline proposals (\acrshort{lp} and \acrshort{fp}) are listed in Table~\ref{tab:Parameters_CLIC_FCC_LCF}. 

The tunnel maximum size has been reduced significantly as compared to the \acrshort{ilc} design (see Section~\ref{sec:ILC_intro}). There is no more a shielding wall separating the main linac and the \acrshort{rf} power sources, similarly to the \acrfull{xfel} tunnel implementation. This has consequences for \acrshort{rf} testing, preventing access to the \acrshort{rf} power sources when the \acrshort{rf} cavities are powered. 
Radiation to electronics might also become an issue in the absence of sufficient shielding.

Figure~\ref{fig:int_lumi_per_TWh} shows the integrated luminosity over electricity consumption for the various $\mathrm{e^+e^-}$ colliders proposed and their upgrades and compares them to \acrshort{lep} and \acrshort{lep2} performance.

\begin{table}[h!]
 \caption{Main parameters for \acrshort{clic}, \acrshort{fccee} and \acrshort{lcf}. The instantaneous and integrated luminosity in-between parentheses includes also the contribution from energies below 99\% of the \acrshort{com}\ energy $\sqrt{s}$.
 }
\centering
\scriptsize
 \begin{threeparttable}
 \begin{tabular}{|l||c|cccc|c|cc|}\hline
& \multicolumn{1}{c|}{\acrshort{clic}} & \multicolumn{4}{c|}{\acrshort{fccee}}  & \multicolumn{3}{c|}{\acrshort{lcf}} \\ 

& \multicolumn{1}{c|}{}& \multicolumn{4}{c|}{}& \acrshort{lp} & \multicolumn{2}{c|}{\acrshort{fp}} \\\hline\hline

Circumference/length collider tunnel~[\SI{}{\km}] & \multicolumn{1}{c|}{12.1} & \multicolumn{4}{c|}{90.7} & \multicolumn{3}{c|}{33.5} \\\hline

Number of experiments (\acrshort{ip}s) & \multicolumn{1}{c|}{2} & \multicolumn{4}{c|}{4}  & \multicolumn{3}{c|}{2}\\\hline

Synchrotron radiation power
per beam~[\SI{}{\MW}] & \multicolumn{1}{c|}{ --- } & \multicolumn{4}{c|}{50} & \multicolumn{3}{c|}{ --- } \\\hline

\acrshort{com}~energy [\SI{}{\GeV}] & \multicolumn{1}{c|}{380} & 91.2 & 160 & 240 & 365 & 250 & 91.2 &250   \\ \hline

Longitudinal polarisation (e$^-$ / e$^+$) [\%] & \multicolumn{1}{c|}{$\pm$80 / 0} &  \multicolumn{4}{c|}{0 / 0\tnotex{tab:Parameters_CLIC_FCC_LCF:1}} & \multicolumn{3}{c|}{$\pm$80 / $\pm$30} \\ \hline

Number of years of operation (total) & 10 & 4 & 2 & 3 & 5 & 5&  1 & 3 \\\hline

Nominal years of operation (equivalent)\tnotex{tab:Parameters_CLIC_FCC_LCF:2} & 8 & 3 & 2 & 3 &  4.5 & 3&  1 & 3 \\\hline

Instantaneous luminosity per~\acrshort{ip}  & \multirow{2}{*}{1.3 (2.2)} & \multirow{2}{*}{140} & \multirow{2}{*}{20} & \multirow{2}{*}{7.5} & \multirow{2}{*}{1.4} & \multirow{2}{*}{1 (1.35)} & \multirow{2}{*}{0.28 (0.28)} & \multirow{2}{*}{2 (2.7)} \\
above $0.99~\sqrt{s}$ (total)~[\(10^{34}\,\SI{}{\per\square\centi\meter\per\second}\)] & & & & & & & & \\\hline

Integrated luminosity above $0.99~\sqrt{s}$ (total) & \multirow{2}{*}{0.32 (0.54)} & \multirow{2}{*}{69} & \multirow{2}{*}{9.6} & \multirow{2}{*}{3.6} &  \multirow{2}{*}{0.67} & \multirow{2}{*}{0.24 (0.32)} &  \multirow{2}{*}{0.067 (0.067)} & \multirow{2}{*}{0.48 (0.65)}\\
over all \acrshort{ip}s per year of nominal operation~[\SI{}{\per\atto\barn}/y]&  &  &  & &  & &  &\\\hline

Integrated luminosity above $0.99~\sqrt{s}$ (total) & \multirow{2}{*}{2.56 (4.4)} & \multirow{2}{*}{205} & \multirow{2}{*}{19.2} & \multirow{2}{*}{10.8} &  \multirow{2}{*}{3.1} & \multirow{2}{*}{0.72 (0.97)} &  \multirow{2}{*}{0.067 (0.067)} & \multirow{2}{*}{ 1.44 (1.94)} \\
 over all \acrshort{ip}s  over the operational phase~[\SI{}{\per\atto\barn}] & & &  & & & & & \\ \hline
 
Peak power consumption~[\SI{}{\mega\watt}] & 166 & 251 & 276 & 297 & 381 & 143 & 123 &182 \\ \hline

Electricity consumption per year   &\multirow{2}{*}{0.82} & \multirow{2}{*}{1.2} & \multirow{2}{*}{1.3} & \multirow{2}{*}{1.4} &
 \multirow{2}{*}{1.9} & \multirow{2}{*}{0.8} & \multirow{2}{*}{0.7} & \multirow{2}{*}{1.0}\\
of nominal operation~[\SI{}{\tera\watt\hour}/y]\tnotex{tab:Parameters_CLIC_FCC_LCF:3}
 & & & & & & & &\\\hline
 
  \end{tabular}
  \begin{tablenotes}
    \item[a] \label{tab:Parameters_CLIC_FCC_LCF:1} Vertical polarisation of at least a few percent for $\sim$200 non-colliding pilot bunches, enabling  precise quasi-continuous measurement of the beam energy. No longitudinal polarisation in the baseline. Residual longitudinal polarisation of colliding bunches should be controlled at the $10^{-5}$ level.
    \item[b] \label{tab:Parameters_CLIC_FCC_LCF:2} This row lists the equivalent number of years of operation at nominal instantaneous luminosity, hence taking into account the luminosity ramp-up.
    \item[c] \label{tab:Parameters_CLIC_FCC_LCF:3} Computed from the peak power consumption and the assumptions on the operational year (see Table~\ref{tab:operational_year}).

  \end{tablenotes}
  \end{threeparttable}
\label{tab:Parameters_CLIC_FCC_LCF}
\end{table}

\subsection{Environmental aspects}
\label{sec:sustainability}

\subsubsection{\acrshort{ghg} emissions associated with construction and operation}

\paragraph*{\acrshort{clic}}

The \acrshort{clic} and \acrshort{ilc} projects have commissioned two \acrshort{lca} \cite{clic_ilc_lca_arup,clic_ilc_lca-accel_arup}. 
These assessments (A1 to A5 modules---see Glossary p.~\pageref{sec:Glossary}) provide an estimate of \acrshort{ghg} emissions expected for \acrshort{clic} construction including:
\begin{itemize}
    \item civil engineering for underground and surface infrastructures for the whole \acrshort{clic} accelerator complex, particularly for the \acrshort{ml} tunnel featuring a \SI{5.6}{\meter} internal diameter;
    \item main hardware components for the whole accelerator complex;
    \item technical infrastructure for the whole accelerator complex;
        \item detectors.
\end{itemize}

The results of the analysis are summarized in Table~\ref{tab:CO2_CLIC_FCC_LCF}. 
Possible reduction opportunities have been identified, particularly for the \acrshort{ce} of the injector linacs. The analysis has revealed that support structures constitute a significant part of the carbon footprint of the accelerator \acrshort{hw} and of the detectors and an appropriate choice of the material and design could lead to an appreciable reduction.

\paragraph*{\acrshort{fccee}}

An \acrshort{lca} for the civil engineering works for the whole \acrshort{fccee} complex has been commissioned~\cite{mauree_2024_13899160}. The \acrshort{lca} for the accelerator systems and \acrshort{ti} components having the largest impact is under way. 
The \acrshort{ghg} emissions for \acrshort{fccee} are presented in Table~\ref{tab:CO2_CLIC_FCC_LCF} and therefore represent a lower bound.

Potential further reductions of the carbon footprint of civil construction have been outlined and might result from a refined design,  use of materials with higher performance and reduced carbon footprint and requiring less steel, local production of construction materials, use of recycled materials and optimisation of the construction process permit. These could further reduce the corresponding minimum value by 5\% for the surface sites and by 16\% for the underground construction works but the implications (e.g.\ on cost, schedule, etc.) have not been assessed.
The estimate of the \acrshort{ghg} emissions for the four detectors is based on the \acrshort{lca} for the \acrshort{clic} detectors provided in Ref.~\cite{clic_ilc_lca-accel_arup}, as we expect similar detector concepts for all \epem collider proposals.

Extensive work is ongoing to improve the efficiency of the \acrshort{rf} sources with the \acrshort{rd} on high-efficiency klystrons that could further increase their efficiency from the presently assumed value of 80\% to 90\%.

\paragraph*{\acrshort{lcf}}
The values of the expected \acrshort{ghg} emissions associated with the project construction (\acrshort{ce}, accelerator \acrshort{hw}, \acrshort{ti}, detectors) are based on the study conducted for the \acrshort{ilc} machine with a \acrshort{com}\ energy of \SI{250}{\GeV}, see Refs.~\cite{clic_ilc_lca_arup,clic_ilc_lca-accel_arup}, and adapted to account for the design differences, in particular a second \acrshort{bds}, a third damping ring for \acrshort{fp} operation, and longer tunnel. No estimate of \acrshort{ghg} emissions associated with the construction of the surface buildings is available. For \acrshort{lcf}, these comprise mainly a detector assembly hall at the \acrshort{ip} location and supply buildings for electricity, cooling and cryogenic plants at the access shaft locations. All injectors and pre-accelerators are located underground in tunnels that are included in the \acrshort{ghg} emission estimate.

\begin{landscape}
\begin{table}[h!]
 \caption{\acrshort{ghg} emissions of the \acrshort{clic}~\cite{clic_ilc_lca-accel_arup}, \acrshort{fccee}~\cite{mauree_2024_13899160} 
 and \acrshort{lcf} construction. The range of values for the \acrshort{ce} works considers the advancements in cement production and concrete constructions as discussed in Section~\ref{sec:criteria_ghgemissions}. 
 In a similar fashion, the range for the \acrshort{ghg} emissions associated with the construction of accelerators, technical infrastructure and detectors provide an indication of the possible reduction achievable by optimisation of the design, choice of the materials and of the construction methods as provided in Ref.~\cite{clic_ilc_lca-accel_arup}. The range of values for carbon intensity of electricity generation is taken from Table~\ref{tab:carbonintensityelectricitygenerationfrance}. The expected electricity consumption of off-line computing is also included. It has been assumed that the electricity consumption per experiment is the same for all the \epem collider proposals. The estimate is based on the expected requirements for \acrshort{fccee}~\cite{Helsens:2021diw} and the energy consumption for the off-line computing for \acrshort{lhc}\cite{Britton:2024rew} and on the expected \acrshort{hw} performance evolution. A range of values is provided to account for different levels of development of the off-line reconstruction algorithms \cite{bib:Campana_OfflineComp}. The present \acrshort{cern} \acrshort{ghg} annual emissions amount to approximately \SI{360}{\kilo\tonne} CO\textsubscript{2} eq./y~\cite{bib:CERNEnvironmentReport}.
 }
\centering
\scriptsize
\begin{threeparttable}
 \begin{tabular}{|L{9cm}||c|c|c|c|c|c|cc|}\hline
 
 & \multicolumn{1}{c|}{\acrshort{clic} (2 \acrshort{ip}s)} & \multicolumn{4}{c|}{\acrshort{fccee} (4 \acrshort{ip}s)}& \multicolumn{3}{c|}{\acrshort{lcf} (2 \acrshort{ip}s)} \\

 & \multicolumn{1}{c|}{} & \multicolumn{4}{c|}{}& \multicolumn{1}{c|}{250 \acrshort{lp}} & \multicolumn{2}{c|}{250 \acrshort{fp}} \\

\acrshort{com}~energy [\SI{}{\GeV}] & 380 & 91.2 & 160 & 240 & 365 & 250 & 91.2 & 250 \\ \hline\hline

\multicolumn{9}{|l|}{\textbf{Civil engineering (A1--A5)}} \\ \hline

\hfill Underground [\SI{}{\kilo\tonne} CO\textsubscript{2} eq.] & \multicolumn{1}{c|}{143--286} & \multicolumn{4}{c|}{480--1000} & \multicolumn{3}{c|}{190--380\tnotex{tab:CO2_CLIC_FCC_LCF:0}} \\ \hline

\hfill Surface sites [\SI{}{\kilo\tonne} CO\textsubscript{2} eq.] & \multicolumn{1}{c|}{59--118} & \multicolumn{4}{c|}{50--184}  & \multicolumn{3}{c|}{N/A} \\ \hline

Total \acrshort{ghg} emissions from \acrshort{ce} [\SI{}{\kilo\tonne} CO\textsubscript{2} eq.] & 202--404 & \multicolumn{4}{c|}{530--1184} & \multicolumn{3}{c|}{>190--380}\\ \hline

Operation period with \acrshort{ce} infrastructure [years] &  \multicolumn{1}{c|}{20\tnotex{tab:CO2_CLIC_FCC_LCF:1}} & \multicolumn{4}{c|}{39\tnotex{tab:CO2_CLIC_FCC_LCF:1}} & \multicolumn{3}{c|}{20\tnotex{tab:CO2_CLIC_FCC_LCF:1}} \\ \hline

\multicolumn{9}{|l|}{\textbf{Accelerators, technical infrastructure, detectors}} \\ \hline

\hfill Accelerators (A1--A3) [\SI{}{\kilo\tonne} CO\textsubscript{2} eq.] & \multicolumn{1}{c|}{105--140 } & \multicolumn{4}{c|}{N/A\tnotex{tab:CO2_CLIC_FCC_LCF:2}} & 169--225 & \multicolumn{2}{c|}{+18--+24\tnotex{tab:CO2_CLIC_FCC_LCF:4}}  \\ \hline

\hfill Technical infrastructure (A1--A3) [\SI{}{\kilo\tonne} CO\textsubscript{2} eq.]  & \multicolumn{1}{c|}{14--19} & \multicolumn{4}{c|}{N/A\tnotex{tab:CO2_CLIC_FCC_LCF:2}} & \multicolumn{3}{c|}{35--46\tnotex{tab:CO2_CLIC_FCC_LCF:7}} \\ \hline

\hfill Detectors (A1--A3) [\SI{}{\kilo\tonne} CO\textsubscript{2} eq.] & \multicolumn{1}{c|}{71--94} & \multicolumn{4}{c|}{142--186\tnotex{tab:CO2_CLIC_FCC_LCF:5}} & \multicolumn{3}{c|}{71--94\tnotex{tab:CO2_CLIC_FCC_LCF:6}}\\ \hline

Total \acrshort{ghg} emissions for accelerator, \acrshort{ti} and detector \acrshort{hw} [\SI{}{\kilo\tonne} CO\textsubscript{2} eq.] & 190--253 & \multicolumn{4}{c|}{N/A} & >275--365 & \multicolumn{2}{c|}{>293--389} \\ \hline

Number of years of physics operation & 10 & 4 & 2 & 3 & 5 & 5 & 1 & 3 \\ \hline

\multicolumn{9}{|l|}{\textbf{Electricity consumption}} \\ \hline

Carbon intensity of electricity generation [\SI{}{\gram} CO\textsubscript{2} eq.\,per \SI{}{\kilo\watt\hour}] & \multicolumn{8}{c|}{14--18} \\ \hline

\hfill Accelerators and detectors [\SI{}{\tera\W\hour}/y] & 0.82 & 1.2 & 1.3 & 1.4 & 1.9 & 0.8 & 0.7 & 1.0 \\ \hline

\hfill Off-line computing [\SI{}{\tera\W\hour}/y] & \multicolumn{1}{c|}{0.07--0.14} & \multicolumn{4}{c|}{0.14--0.28} & \multicolumn{3}{c|}{0.07--0.14} \\\hline 

\acrshort{ghg} emissions/year of physics operation [\SI{}{\kilo\tonne} CO\textsubscript{2} eq./y] & 12--17 & 18--26 & 20--29 & 22--31 & 29--40 & 13--17 & 10--14 & 15--20 \\ \hline

 \end{tabular}
 \begin{tablenotes}
    \item[a] \label{tab:CO2_CLIC_FCC_LCF:0} Based on a rescaling of the \acrshort{ce} carbon footprint of \acrshort{clic}.
    \item[b] \label{tab:CO2_CLIC_FCC_LCF:1} Include the expected physics operation of \acrshort{fcchh} for 25 years, of \acrshort{clic}--\SI{1.5}{\TeV} for 10 years and of \acrshort{lcf}--\SI{550}{\GeV} for 10 years.
    \item[c] \label{tab:CO2_CLIC_FCC_LCF:2} The \acrshort{lca} for the systems having the largest impact is under way.
    \item[d] \label{tab:CO2_CLIC_FCC_LCF:4} \acrshort{ghg} emissions for the injectors' upgrade to run at \acrshort{fp}.
   \item[e] \label{tab:CO2_CLIC_FCC_LCF:7} Based on the estimate for Main Linac services, scaled up to cover injectors and \acrshort{bds}.
   \item[f] \label{tab:CO2_CLIC_FCC_LCF:5} We assume that \acrshort{fccee} will use detectors very similar to the \acrshort{lc} ones, so we scale here the emissions on the base of four detectors for \acrshort{fccee}.
   \item[g] \label{tab:CO2_CLIC_FCC_LCF:6} A1 estimate provided~\cite{clic_ilc_lca-accel_arup} for the main materials in \acrshort{lcf} detectors, not including supports, experimental infrastructure or services. 
 \end{tablenotes}      
 \end{threeparttable}
\label{tab:CO2_CLIC_FCC_LCF}
\end{table}
\end{landscape}

\subsubsection{Consumption of land}

\paragraph*{\acrshort{clic}}
The tunnel of the baseline \SI{380}{\GeV} collider includes four access shafts, two of which are serving the \acrshort{ip} and are located in land attributed to \acrshort{cern} close to the Pr\'evessin site, and two are located at the extremities of the tunnel. An energy upgrade to \SI{1.5}{\TeV} would require the construction of additional four shafts. The orientation and depth of the tunnel are adapted to the maximum operating energy of 1.5~TeV. The exact location of the shafts needs detailed studies in the next phase of the project. The main parameters defining the consumption of land are summarized in Table~\ref{tab:CLIC_FCC_LCFlandoccupation}.

\paragraph*{\acrshort{fccee}}
Compared to an initial scenario of \SI{97.75}{\km} length and twelve surface sites considered in the 2019 Conceptual Design Report~\cite{bib:FCC:2018evy}, the reference scenario is now a \SI{90.65}{\km} long circular collider with eight surface sites. The space requirements for the surface sites have also been gradually reduced. Except for the damping ring (located in land attributed to \acrshort{cern}), the injector is entirely located inside the fenced area of the \acrshort{cern} Pr\'{e}vessin site, thus not consuming any additional land. One experiment site (\acrshort{fccpa}) leverages synergies with the existing \acrshort{lhc} Point 8 site in Ferney-Voltaire.

With eight surface sites, \acrshort{fccee} requires approximately~\SI{0.5}{\kilo\meter\squared} of land, down from~\SI{1.1}{\kilo\meter\squared} in the initial scenario. In total, about \SI{0.26}{\kilo\meter\squared} of this land would be constructed. About ~\SI{0.08}{\kilo\meter\squared} involve areas with ecological value. The indicated values include all the space necessary for a subsequent \acrshort{fcchh} phase and include all the buffer zones required for landscape integration. All sites are located close to roads and highways. A total of about \SI{3}{\km} of new road construction are required. The main parameters defining the consumption of land are summarized in Table~\ref{tab:CLIC_FCC_LCFlandoccupation}.

\paragraph*{\acrshort{lcf}}

The \SI{33.5}{\km} long tunnel includes three access shafts, for accelerator and detectors, at the \acrshort{ip}. These are located in land attributed to \acrshort{cern} close to Pr\'evessin site. Along the tunnel there are four access points on each side. As for \acrshort{clic} the exact location of the shafts needs detailed studies in the next phase of the project. For \acrshort{lcf} it is possible to move the \acrshort{ip} infrastructure inside the \acrshort{cern} fence if this is advantageous, as there are no conflicts with a drive beam. The main parameters defining the consumption of land are summarized in Table~\ref{tab:CLIC_FCC_LCFlandoccupation}.

\begin{table}[h!]
\caption{Area of land consumed by the \acrshort{clic}, \acrshort{fcc} and \acrshort{lcf} projects, i.e.\ lying outside the fenced \acrshort{cern} site. The fraction of the area that lies outside the land attributed to \acrshort{cern} is indicated in parentheses (the rest lies outside the \acrshort{cern} fence but still within \acrshort{cern}-attributed land).}
\scriptsize
\centering
\begin{threeparttable}
\begin{tabular}{|l||c|c|c|}\hline
~ &\acrshort{clic} \SI{380}{\GeV}& \acrshort{fccee} & \acrshort{lcf} \\\hline\hline
Number of new access shafts & 4 &  12 & 11  \\\hline
Number of new surface sites & 3 & 8 & 9 \\\hline
Area of land permanently consumed (fraction outside \acrshort{cern}-attributed land) ~[\SI{}{\square\km}] & 0.37 (14\%) & 0.5 (100\%) & 0.34 (62\%) \\\hline
Area of surface constructions (fraction outside \acrshort{cern}-attributed land)~[\SI{}{\square\km}] & 0.118 (5\%) & 0.26 (100\%) & 0.073 (55\%) \\ \hline
\end{tabular}
\end{threeparttable}

\label{tab:CLIC_FCC_LCFlandoccupation}
\end{table}

\subsubsection{Other measures to reduce the environmental impact during construction and  operation}

Currently, approximately 90\% of \acrshort{cern}’s direct~\acrshort{ghg} emissions come from its experiments. These use a wide range of gas mixtures for particle detection and detector cooling, including fluorinated gases (F-gases) which have a high \gls{gwpgl}~(\acrshort{gwp}) and therefore account for about 78\% of \acrshort{cern} direct emissions~\cite{bib:CERNEnvironmentReport}. Design of detectors with improved gas tightness and \acrshort{rd} towards detector gases with lower~\acrshort{gwp} is pursued~\cite{bib:DetectorRDRoadmap} 
within the \acrfull{drd} Collaborations 
to reduce significantly and/or possibly eliminate this contribution.

\paragraph*{\acrshort{clic}}

Means of reducing the \acrshort{ghg} emission and cost associated with the electricity required to operate the complex by optimizing the machine schedule have been considered and the possibility of reusing the waste heat has been explored~\cite{CLIC_Fraunhofer}.
A more comprehensive approach, as followed by \acrshort{fcc}, would be part of future environmental studies.

\paragraph*{\acrshort{fccee}}

The \acrshort{fcc} project takes a comprehensive approach to reduce its environmental impact across construction, operation, and long-term sustainability. Key strategies outlined in Ref.~\cite{Benedikt:2928194} include, among others: 
\begin{itemize}
    \item reuse of excavated materials in construction, agriculture, and ecological restoration;
    \item waste heat recovery;
    \item gas management in detectors;
    \item ecosystem and biodiversity protection with site-specific plans to minimize disruption to local wildlife and ecosystems;
    \item water conservation to maintain the operational demands within the current \acrshort{cern} levels.
\end{itemize}

Following an analysis of the demand potential for low-grade heat in the vicinity of the surface sites~\cite{ginger_burgeap_2024_11192000,ginger_burgeap_2024_11192180}, \acrshort{fcc} will integrate the recovery and supply of heat. 
The concept has been successfully demonstrated with an installation in Ferney-Voltaire (France) in relation to the \acrshort{lhc} Point~8. The supply of heat allows reducing the consumption of electricity and water of the cooling system because of the lower demand for operating the evaporation towers.

Depending on the mode of operation, between \SIrange[range-units=single,range-phrase={ and }]{300}{424}{\giga\watt\hour} of heat can be supplied per year on a perimeter of \SI{5}{\km} with a minimum of \SI{60}{\giga\watt\hour} during shutdown periods. Supplying these amounts of heat directly results in a reduction of 24\% to 30\% of the expected water intake per year, which varies from \SIrange[range-units=single,range-phrase={ to }]{1.5}{2.8}{million~\cubic\m} according to the mode of operation, and of the power consumption of the cooling circuits.

\subsection{Technical readiness and \acrshort{rd} requirements}\label{sec:RandD}

\paragraph*{\acrshort{clic}}

The recent review of the \acrshort{ldg} accelerator \acrshort{rd} programme has analysed, among others, the status of the \acrshort{rd} of the main \acrshort{rf} components of the large accelerator projects~\cite{bib:LDG_RD_review_RF, bib:LDG_RD_review_ESPP2026}. The \acrshort{clic} design relies on the achievement of gradients >\SI{72}{\mega\volt\per\meter} over a large number of X-band high-gradient structures. This gradient has been exceeded, but on a limited number of structures. Demonstration of reproducibility of performance over larger samples and with acceptable conditioning times as well as industrialization of the production are key goals for the next phase of the project.

The drive-beam linac is the main post contributing to the overall \acrshort{clic} electricity consumption. An efficiency of 82\% for the L-band klystrons powering it is targeted; 70\% has been reached with industrial prototypes, but prototyping of new cost-effective designs for higher efficiency is needed~\cite{bib:LDG_RD_review_ESPP2026,bib:Nuria_privatecomm}.

\acrshort{lc}s require larger rates of positron production as compared to circular colliders where the beams are stored and only losses need to be compensated for. \acrshort{clic} average positron intensity is a factor 16 higher than that required for \acrshort{fccee}, which exceeds by a factor two the maximum intensity achieved so far at \acrshort{slc}. Demonstration of such performance for \acrshort{clic} is for the time being at component level (\acrshort{trl}~5)~\cite{bib:SnowMass2021}.

\acrshort{clic} performance relies on the capability to preserve the emittance of the beams delivered by the damping rings all through the main linac and the \acrshort{bds} and to minimize the chromatic aberrations resulting from the extremely small optics waist at the \acrshort{ip}. This implies achieving an excellent control of the wakefields in the accelerator and of the optics at the final focus to maintain in collision nanometric vertical beam sizes (a factor 4 smaller than for \acrshort{lcf} and at least 20 than for \acrshort{fccee}) at the \acrshort{ip} in a reproducible fashion demanding complex feed-forward systems to minimize the effects of ground motion and temperature variations. Though quite some experience has been gained at the \acrfull{fftb} and \acrfull{facet} at the \acrfull{slac} and at the \acrfull{atf} at \acrshort{kek}, the demonstration of wakefield minimization in long accelerating sections and beam spot size control and stability at the \acrshort{ip} under realistic conditions remain major challenges that require further validation with beam tests for the \acrshort{clic} proposal. These challenges and the \acrshort{trl} levels of the associated technologies are discussed in Ref.~\cite{bib:SnowMass2021}.

The five most critical/high-risk systems/components for which \acrshort{rd} effort is still required before reaching a sufficient level of confidence for construction are listed in Table~\ref{tab:CLICTRL} in order of increasing \acrshort{trl}.

 \begin{table}[h!]
   \caption{Technical readiness and \acrshort{rd} requirements for~\acrshort{clic}. The colour code corresponds to the code used in Ref.~\cite{bib:LDG_RD_review_ESPP2026}.
  } 
 \centering
 \scriptsize
 \begin{tabular}{|l|c|C{2cm}|c|c|c|c|}
 \hline

 \multirow{3}{*}{Component/Sub-system} & \multirow{3}{*}{\acrshort{trl}} & \multirow{2}{*}{Main parameter}   & \multirow{3}{*}{Improvement factor} & \multicolumn{3}{c|}{~\acrshort{rd} effort}  \\ 
 ~&  & \multirow{2}{*}{to be improved} & & Personnel  & Material  & Timescale \\
~ & ~ &  & ~ & [\acrshort{ftey}] & [M\acrshort{chf}] &  [years] \\ \hline\hline

High-efficiency klystrons for drive-beam linac &  \cellcolor{yellow}4 & Efficiency& 1.2  & 6 & 3 & 4 \\ \hline 

X-band high-gradient structures & \cellcolor{yellow}5&  Industry yield / gradient & Verification & 36 & 12 & 6\\ \hline

Positron source  &\cellcolor{yellow}5 & e$^+$ rate & 33  & 8 & 3 & 6\\ \hline

\acrshort{ip} spot size/stability  & \cellcolor{yellow}5 & \acrshort{bds} & 1.3 &  18 & 4 & 6\\ \hline

Emittance preservation  & \cellcolor{yellow}6 & \acrshort{ml} & 1.3 & 6 & 2 & 4\\ \hline

 \end{tabular}
 \label{tab:CLICTRL}
 \end{table}

The \acrshort{rd} items described above are part of a larger six year preparation phase programme, which also covers site specific studies and further design and parameter optimisation, described in Ref.~\cite{adli2025compactlineareecollider}. The klystron efficiency has a softer impact than the others.
The performance improvements that can be expected within the \acrshort{rd} programme described cover improvements in damping emittances and emittance preservation methodology, as well as a potential shortening of the beam-delivery system.  These improvements are currently foreseen to increase the robustness of the design, but can also, combined with technical studies, potentially reduce costs and power consumption of the collider in the 3--10\% range.

\paragraph*{\acrshort{fccee}}

The power consumption estimated for \acrshort{fccee} relies on the availability of high-efficiency klystrons with efficiencies of 85\% in the range of frequency from \SIrange[range-units=single]{400}{800}{\MHz}. Significant progress has been made in this technology, also boosted by the need of replacing the \acrshort{lhc} klystrons, indicating a \acrshort{trl} 4~\cite{bib:Nuria_privatecomm,bib:LDG_RD_review_ESPP2026,bib:LDG_RD_review_RF}. Further improvement in efficiency could be obtained with multi-beam Tristrons (not considered in the baseline for the time being) though these have a lower \acrshort{trl} (2 to 3)~\cite{bib:LDG_RD_review_ESPP2026}.
The analysis presented in Refs.~\cite{bib:LDG_RD_review_ESPP2026,bib:LDG_RD_review_RF} indicates a \acrshort{trl} 4 for both the \SI{800}{\MHz} \acrshort{rf} cryomodules operated at \SI{2}{\kelvin} and the \SI{400}{\MHz} \acrshort{rf} cryomodule operated at \SI{4.5}{\kelvin}.
The positron source performance is very close to the \acrshort{slc} positron source one and significantly less demanding than those of the \acrshort{clic} and \acrshort{lcf} sources. 
The \acrshort{fccee} vacuum system performance relies on distributed \acrfull{neg} pumping with fast conditioning of localised synchrotron radiation absorbers. The \acrshort{rd} of several critical technologies to fulfil \acrshort{fccee} requirements and/or to reduce costs is in full swing for the pre-\acrshort{tdr} phase. The main challenge is to minimise costs by making these technologies scalable, industrially viable, and capable of operating with low maintenance and operational costs.
The five most critical/high risk systems/components for which \acrshort{rd} effort is still required before reaching a sufficient level of confidence for construction are listed in Table~\ref{tab:FCCeeTRL} in order of increasing \acrshort{trl}.

\begin{table}[h!]
  \caption{Technical readiness and \acrshort{rd} requirements for~\acrshort{fccee}. The colour code corresponds to the code used in Ref.~\cite{bib:LDG_RD_review_ESPP2026}.
}
\centering
\scriptsize
\begin{threeparttable}
\begin{tabular}{|l|c|C{2cm}|c|c|c|c|}
 \hline
 \multirow{3}{*}{Component/Sub-system} & \multirow{3}{*}{\acrshort{trl}} & \multirow{2}{*}{Main parameter}   & \multirow{3}{*}{Improvement factor} & \multicolumn{3}{c|}{~\acrshort{rd} effort}  \\ 
 ~&  & \multirow{2}{*}{to be improved} & & Personnel  & Material  & Timescale \\
~ & ~ &  & ~ & [\acrshort{ftey}] & [M\acrshort{chf}] &  [years] \\ \hline\hline
\acrshort{rf} power sources & \cellcolor{yellow}4 & Efficiency  & 1.3 & $\sim$10 & $\sim$6 & 5--6 \\ \hline
\SI{800}{\MHz} \acrshort{rf} cavities\tnotex{tab:FCCeeTRL:1} & \cellcolor{yellow}4 & $\mathrm{Q_0}$ & 6
& $\sim$10 & 6 & 5--6 \\ \hline
\SI{400}{\MHz} \acrshort{rf} cavities\tnotex{tab:FCCeeTRL:2} & \cellcolor{yellow}4 & Gradient & 2 & $\sim$12 & 6 & 5--6 \\ \hline
Vacuum system & \cellcolor{yellow}6 & Cost \& industrialisation & 2 (on cost) & 10 & 10 & 5--6 \\ \hline
Positron source  & \cellcolor{yellow}6 & e$^+$ rate  & 2  & 20 & 2 & 4  \\ \hline
 \end{tabular}
\begin{tablenotes}
 \scriptsize
    \item[a] \label{tab:FCCeeTRL:1} Bulk-Nb \SI{800}{\MHz} \acrshort{rf} cavities operated at \SI{22.5}{\mega\volt\per\m} with $\mathrm{Q_0}=3.5\cdot10^{10}$~\cite{bib:FCCee_ESPP2026}. 
     \item[b] \label{tab:FCCeeTRL:2} Nb-coated copper \SI{400}{\MHz} \acrshort{rf} cavities operated at \SI{11.8}{\mega\volt\per\m} with $\mathrm{Q_0}=2.7\cdot10^{9}$~\cite{bib:FCCee_ESPP2026}. 
\end{tablenotes}
\end{threeparttable}
 \label{tab:FCCeeTRL}
 \end{table}

\paragraph*{\acrshort{lcf}}

As compared to~\acrshort{ilc} the \SI{1.3}{\GHz} cavities are assumed to reach $\mathrm{Q_0} \geq 2\times10^{10}$ in the cryomodules, twice the value considered for \acrshort{ilc}, in order to double the operational repetition rate. Given the large number of items of this type, this must be re-assessed after industrial level studies for the proposed new surface treatments are available~\cite{bib:LCF4CERN_backup,bib:SnowMass2021}. The \acrshort{lcf} baseline design foresees a 6\% margin of additional cryomodules
in the \SI{250}{\GeV} configuration, and would reach \SI{240}{\GeV} c.o.m. energy with an average gradient of \SI{28.5}{\mega\volt\per\meter} in the main linac. Furthermore, the associated risk can be alleviated by installing larger number of modules than the minimum required based on the experience during production, or by an increase of the cryogenic cooling. This is possible thanks to the length of the tunnel designed for a \acrshort{com}\ energy of \SI{550}{\GeV}.

As mentioned earlier, linear colliders have demanding positron rates, in particular the \acrshort{lcf} \acrshort{fp} requires a factor around 90 higher production rate as compared to that achieved at the \acrshort{slc}.

Emittance preservation and \acrshort{ip} spot/size stability remain critical aspects also for \acrshort{lcf}, though to a lesser extent as compared to \acrshort{clic}, due to the larger aperture of the L-band \acrshort{rf} cavities. \acrshort{bds}s for two experiments were extensively studied in the past but need to be improved and adapted to the machine layout at \acrshort{cern}. Beam sharing between two experiments provides additional reproducibility/stability challenges for the beam position at the \acrshort{ip}. The \acrshort{lcf} requirements have been partially demonstrated at the \acrshort{atf} at \acrshort{kek}.

The main dumps and their entry windows are a critical item as they need to be able to absorb \SI{17}{\MW} beam power. Their design, based on the \SI{2}{\MW} \acrshort{slac} water dump, needs to be validated. 

An efficiency of 80\% is assumed for the klystrons powering the main linac \acrshort{rf} cavities, based on the same two-stage design as considered for \acrshort{clic} and \acrshort{fccee}. Present devices have an efficiency of about 65\%. The \acrshort{trl} for the required power sources has been assessed to be 4~\cite{bib:LDG_RD_review_RF, bib:Nuria_privatecomm}. Failure to achieve the target efficiency will imply more powerful modulators and cooling systems and therefore higher costs and larger power consumption though these are considered to remain in the percent range.

Table~\ref{tab:LCFTRL} summarizes the most critical elements above described.

\begin{table}[h!]
  \caption{Technical readiness and \acrshort{rd} requirements for~\acrshort{lcf}. The colour code corresponds to the code used in  Ref.~\cite{bib:LDG_RD_review_ESPP2026}.
}
\centering
\scriptsize
\begin{threeparttable}
\begin{tabular}{|l|c|C{2cm}|c|c|c|c|}
 \hline
 \multirow{3}{*}{Component/Sub-system} & \multirow{3}{*}{\acrshort{trl}} & \multirow{2}{*}{Main parameter}   & \multirow{3}{*}{Improvement factor} & \multicolumn{3}{c|}{~\acrshort{rd} effort}  \\ 
 ~&  & \multirow{2}{*}{to be improved} & & Personnel  & Material  & Timescale \\
~ & ~ &  & ~ & [\acrshort{ftey}] & [M\acrshort{chf}] &  [years] \\ \hline\hline
\acrshort{rf} power sources & \cellcolor{yellow}4 & Efficiency  & 1.25 & $ $10 & 3  & 3 \\ \hline
\multirow{3}{*}{Positron source}  & \cellcolor{yellow}{} & \multirow{1}{*}{Target cooling} & \multirow{3}{*}{N/A} & \multirow{3}{*}{15} &\multirow{3}{*}{5} & \multirow{3}{*}{3} \\ 
& \cellcolor{yellow}{5} & Pulsed solenoid & & & & \\
 & \cellcolor{yellow}{} & peak field & & & & \\ \hline
Main dump & \cellcolor{yellow}{5} & Maximum power & 8 & 4 & 2 & 5 \\ \hline
\acrshort{ip} spot size/stability  & \cellcolor{yellow}6 & Vertical beam size at nominal bunch population & 10\% & 18 & 4 & 6\\ \hline
\SI{1.3}{\GHz} \acrshort{rf} cavities/cryomodules  & \cellcolor{green}7 & $\mathrm{Q_0}$ & 2 & 30 & 10 & 3 \\ \hline

 \end{tabular}
\end{threeparttable}
 \label{tab:LCFTRL}
 \end{table}

\subsection{Construction and installation costs}

\paragraph*{\acrshort{clic}}

The capital cost for construction of 
\acrshort{clic} is summarised in Table~\ref{tab:CLIC_cost}.
This cost includes construction of the entire new infrastructure and all equipment for operation up to \SI{380}{\GeV}. Collider includes main linac and decelerators in the main tunnel as well as \acrfull{bds}, final focus and post collision lines and dumps. The cost of the experiments includes the full cost of the detector and the required infrastructure.

 \begin{table}[h!]
    \caption{Cost summary table in 2024 M\acrshort{chf} for the \acrshort{clic} accelerator operating up to a maximum~\acrshort{com}\ energy of~\SI{380}{\GeV}---with two experiments~\cite{adli2025compactlineareecollider}. Cost classes are defined in Table~\ref{tab:AACEcostScheme}. The total cost of the experiments is listed here---it is expected that \acrshort{cern} contribution to the detector construction plus host laboratory responsibility is 146 M\acrshort{chf}. Costs related to land, roads, electricity and water connections as well as for administrative processes and underground rights-of-way are not included, though these are expected to represent only a minor contribution to the total cost (at the percent level).
  }
  \centering
  \scriptsize
  \begin{threeparttable}
  \begin{tabular}{|l|c|c|}
    \hline
    \multirow{2}{*}{Domain}& \multicolumn{2}{c|}{\acrshort{clic} -- \SI{380}{\GeV}}\\
     & Cost & Cost class\\
    ~& [M\acrshort{chf}] & ~ \\ \hline
    Collider & 2471 & 3--4  \\ \hline
    Main beam production and transfer lines & 1046 & 3--4  \\ \hline
    Drive beam production and transfer lines & 1060 & 3--4 \\ \hline
    Civil engineering & 1403\tnotex{tab:CLIC_cost:1} & 4 \\ \hline
    Technical infrastructures & 1361 & 3--4 \\ \hline
    Experiments & 795 & 4 \\ \hline
    \textbf{TOTAL} & \textbf{8136} &\\ \hline
  \end{tabular}
  \begin{tablenotes}
    \item[a] \label{tab:CLIC_cost:1} The cost for spoil removal (estimated at 100~M\acrshort{chf}) is included here.
\end{tablenotes}
\end{threeparttable}
  \label{tab:CLIC_cost}
\end{table}

\paragraph*{\acrshort{fccee}}

The main cost items for the construction of 
\acrshort{fccee} are listed in \autoref{tab:FCCee_cost}.
This cost includes construction of the entire new infrastructure and all equipment for operation up to \SI{240}{\GeV}. Operation at higher energies will require later installation of additional \acrshort{rf} cavities and associated cryogenic cooling infrastructure with a corresponding total cost of 1260 M\acrshort{chf}.
The cost estimate has not varied significantly as compared to that presented in the 2024 mid-term report
. The latter was the subject of an extensive external review whose conclusions and recommendations have been reviewed by the \acrshort{cern} Finance committee and Council.

\begin{table}[h!]
  \caption{Cost summary table in 2024 M\acrshort{chf} for the \acrshort{fccee} accelerator operating up to a maximum~\acrshort{com}\ energy of~\SI{240}{\GeV}---with four experiments~\cite{bib:FCCee_ESPP2026}. Cost classes are defined in Table~\ref{tab:AACEcostScheme}. The total cost of the experiments is listed here---it is expected that \acrshort{cern} contribution to the detector construction plus host laboratory responsibility is 292 M\acrshort{chf}.
  }
  \centering
  \scriptsize
  \begin{tabular}{|l|c|c|}
    \hline
    \multirow{2}{*}{Domain}& \multicolumn{2}{c|}{\acrshort{fccee} -- \SI{240}{\GeV}}\\
     & Cost & Cost class\\
    ~& [M\acrshort{chf}] & ~ \\ \hline
    Booster and collider  & 4140  & 3  \\ \hline
    Pre-injectors and transfer lines & 590 & 3 \\ \hline
    Civil engineering  & 6160   &  3 \\ \hline
    Technical infrastructures & 2840  & 3 \\ \hline
    Experiments & 1590 & 4 \\ \hline
    \textbf{TOTAL} & \textbf{15320} &\\ \hline
  \end{tabular}
  \label{tab:FCCee_cost}
\end{table}

\paragraph*{\acrshort{lcf}}

The \acrshort{lcf} cost estimates are based on the \acrshort{ilc} \acrshort{tdr}~\cite{bib:ILC_TDR_vol3_II}, updated in 2017~\cite{Evans:2017rvt} and in 2024 by the 
\acrfull{idt}~\cite{bib:IDT_ESPP2026, bib:LCF4CERN_ESPP2026,bib:Yamamoto_cost_update} and on a new evaluation of the construction costs for the \acrshort{cern} site from 2025~\cite{bib:CERN_CFS_Cost_Update}.
The updated 2024 cost estimate was reviewed by an international expert panel in December 2024~\cite{bib:IDT_ESPP2026}.
The total cost for the baseline \SI{250}{\GeV} \acrshort{lcf} \acrshort{lp} configuration amounts to \SI{9287}{M\acrshort{chf}} in 2024 prices, as presented in Table~\ref{tab:LCF_cost}, the additional cost of the upgrade to \acrshort{fp} is \SI{770}{M\acrshort{chf}}.

\begin{table}[h!]
   \caption{Cost summary table in 2024 M\acrshort{chf} for the \acrshort{lcf} accelerator operating up to a maximum~\acrshort{com}\ energy of~\SI{250}{\GeV}---with two experiments~\cite{bib:LCF4CERN_backup}. Cost classes are defined in Table~\ref{tab:AACEcostScheme}. Land acquisition and site activation (external roads, water supplies, power lines) are not costed, though these items are expected to represent a minor contribution to the total cost (at the percent level). The total cost of the experiments is listed here---it is expected that \acrshort{cern} contribution to the detector construction plus host laboratory responsibility is 146 M\acrshort{chf}.
  }
  \centering
  \scriptsize
  \begin{threeparttable}
  \begin{tabular}{|l|c|c|}
    \hline
    \multirow{2}{*}{Domain}& \multicolumn{2}{c|}{\acrshort{lcf} \acrshort{lp} -- \SI{250}{\GeV}}\\
     & Cost & Cost class\\
    ~& [M\acrshort{chf}] & ~ \\ \hline
    Collider  & 3864  & 3  \\ \hline
    Injectors and transfer lines    & 1181 & 3 \\ \hline
    Civil Engineering               & 2338\tnotex{tab:LCF_cost:1}    &  4 \\ \hline
    Technical infrastructures       & 1109  & 4 \\ \hline
    Experiments & 795 & 4 \\ \hline
    \textbf{TOTAL} & \textbf{9287} &\\ \hline
  \end{tabular}
  \begin{tablenotes}
 \scriptsize
    \item[a] \label{tab:LCF_cost:1} The cost for spoil removal (estimated at 200 M\acrshort{chf}) is included here.
\end{tablenotes}
\end{threeparttable}
  \label{tab:LCF_cost}
\end{table}

\subsection{Accelerator construction and installation: human resources}\label{sec:HR_resources}

\paragraph*{\acrshort{clic}}

According to Eq.~\eqref{eq:FTEvsCAPEXconstruction}, approximately \SI{10500}{\acrshort{ftey}} of explicit labour would be required for the construction of \acrshort{clic} and its technical infrastructure.

\paragraph*{\acrshort{fccee}}

An integrated total of approximately \SI{15000}{\acrshort{ftey}} is required for the accelerator construction and installation. This estimate was obtained through various approaches that converge on a similar result:
\begin{itemize}
    \item an overall fit based on scaling from comparable accelerator infrastructures;
    \item scaling by individual equipment type;
    \item a bottom-up approach.
\end{itemize}

According to Eq.~\eqref{eq:FTEvsCAPEXconstruction}, approximately \SI{13100}{\acrshort{ftey}} would be required for the construction of the accelerators and their technical infrastructure (the latter including that of the experiments).

\paragraph*{\acrshort{lcf}}

The personnel required for the construction of the \SI{250}{GeV} baseline \acrshort{lp} configuration is estimated in Ref.~\cite{Evans:2017rvt} and it amounts to \SI{10120}{\acrshort{ftey}}, to be compared with \SI{10900}{\acrshort{ftey}} estimated according to Eq.~\eqref{eq:FTEvsCAPEXconstruction}.

\subsection{Project timeline}

\paragraph*{\acrshort{fccee}}

Preparatory placement studies for the \acrshort{fcc} are well advanced. The further timeline is mainly determined by the preparation of the civil engineering (subsurface investigations, civil engineering design and tendering, and the project authorisation processes with the host states, planned to advance in parallel), and, afterwards, by the civil construction and the subsequent installation of technical infrastructure and accelerator components. Key dates  are summarized in Table \ref{tab:timeline_FCCee}, including work already accomplished.

\begin{table}[h!]
 \caption{Timeline for design, construction, and operation of \acrshort{fccee}.} 
\centering
\footnotesize
\begin{threeparttable}
\begin{tabular}{|L{10cm}||C{4cm}|}\hline
Milestone & \acrshort{fccee} \\ \hline\hline
Conceptual Design Study & 2014--2018 \\
Definition of the placement scenario
& 2022 \\
Preliminary implementation with the Host states & 2024--2025 \\
Feasibility Report ready & 2025 \\
Earliest Project Approval\tnotex{tab:timeline_FCCee:2} & 2028 \\
Environmental evaluation \& project authorisation processes & 2026--2031 \\
Main technologies \acrshort{rd} completion\tnotex{tab:timeline_FCCee:3}& 2031 \\ 
Technical Design Report ready\tnotex{tab:timeline_FCCee:8} & 2032 \\
Civil engineering & 2033--2041 \\
\acrshort{ti} installation & 2039--2043 \\
Accelerator installation & 2041--2045 \\
\acrshort{hw} commissioning & 2042--mid 2046 \\
Beam commissioning -- collider & mid 2046--2047 \\
Physics operation start &  2048 \\
\hline
\end{tabular}
 \begin{tablenotes}
 \scriptsize
    \item[a] \label{tab:timeline_FCCee:2} This is the approval by the Council and it must follow the process of the Update of the   European Strategy for Particle Physics---assumed to take at least two years once the feasibility report is delivered.
    \item[b] \label{tab:timeline_FCCee:3} Based on the information in Table~\ref{tab:FCCeeTRL}.
    \item[c] \label{tab:timeline_FCCee:8}The \acrshort{tdr} can be delivered only once the environmental evaluation, project authorization processes and main technologies \acrshort{rd} is completed.
 \end{tablenotes}
 \end{threeparttable}
\label{tab:timeline_FCCee}
\end{table}

\paragraph*{\acrshort{clic}}

A timeline for the implementation of the \acrshort{clic} project has been outlined in Ref.~\cite{adli2025compactlineareecollider}. This is summarized in Table~\ref{tab:timeline_CLIC_LCF} where the main phases are sketched, including those already achieved. The timeline suggests initial preparatory phase studies to finalise the collider design and optimise its placement, in parallel with technical studies. It is followed by a extended preparatory phase for implementation studies with the host states, environmental studies and industrialization.  
The technically-limited seven year construction timeline assumes~\cite{adli2025compactlineareecollider}:
\begin{itemize}
    \item three years for the civil engineering works (drive beam, injectors, collider, surface infrastructures);
    \item four years for the installation of the drive beam, injector and collider \acrshort{hw};
    \item one year of \acrshort{hw} commissioning of the collider followed by one year of beam commissioning of the collider.
\end{itemize}
Some of these activities overlap in time as they affect different areas (e.g. injectors and collider).

While the initial phase requires limited resources and can be implemented as a potential outcome of the strategy update, specific decisions processes are needed for entering in an extended preparatory phase and construction. This is indicated in Table~\ref{tab:timeline_CLIC_LCF} by starting these phases at T$_0$ and T$_1$, and indicating their length with respect to these starting times. 
To accommodate the fact the transition between the phases might require some time, and also to avoid the clear conflict between operating \acrshort{hllhc} and beam commissioning of a new collider, the construction phase is increased in the table to take ten years before beam commissioning, instead of the technically-limited seven years.

\begin{table}[h!]
 \caption{Timeline of essential development and construction steps of the \acrshort{clic} and \acrshort{lcf} projects. T$_0$ is determined by a process in 2028--29 to validate the progress and promise of the project for a further development towards implementation. T$_1$ following Preparation Phase 2 will be determined by the processes needed, by the \acrshort{cern} Council and with host-states, for project approval and to start construction. The construction phase is extended as explained in the text with respect to the technically-limited schedules to allow a transfer time into construction, and to avoid the resource conflict between \acrshort{hllhc} operation and initiating beam commissioning for a next collider. While many aspects can initially be studied together for the two options, it will be necessary to prioritize one of the two \acrshort{lc} options for detailed implementation studies during Phase 1.
 }
\footnotesize
\centering
\begin{tabular}{|L{8cm}||C{3cm}|C{3cm}|}\hline
Milestone & \acrshort{clic} & \acrshort{lcf} \\ \hline\hline
Conceptual/Reference Design Report & 2004--2012 & 2002--2007\\
Site-independent \acrshort{tdr} for \acrshort{ilc} \SI{500}{\GeV} & & 2007--2013 \\ 
Project Implementation Plan, Readiness Report  & 2013--2025 & \\
\acrshort{ilc} 250 GeV reports and Prelab planning &       & 2013--2025 \\

\hline
\hline
Project Preparation Phase 1 & \multicolumn{2}{|c|}{2026--2028} \\
\hline
\multicolumn{3}{|c|}{Definition of the placement scenario}  \\

\multicolumn{3}{|c|}{Design optimisation and finalization}  \\

\multicolumn{3}{|c|}{Main technologies \acrshort{rd} conclusions}  \\

\multicolumn{3}{|c|}{Technical Design Report---two \acrshort{ip}s at \acrshort{cern}} \\

\hline
\hline
Project Preparation Phase 2 & \multicolumn{2}{|c|}{T$_0$--(T$_0$+5)} \\
\hline 

\multicolumn{3}{|c|}{Site investigation and preparation}  \\
\multicolumn{3}{|c|}{Implementation studies with the host states}  \\
\multicolumn{3}{|c|}{Environmental evaluation \& project authorisation processes} \\
\multicolumn{3}{|c|}{Industrialisation of key components}  \\
\multicolumn{3}{|c|}{Engineering design completion}  \\

\hline
\hline
Construction phase (from ground breaking) & \multicolumn{2}{|c|}{T$_1$--(T$_1$+10)} \\
\hline
\multicolumn{3}{|c|}{Civil engineering}  \\
\multicolumn{3}{|c|}{Construction of components}  \\
\multicolumn{3}{|c|}{Installation and hardware commissioning}  \\
\hline
\hline
Beam commissioning and physics operation start & \multicolumn{2}{|c|}{T$_1$+11} \\
\hline
\end{tabular}
\label{tab:timeline_CLIC_LCF}
\end{table}

\paragraph*{\acrshort{lcf}}

The implementation studies for \acrshort{clic} and \acrshort{lcf} have been done in common over the last two years. The timeline for~\acrshort{lcf} is very similar and organised in the same phased structure: Preparation phases 1 and 2 are followed by construction, as for \acrshort{clic} in Table \ref{tab:timeline_CLIC_LCF}.

The \acrshort{lcf} \acrshort{rf} technology is industrially more mature and requires less development resources than for \acrshort{clic}. 
The construction, installation and commissioning timeline is mostly based on the \acrshort{ilc} one~\cite{bib:ILC_TDR_vol3_II}:
\begin{itemize}
    \item four years for the civil engineering works (drive beam, injectors, collider, surface infrastructures);
    \item four years for the installation of the~\acrshort{ti} and of the accelerator \acrshort{hw};
    \item one year of \acrshort{hw} commissioning of the collider followed by one year of beam commissioning of the collider.
\end{itemize}
This schedule has some overlapping activities as work affects different areas. The experience from construction and installation of existing \acrshort{srf} linacs has been folded into this schedule providing important guidelines for resources needed and schedules to use. Nevertheless, as for \acrshort{clic}, to accommodate the fact the transition between the phases might require some time, and also to avoid the clear conflict between operating \acrshort{hllhc} and beam commissioning of a new collider, the construction phase is increased in Table \ref{tab:timeline_CLIC_LCF} from eight to ten years before commissioning is started.

\subsection{Collider complex operation: resources requirements}

Table~\ref{tab:operation_CLICFCCeeLCF} summarizes the resources required for the operation of the \acrshort{clic}, \acrshort{fccee} and \acrshort{lcf} colliders:

\begin{itemize}
    \item annual maintenance and replacement material costs estimated according to the methodology defined in Section~\ref{sec:criteriaandmetrics_opresources} by individually categorising each item of equipment according to its nature;
    \item annual electricity cost range according to the operation energy (or beam power for \acrshort{lcf}) based on the annual electricity consumption in Table~\ref{tab:Parameters_CLIC_FCC_LCF};
    \item personnel required for the operation of the collider complex and its experiments' technical infrastructure.
\end{itemize}

\begin{table}[h!]
 \caption{Annual resource requirements for the operation of the accelerator complex of the proposed \epem colliders at~\acrshort{cern}.} 
\centering
\footnotesize
\begin{threeparttable}
\begin{tabular}{|l||c|c|c|}\hline
 & \acrshort{clic} & \acrshort{fccee} & \acrshort{lcf} \\ 
  & \SI{380}{\GeV} & \SIrange[range-units=single,range-phrase={--}]{91.2}{365}{\GeV}& \SI{250}{\GeV} \\ 
  & & & \acrshort{lp}--\acrshort{fp} \\\hline\hline
Maintenance and replacement material cost [M\acrshort{chf}/y] & 137 & 200\tnotex{tab:operation_CLICFCCeeLCF:1}& 170\tnotex{tab:operation_CLICFCCeeLCF:2}\\ \hline
Electricity cost [M\acrshort{chf}/y] & 63 & 92--146 & 61--77 \\ \hline
Personnel for operation of the future collider accelerator complex [\acrshort{fte}/y]& 650 & 950 & 800\\ \hline
\hline
\end{tabular}
  \begin{tablenotes}
  \item[a] \label{tab:operation_CLICFCCeeLCF:1} Average of the costs for the different energy configurations.
  \item[b] \label{tab:operation_CLICFCCeeLCF:2} Average of the costs for the \acrshort{lp} and \acrshort{fp} configurations.
  \end{tablenotes}
 \end{threeparttable}
\label{tab:operation_CLICFCCeeLCF}
\end{table}


%% file: include/04-Beyond2050/Beyond2050.tex
\section{Future collider options at \acrshort{cern} for operation beyond 2050: \epem colliders}

\subsection{Main parameters and performance}

\paragraph*{\acrshort{clic} \SI{1.5}{\TeV}}

After ten years of operation with the baseline configuration it is proposed to upgrade the collider to provide collisions to a~\acrshort{com}\ energy of \SI{1.5}{\TeV}. This is within the maximum achievable energy with a single drive-beam accelerator whose energy will have to be increased from \SI{1.91}{\GeV} to \SI{2.4}{\GeV}. 
The number of colliding bunches per pulse will be reduced from 352 to 312 and their population will be reduced from $5.2\times10^9$ to $3.7\times10^9$ to permit the acceleration to higher gradients. The repetition rate will be reduced from \SIrange{100}{50}{\Hz} to keep the power below \SI{300}{\MW}. Only one \acrshort{ip} will be served. The upgrade to higher energies requires an increase of the length of the tunnel and of the main linacs, connecting new tunnels to existing tunnels, moving the existing modules providing accelerating gradients of~\SI{72}{\mega\volt\per\m} at the extremities of the new tunnels and adding new, higher-gradient (\SI{100}{\mega\volt\per\m}) modules. The length of the \acrshort{bds} needs to be increased and new magnets installed. The main-beam production complex needs only minor modifications while the drive-beam complex needs to be extended to increase the drive-beam energy. The overall length of the collider tunnel will be \SI{29.6}{\km}. New detectors are currently not considered, but upgrades and improvements will be important. The main parameters after the upgrade are listed in Table~\ref{tab:Param_CLIC_LCF_upgrades}.

An alternative upgrade to \SI{550}{\GeV}, maintaining the repetition rate at \SI{100}{Hz} and two \acrshortpl{ip}, is also considered. A possible implementation would be to start with a longer tunnel by \SI{5}{\km} from the beginning adding about 25\% to the initial cost of the \acrshort{ce} of the baseline \SI{380}{\GeV} machine. CLIC at \SI{550}{\GeV} provides $\sim$45\% higher luminosity than the \SI{380}{\GeV} collider, at $\sim$30\% higher cost and power~\cite{adli2025compactlineareecollider}.

\paragraph*{\acrshort{lcf} \SI{550}{\GeV}}

The upgrade in energy from \SIrange{250}{550}{\GeV} is proposed to take place after ten years of operation at~\SI{250}{\GeV} and it would require a \acrshort{ls} of two years assuming that the upgrade from \acrshort{lp} to \acrshort{fp} has already taken place in the previous phase.  It does not require any additional \acrshort{ce} as the length of the collider tunnel has been defined based on the requirements for a \acrshort{com}\ energy of \SI{550}{\GeV} assuming nominal parameters for the \acrshort{srf} cryomodules.
The bulk of the upgrade will consist in the installation of additional cryomodules. Technology improvements could reduce the number of modules needed. As for \acrshort{clic} new detectors are not required, but upgrades and improvements will be important. 
The main parameters after the upgrade are listed in Table~\ref{tab:Param_CLIC_LCF_upgrades}.

\begin{table}[h!]
 \caption{Main parameters for \acrshort{clic} and \acrshort{lcf} after upgrade. The instantaneous and integrated luminosity in-between parentheses includes also the contribution from energies below 99\% of the \acrshort{com}\ energy $\sqrt{s}$.
 }
\centering
\footnotesize
 \begin{threeparttable}
 \begin{tabular}{|l||c|c|}\hline
& \multicolumn{1}{c|}{\acrshort{clic}}  & \multicolumn{1}{c|}{\acrshort{lcf}~\acrshort{fp}} \\\hline\hline

Length collider tunnel~[\SI{}{\km}] & \multicolumn{1}{c|}{29.6} &  \multicolumn{1}{c|}{33.5} \\\hline

Number of experiments (\acrshort{ip}s) & \multicolumn{1}{c|}{1}   & \multicolumn{1}{c|}{2}\\\hline

\acrshort{com}~energy [\SI{}{\GeV}] & \multicolumn{1}{c|}{1500} &550   \\ \hline

Longitudinal polarisation (e$^-$ / e$^+$) [\%] & \multicolumn{1}{c|}{$\pm$80 / 0} & \multicolumn{1}{c|}{$\pm$80 / $\pm$60} \\ \hline

Number of years of operation (total) & 10 & 10 \\\hline

Nominal years of operation (equivalent)\tnotex{tab:Param_CLIC_LCF_upgrades:1} & 9\tnotex{tab:Param_CLIC_LCF_upgrades:2} & 9\tnotex{tab:Param_CLIC_LCF_upgrades:2} \\\hline

Instantaneous luminosity per~\acrshort{ip} above $0.99~\sqrt{s}$ (total)~[\(10^{34}\,\SI{}{\per\square\centi\meter\per\second}\)] & 1.4 (3.7)\tnotex{tab:Param_CLIC_LCF_upgrades:3} & 2.25 (3.85) \\\hline

Integrated luminosity above $0.99~\sqrt{s}$ (total) over all \acrshort{ip}s per year of nominal operation~[\SI{}{\per\atto\barn}/y] & 0.17 (0.44)\tnotex{tab:Param_CLIC_LCF_upgrades:3} & 0.54 (0.92) \\\hline

Integrated luminosity above $0.99~\sqrt{s}$ (total) over all \acrshort{ip}s  over the full programme~[\SI{}{\per\atto\barn}] & 1.51 (3.96)\tnotex{tab:Param_CLIC_LCF_upgrades:3} & 4.85 (8.32) \\ \hline

Peak power consumption~[\SI{}{\mega\watt}] & 287 &322 \\ \hline

Electricity consumption per year of nominal operation~[\SI{}{\tera\watt\hour}/y]\tnotex{tab:Param_CLIC_LCF_upgrades:4}  & 1.4 & 1.8\\\hline
 
  \end{tabular}
  \begin{tablenotes}
    \item[a] \label{tab:Param_CLIC_LCF_upgrades:1} This row lists the equivalent number of years of operation at nominal instantaneous luminosity, hence taking into account the luminosity ramp-up.
    \item[b] \label{tab:Param_CLIC_LCF_upgrades:2} A luminosity ramp-up is assumed with a luminosity corresponding to 25\% and 75\% over the first two years, respectively.
    \item[c] \label{tab:Param_CLIC_LCF_upgrades:3} The improvement in luminosity obtained recently for \acrshort{clic} \SI{380}{\GeV}, has not been included for \acrshort{clic} at \SI{1.5}{\TeV}, because the start-to-end studies have not yet been  been completed at that energy.
    \item[d] \label{tab:Param_CLIC_LCF_upgrades:4} Computed from the peak power consumption and the assumptions on the operational year (see Table~\ref{tab:operational_year}).
  \end{tablenotes}
  \end{threeparttable}
\label{tab:Param_CLIC_LCF_upgrades}
\end{table}

\subsection{Environmental aspects}

In the following we identify the main origins of the additional \acrshort{ghg} emissions but no detailed estimate is available to provide quantitative values. The corresponding contributions have even larger uncertainties than those provided for the baseline configuration also considering that low carbon technologies might have further progressed by the time of the construction (expected to take place at the earliest at the end of the 50s) and low- or zero- carbon electricity might be available.
Considerations on the consumption of land are also added.

\paragraph*{\acrshort{clic}~\SI{1.5}{\TeV}}

The additional~\acrshort{ghg} emissions associated to the energy upgrade are related to:
\begin{itemize}
    \item the \acrshort{ce} work required to the extension of the tunnel, the construction of four additional shafts outside of the~\acrshort{cern} fence, the associated surface buildings, and the construction of two additional underground turn-around loops to transport the drive beam;
    \item the construction, transport and installation of the accelerator \acrshort{hw} and the associated technical infrastructure;
    \item generation of the electricity required for the operation of the collider.
\end{itemize}

The construction of the four additional shafts and corresponding surface sites will require to consume additional land outside of the \acrshort{cern} fence.

\paragraph*{\acrshort{lcf} \SI{550}{\GeV}}

The additional~\acrshort{ghg} emissions associated to the energy upgrade are related to:
\begin{itemize}
    \item the construction, transport and installation of the accelerator \acrshort{hw} and the associated technical infrastructure;
    \item generation of the electricity required for the operation of the collider.
\end{itemize}

No additional \acrshort{ce} or land consumption are necessary.

\subsection{Technical readiness and \acrshort{rd} requirements} 

\paragraph*{\acrshort{clic} \SI{1.5}{\TeV} and \acrshort{lcf} \SI{550}{\GeV}}

Both upgrades build on the technologies developed for the baseline configuration though pushing further the requirements on gradients and on stabilization due to the reduced beam size at the~\acrshort{ip}. The necessary technologies are expected to be mature by the time of the upgrades also considering the experience gained in the first low-energy phase of operation of the colliders.

\subsection{Construction and installation costs}

\paragraph*{\acrshort{clic} \SI{1.5}{\TeV}}

The cost of the upgrade of \acrshort{clic} from \SI{380}{\GeV} to \SI{1.5}{\TeV} is expected to amount to  7116~M\acrshort{chf}. The breakdown in the main items is presented in Table~\ref{tab:clic1500cost}. The \acrshort{ce} costs for the tunnel extension will also need to consider costs related to land, roads, electricity and water connections as well as for administrative processes and underground rights-of-way, as for the initial stage. The cost uncertainty for the upgrade is similar to that for the construction of the baseline version as both versions share the same technology.

 \begin{table}[h!]
    \caption{Cost summary table in 2024 M\acrshort{chf} for the upgrade of the \acrshort{clic} accelerator from~\SI{380}{\GeV} to \SI{1.5}{\TeV} with one \acrshort{ip}~\cite{adli2025compactlineareecollider}. Cost classes are defined in Table~\ref{tab:AACEcostScheme}. The cost of possible detector upgrades is not included.}
  \centering
  \footnotesize
  \begin{tabular}{|l|c|c|}
    \hline
    \multirow{2}{*}{Domain}& \multicolumn{2}{c|}{\acrshort{clic} -- \SI{1.5}{\TeV}}\\
     & Cost & Cost class\\
    ~& [M\acrshort{chf}] & ~ \\ \hline
    Collider & 4684  & 3--4  \\ \hline
    Main beam production and transfer lines & 23 & 3--4  \\ \hline
    Drive beam production and transfer lines &  302 & 3--4 \\ \hline
    Civil engineering & 703 & 4 \\ \hline
    Technical infrastructures & 1404 & 3--4 \\ \hline
    Experiments & \multicolumn{2}{c|}{N/A} \\ \hline
    \textbf{TOTAL} & \textbf{7116} &\\ \hline
  \end{tabular}
  \label{tab:clic1500cost}
\end{table}

\paragraph*{\acrshort{lcf} \SI{550}{\GeV}}

The cost of the upgrade of \acrshort{lcf} from \SI{250}{\GeV} to \SI{550}{\GeV} is expected to amount to 5464~M\acrshort{chf}. The breakdown in the main items is presented in Table~\ref{tab:LCF550_cost}. Here it is assumed that the upgrade from \acrshort{lp} to \acrshort{fp} has taken place during the previous phase. The cost uncertainty for the upgrade is similar to that for the construction of the baseline version as both versions  share the same technology.

\begin{table}[h!]
   \caption{Cost summary table in 2024 M\acrshort{chf} for the upgrade of the \acrshort{lcf} \acrshort{fp} collider from~\SIrange{250}{550}{\GeV}---with two \acrshortpl{ip}~\cite{bib:LCF4CERN_backup}. Cost classes are defined in Table~\ref{tab:AACEcostScheme}. The cost of  possible detector upgrades is not included.
  }
  \centering
  \footnotesize
  \begin{threeparttable}
  \begin{tabular}{|l|c|c|}
    \hline
    \multirow{2}{*}{Domain}& \multicolumn{2}{c|}{\acrshort{lcf} \acrshort{fp} -- \SI{550}{\GeV}}\\
     & Cost & Cost class\\
    ~& [M\acrshort{chf}] & ~ \\ \hline
    Collider  & 4204  & 3  \\ \hline
    Injectors and transfer lines    & 86 & 3 \\ \hline
    Civil Engineering               & 0     &  \\ \hline
    Technical infrastructures       & 1174  & 4 \\ \hline
    Experiments & \multicolumn{2}{c|}{N/A} \\ \hline
    \textbf{TOTAL} & \textbf{5464} &\\ \hline
  \end{tabular}
\end{threeparttable}
  \label{tab:LCF550_cost}
\end{table}

\subsection{Project timeline}

\paragraph*{\acrshort{clic} \SI{1.5}{\TeV}}

The construction and installation of the upgrade extend over a period of 4.5 years and the decision about the next higher energy stage would need to be taken after 4--5 years of data taking in the baseline configuration. Most of the construction and installation work can be carried out in parallel with the data-taking at \SI{380}{\GeV}. According to the proponents a \acrshort{ls} of two years is needed to make the connection between the existing machine and its extensions, to reconfigure the modules, and to modify the \acrshort{bds}. This timescale is ambitious and needs to be reevaluated based on experience from the initial stage.

\paragraph*{{\acrshort{lcf} \SI{550}{\GeV}}}

Differently from \acrshort{clic} preparation work in the tunnel for the upgrade will be more limited as the additional \acrshort{srf} cryomodules will be installed in the same tunnel of the \SI{250}{\GeV} machine, but no \acrshort{ce} is required. According to the proponents, the activities will take place during a \acrshort{ls} of two years. As for \acrshort{clic} this timeline will need further reevaluation based on the initial stage experience. 



\newpage
\section{Future collider options at \acrshort{cern} for operation beyond 2050: \acrshort{fcchh}}

\subsection{Main parameters and performance}

\begin{figure}[h]
\centering
\includegraphics[width=0.7\textwidth]{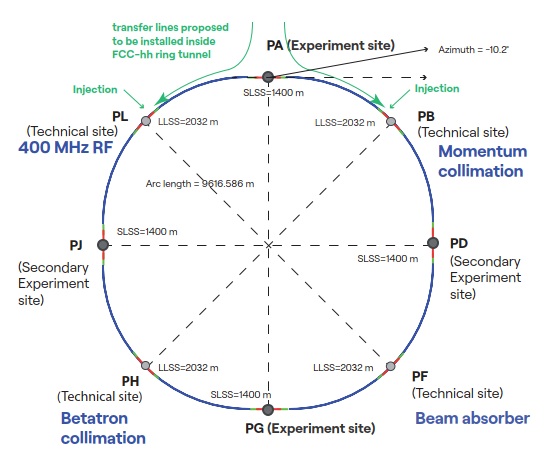}
\caption{Schematic diagram of the \acrshort{fcchh} collider.}
\label{fig:FCChh_layout}
\end{figure}

\acrshort{fcchh} will be installed in the same tunnel of \acrshort{fccee} and the current baseline \acrshort{com}\ energy is approximately \SI{85}{\TeV}. The present design allows four collision points and experiments. Two multi-purpose detectors are operated at high-luminosity and the other two specialised detectors are operated at lower luminosity. In addition to colliding protons with protons, also proton-ion and ion-ion collisions are envisaged. A schematic diagram of \acrshort{fcchh} is shown in Fig.~\ref{fig:FCChh_layout}.
\acrshort{fcchh} shares much of  the \acrshort{ce} and  \acrshort{ti} of \acrshort{fccee}~\cite{bib:FCCee_ESPP2026}. 
The \acrshort{fcchh} baseline assumes \SI{14}{\tesla} magnets whose performance is closer to the state of the art of $\mathrm{Nb_3Sn}$ technology and are a relatively lower-risk and potentially fast-tracked option~\cite{bib:FCCee_ESPP2026}
. \acrshort{rd} towards \acrshort{hts} magnets is pursued and could enable higher collision energies.

The main parameters for \acrshort{fcchh} are compiled in Table~\ref{tab:paramfcchh}. The table illustrates how the synchrotron radiation strongly increases with higher magnetic field. The synchrotron-radiation heat, which must be extracted from inside the cold magnets, is a major contribution to the thermal load on the cryogenic system. To limit the latter and also because the radiation damping during the store is significant, the beam current and the bunch population are relaxed compared to those of the \acrshort{hllhc}. The synchrotron radiation power has been significantly reduced by reducing the magnetic field from \SIrange{16}{14}{\tesla} as compared to the~\acrshort{cdr}~\cite{FCC:2018vvp}.

\begin{table}[h!]
\caption{Parameters of \acrshort{fcchh} compared with the nominal \acrshort{hllhc} and \acrshort{lhc} designs. The plan to operate the $\mathrm{Nb_3Sn}$ magnets at \SI{4.5}{\kelvin}, being studied, would further reduce the annual electricity consumption to \SI{1.8}{\tera\watt\hour}/y.}
  \centering
  \footnotesize
  \begin{threeparttable}
\begin{tabular}{|L{10cm}||c|c|c|}
    \hline
 & \acrshort{fcchh} & \acrshort{hllhc} & \acrshort{lhc} \\ \hline\hline
\acrshort{com}~energy [\SI{}{\TeV}] & 84.6 & \multicolumn{2}{c|}{14} \\ 
Circumference [\SI{}{\km}] & 90.7 & \multicolumn{2}{c|}{26.7} \\
Arc dipole field [\SI{}{\tesla}] & 14 & \multicolumn{2}{c|}{8.33} \\
Beam current [\SI{}{\ampere}] & 0.5 &  1.1 & 0.58 \\
Synchrotron radiation power per beam [\SI{}{\kilo\watt}] & 1200 & 7.5 & 3.5  \\
Peak luminosity per \acrshort{ip}~[\(10^{34}\,\SI{}{\per\square\centi\meter\per\second}\)] & 30 &  5 (levelling) & 1 \\
Peak number of events per bunch crossing & 1000  & 132 & 27 \\
Integrated luminosity per \acrshort{ip} per year of nominal operation~[\SI{}{\per\atto\barn}/y]  & 0.9 & 0.25 & 0.05 \\
Electricity consumption per year of nominal operation~[\SI{}{\tera\watt\hour}/y]\tnotex{tab:paramfcchh:1}  & 2.0\tnotex{tab:paramfcchh:2} & 0.8 & 0.7  
\\\hline
\end{tabular}
  \begin{tablenotes}
    \item[a] \label{tab:paramfcchh:1} The power consumption of the injectors is not included.
    \item[b] \label{tab:paramfcchh:2} Computed from the peak power consumption and the assumptions on the operational year (see Table~\ref{tab:operational_year_FCChhMuC}).
\end{tablenotes}
 \end{threeparttable}
\label{tab:paramfcchh}
\end{table}

Three options for an \acrshort{fcchh} injector exist. All options can deliver the main beam parameters required, such as intensity, emittance, and bunch spacing. 
The main difference is the implied \acrshort{fcchh} injection energy. 
An attractive option for a high-energy injector is a new \SI{4}{T} machine in the \acrshort{lhc} tunnel. 
 Another possibility, with intermediate beam energy, is a \SI{2}{T} superferric machine. 
If a collider injection energy of \SI{1.3}{TeV} is acceptable, a \acrshort{sc} ring in the \acrshort{sps} tunnel (\acrlong{scsps}---\acrshort{scsps}) becomes attractive.

\subsection{Environmental aspects}

Efforts are ongoing to minimize the power consumption of the collider. The introduction of an \textit{eco-mode} for the cryogenic system during shutdown periods reduces the power consumption during non-operational periods. In addition the possibility of operating the $\mathrm{Nb_3Sn}$ magnets at \SI{4.5}{\kelvin} is being investigated and it would reduce the annual electricity consumption to \SI{1.8}{\tera\watt\hour}/y.

Most of the \acrshort{ce} and land consumption for \acrshort{fccee} have been conceived considering the needs for \acrshort{fcchh}. The only additional underground structures that will have to be built for \acrshort{fcchh}~\cite{bib:FCCee_ESPP2026
} are:
\begin{itemize}
    \item two beam dump tunnels (approximately~\SI{2}{\km} long in total) and caverns with corresponding junction tunnels;
    \item two transfer line tunnels with  a total length varying between 2.5 and \SI{5.7}{\km}, depending on the selected option for the injector. 
\end{itemize}

Additional surface \acrshort{ce} will be required for \acrshort{fcchh} as compared to FCC-ee 
for the assembly halls of at least the two largest \acrshort{fcchh} experiments and to provide additional space for additional cryogenics and cooling plants. Space reservations will already be made during the \acrshort{fccee} design phase.

\subsection{Technical readiness and \acrshort{rd} requirements} \label{sec:fcchhtrl}

The key challenges for the construction of \acrshort{fcchh} are the \acrshort{hfm} technology and power consumption in the presence of strong synchrotron radiation.  The technologies needed for delivery that are still under development, and the target performance parameters of each development are summarised in Table~\ref{tab:FCChhTRL}.


\begin{table}[h!]
  \caption{Technical readiness and \acrshort{rd} requirements for~\acrshort{fcchh}. The material and personnel estimates refer to \acrshort{cern} contribution, the contributions from collaborations is not included.
}
\centering
\scriptsize
\begin{threeparttable}
\begin{tabular}{|l|c|C{3cm}|c|c|c|c|}
 \hline
 \multirow{3}{*}{Component/Sub-system} & \multirow{3}{*}{\acrshort{trl}} & \multirow{2}{*}{Main parameter}   & \multirow{3}{*}{Improvement factor} & \multicolumn{3}{c|}{~\acrshort{rd} effort}  \\ 
 ~&  & \multirow{2}{*}{to be improved} & & Personnel  & Material  & Timescale \\
~ & ~ &  & ~ & [\acrshort{ftey}] & [M\acrshort{chf}] &  [years] \\ \hline\hline

Nb$_3$Sn conductor\tnotex{tab:FCChhTRL:1} & \cellcolor{green} 7 & Cost \& industrialisation & 3 & 10 & 30 & 7 \\ 
 & \cellcolor{green} &  (more manufacturers) &  &  &  & \\ \hline\hline
 
Nb$_3$Sn magnet short model    & \cellcolor{yellow}6 & Max. field\tnotex{tab:FCChhTRL:2}  & 
20\% & 100 & 20 & 5\\\hline

Nb$_3$Sn scaling to \SI{5}{\m}  & \cellcolor{yellow}5 & Length & 3 & 100 & 40 & 10 \\ \hline

Nb$_3$Sn scaling to \SI{15}{\m}  & \cellcolor{yellow}4 & Length & 3 & 100 & 100 & 15 \\ \hline\hline

\acrshort{hts} magnet short model \SI{20}{\tesla} & \cellcolor{red}3 &  Field\tnotex{tab:FCChhTRL:3} ~and all accelerator magnets' features & 5--10 & 100 &  50 & 10  \\ 
\hline
\acrshort{hts} long magnet \SI{20}{\tesla} & \cellcolor{red}1 & Length & 15 & 100 & 150 & 25 \\ \hline

 \end{tabular}
\begin{tablenotes}
 \scriptsize
    \item[a] \label{tab:FCChhTRL:1} This includes only contracts to lower the price and to have more manufacturers, not the conductor cost for the model and prototype that is included in the next lines. 
     \item[b] \label{tab:FCChhTRL:2} Target of \SI{15}{\tesla}, for operation at \SI{14}{\tesla}. 
     \item[c] \label{tab:FCChhTRL:3} \SIrange{2}{4}{\tesla} already achieved, but not with all requirements. 
\end{tablenotes}
\end{threeparttable}
\label{tab:FCChhTRL}
 \end{table}

According to the reviewers of the European Accelerator \acrshort{rd} platform~\cite{bib:LDG_RD_review_ESPP2026}, the maturity gap between \acrshort{lts} and \acrshort{hts} implies that the decision on \acrshort{fcchh} magnet technology cannot be taken before 2035 and \acrshort{hts} magnets do not appear compatible with a ``fast track'' \acrshort{fcchh}.

\acrshort{hts} can generate magnetic fields in excess of \SI{20}{\tesla}, well above the levels allowed by Nb$_3$Sn and can operate at temperatures of the order of \SI{20}{\kelvin} with corresponding energy savings in the cryogenic systems.
\acrshort{hts} technology has been proven for solenoids operating above~\SI{20}{\tesla} in steady state, but no high-field dipoles with the requirements needed for accelerators have been built and tested successfully so far. In particular, the field quality across the whole operational range and with ramping times compatible with \acrshort{fcchh} operation as well as the protection aspects have not been mastered, yet. The \acrshort{hfm} programme aims to demonstrate $\mathrm{Nb_3Sn}$ magnet technology for large-scale deployment pushing it to its limits in terms of maximum field and production scale and to demonstrate the suitability of \acrshort{hts} for accelerator magnets providing a proof-of-principle of \acrshort{hts} magnet technology beyond the reach of $\mathrm{Nb_3Sn}$~\cite{bib:HFM_ESPP2026}.

Among \acrshort{hts}, \acrshort{rebco} superconductor does not need reaction after winding, allowing a simplified process of magnet manufacturing with respect to Nb$_3$Sn, but it is available only in tapes and the architecture of a \acrshort{rebco} accelerator magnet cable is one of the main technological challenges. 
\acrshort{bscco} is available in round strands, but the manufacturing process involves a reaction of the coil at temperature higher than Nb$_3$Sn (>\SI{800}{\celsius}), and in high-pressure oxygen~\cite{bib:HFM_ESPP2026}. 
\acrshort{ibs} offer promise for larger cost 
reduction than \acrshort{rebco} and \acrshort{bscco}, although their technological readiness currently lags behind both of them~\cite{bib:HFM_ESPP2026}.

\subsection{Construction and installation costs}

The capital cost for \acrshort{fcchh} construction, as a second stage of the integrated \acrshort{fcc} programme, is summarized  
 in Table~\ref{tab:fcchh_cost}. 
The total construction cost amounts to 19080~M\acrshort{chf} and it is dominated by cost of the the collider magnets, amounting to approximately 10000~M\acrshort{chf} .

The main additional \acrshort{ce} structures required for \acrshort{fcchh} as second stage are the beam dump tunnels and the transfer-line tunnels. Most of the electrical, cooling and ventilation installations are reused. The cryogenics infrastructure for the main magnet cooling, therefore, drives the capital cost. The cost of the \acrshort{fcchh} injector and transfer lines is included but not that required by the possible upgrade of the pre-injectors. The final cost will depend on which injector option is chosen. At the time of the \acrshort{fcc} \acrshort{cdr}~\cite{FCC:2018vvp}, the combined cost estimate for modifying the existing \acrshort{lhc} and for the new transfer lines was 615~M\acrshort{chf}.   
The cost of the four \acrshort{fcchh} experiments can be estimated only at a later stage, once more detailed designs of the detectors exist.

\begin{table}[h!]
   \caption{Cost summary table in 2024 M\acrshort{chf} for the construction of \acrshort{fcchh}. Cost classes are defined in Table~\ref{tab:AACEcostScheme}. 
  }
  \centering
  \footnotesize
  \begin{threeparttable}
  \begin{tabular}{|l|c|c|}
    \hline
    \multirow{2}{*}{Domain}& \multicolumn{2}{c|}{\acrshort{fcchh}}\\
     & Cost & Cost class\\
    ~& [M\acrshort{chf}] & ~ \\ \hline
    \acrshort{fccee} dismantling & 200 & 4 \\ \hline
    Collider  & 13400  & 4  \\ \hline
    Injectors and transfer lines    & 1000 & 4 \\ \hline
    Civil Engineering               & 520    &  4 \\ \hline
    Technical infrastructures       & 3960  & 4 \\ \hline
    Experiments                     & \multicolumn{2}{c|}{N/A}  \\ \hline

    \textbf{TOTAL} & \textbf{19080} & \\ \hline
  \end{tabular}
\end{threeparttable}
  \label{tab:fcchh_cost}
\end{table}

\subsection{Project timeline}

The timelines of \acrshort{fcchh} as a second phase of the \acrshort{fcc} integrated programme or as a stand-alone project are summarised in Tables~\ref{tab:timeline_FCChh_after_FCCee} and \ref{tab:timeline_FCChh_standalone}, respectively.

\begin{table}[h!]
 \caption{\acrshort{fcchh} timeline as a second phase after \acrshort{fccee}.} 
\centering
\footnotesize
\begin{threeparttable}
\begin{tabular}{|l||c|}\hline
Milestone & \acrshort{fcchh}  \\ \hline\hline
Conceptual Design Study & 2014--2018 \\  
Definition of the placement scenario & 2022 \\
Feasibility Report ready
& 2025  \\
Main technologies \acrshort{rd} completion & 2054  \\ 
Technical Design Report ready & 2054 \\
Latest Project Approval &  2054 \\
Environmental evaluation \& project authorisation processes & 2054--2058 \\
Industrialization \& magnet production & 2054--2069 \\
Civil engineering -- collider & 2060\tnotex{tab:timeline_FCChh_after_FCCee:1}--2068 \\
\acrshort{fccee} dismantling & 2063--2064
\\ 
\acrshort{ti} installation -- collider & 2065--2069 \\
Accelerator installation -- collider
& 2068--2072 \\
\acrshort{hw} commissioning -- collider & 2071--2073   \\
Beam commissioning -- collider & 2073  \\
Physics operation start & 2074   \\
\hline
\end{tabular}
 \begin{tablenotes}
 \scriptsize
    \item[a] \label{tab:timeline_FCChh_after_FCCee:1} The starting date corresponds to the start of the surface \acrshort{ce} works.
 \end{tablenotes}
 \end{threeparttable}
\label{tab:timeline_FCChh_after_FCCee}
\end{table}

\begin{table}[h!]
 \caption{Fastest possible \acrshort{fcchh} timeline as a stand-alone project.} 
\centering
\footnotesize
\begin{threeparttable}
\begin{tabular}{|l||c|}\hline
Milestone & \acrshort{fcchh}  \\ \hline\hline
Conceptual Design Study & 2014--2018 \\
Definition of the placement scenario & 2022 \\
Feasibility Report ready
& 2025  \\
Latest Project Approval~ &  2033 \\
Environmental evaluation \& project authorisation processes & 2026--2035 \\
Main technologies \acrshort{rd} completion & 2037\tnotex{tab:timeline_FCChh_standalone:1}  \\
Technical Design Report ready & 2037 \\
Industrialization \& magnet production & 2038--2053 \\
Civil engineering -- collider & 2037--2046 \\
\acrshort{ti} installation -- collider
& 2043--2050 \\
Accelerator installation -- collider
& 2046--2052 \\
\acrshort{hw} commissioning -- collider & 2049--2053   \\
Beam commissioning -- collider & 2054    \\
Physics operation start & 2055   \\
\hline
\end{tabular}
 \begin{tablenotes}
 \scriptsize
    \item[a] \label{tab:timeline_FCChh_standalone:1} Assuming an accelerated \acrshort{rd} programme as compared to that outlined in Table~\ref{tab:FCChhTRL}.
 \end{tablenotes}
 \end{threeparttable}
\label{tab:timeline_FCChh_standalone}
\end{table}



\newpage
\section{Future collider options at \acrshort{cern} for operation beyond 2050: Muon Collider}

\subsection{Main parameters and performance}

The muon collider is an innovative concept to reach multi-TeV lepton collisions. The associated design and the technologies are less mature than for some other approaches and require a significant \acrshort{rd} to reach a maturity level to make informed decisions.

\begin{figure}[h]
\includegraphics[width=\textwidth]{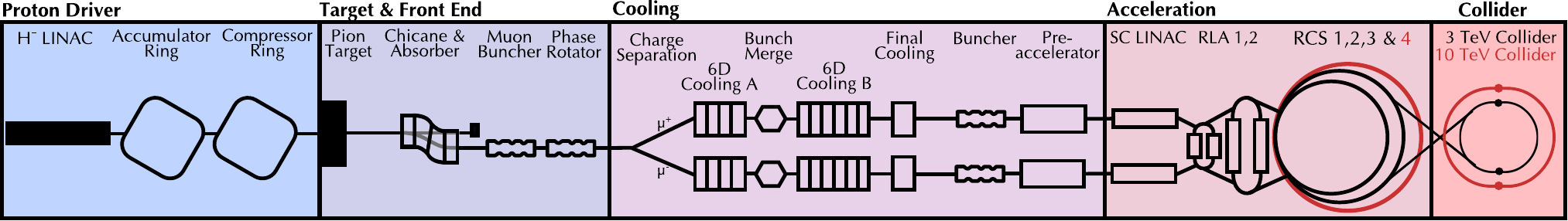}
\caption{Conceptual layout of the \SI{3}{\TeV} (black line) and \SI{10}{\TeV} (red line) \acrshort{mc} complex~\cite{bib:imcc_esppu2026}.}
\label{fig:MC_layout}
\end{figure}

The baseline muon collider design is a green-field, \SI{10}{\tera\electronvolt} \acrshort{com}\ collider \cite{bib:Accettura:2023ked,bib:imcc_backup_esppu2026} based on a concept which was developed by the \acrshort{us} \acrfull{map} until 2017 \cite{bib:map_overview}.
The design is now being progressed by the \acrfull{imcc} \cite{bib:InternationalMuonCollider:2024jyv}. A schematic layout of the collider is shown in Fig.~\ref{fig:MC_layout} and
contains the following key areas:
\begin{enumerate}
    \item The proton driver (blue box in the diagram) produces a short, high-intensity proton pulse (e.g. a \SI{2}{\MW} beam at \SI{5}{\GeV}, with a repetition rate of~\SI{5}{\Hz}~\cite{bib:imcc_backup_esppu2026}).
    \item This pulse hits the target (indigo) and produces pions. The decay channel guides the pions and forms a beam with the resulting muons via a buncher and phase rotator system.
    \item Several cooling stages (purple) reduce the longitudinal and transverse emittance of the beam using a sequence of absorbers and \acrshort{rf} cavities in a high magnetic field.
    \item A system of a linac and two recirculating linacs accelerate (light red) the beams to \SI{63}{\GeV}, followed by a sequence of high-energy accelerator rings (\acrlong{rcs}---\acrshort{rcs}) which reach \SI{1.5}{\tera\electronvolt} or \SI{5}{\tera\electronvolt}.
    \item Finally the beams are injected at full energy into the collider ring (red). Here, they will circulate and collide within the detectors until they decay. The muon collider ring has to provide sufficient luminosity, leading to a challenging design with $\beta^*$ and RMS bunch length of a few~\SI{}{\mm}, while the RMS momentum spread remains large at around $10^{-3}$~\cite{bib:imcc_backup_esppu2026}.
\end{enumerate} 

An implementation at \acrshort{cern} is under consideration and could reuse the existing \acrshort{sps} and \acrshort{lhc} tunnels to accelerate the muon beam. This design could reach a \acrshort{com}\ collision energy of up to \SI{7.6}{\tera\electronvolt} and a practical initial energy stage could reach up to \SI{3.2}{\tera\electronvolt}, depending on the layout. The significant reduction of \acrshort{ce} and associated environmental impact might justify the reduction in physics scope with respect to the \SI{10}{\TeV} machine, though the feasibility of integrating the \acrshort{rcs}s in the \acrshort{sps} and \acrshort{lhc} tunnels has not been demonstrated.

Example parameters for a \acrshort{cern} implementation are given in Table~\ref{tab:Parameters_MC}, assuming in a first option that a single new collider ring tunnel is constructed and used for both energy stages, which is consistent with the use of \SI{11}{\tesla} Nb$_3$Sn magnets at full energy. This scenario assumes magnets whose performance is state of the art but development towards industrialization is required. 
It also allows to implement a \SI{3.2}{\tera\electronvolt} collider using
very mature NbTi technology. A second scenario is also shown, in which two independent collider rings at \acrshort{cern} are used with Nb$_3$Sn dipoles for the \SI{3.2}{\tera\electronvolt} and \acrshort{hts} for the \SI{7.6}{\tera\electronvolt} stage. The luminosities would then increase to \(2\times10^{34}\,\SI{}{\per\square\centi\meter\per\second}\) and \(10.1\times10^{34}\,\SI{}{\per\square\centi\meter\per\second}\), respectively. Both scenarios are shown in Table~\ref{tab:Parameters_MC}, and the high-level parameters of the corresponding \acrshort{rcs} in Table~\ref{tab:Parameters_MC_RCS}.

\begin{table}[h!]
 \caption{Tentative high-level parameters for an \acrshort{mc} implementation at \acrshort{cern} using the \acrshort{sps} and \acrshort{lhc} tunnels~\cite{bib:imcc_esppu2026,bib:imcc_addendum_esppu2026}. The first scenario uses a common collider ring tunnel for both energy stages. The second uses an optimised collider ring tunnel for each collision energy. Luminosity is assumed to ramp up over the first three years as 5\%, 25\%, 70\%, respectively. In the table the number of years of operation is chosen so to achieve an integrated luminosity of at least \SI{1}{\per\atto\barn} (\SI{10}{\per\atto\barn}) at \SI{3.2}{\TeV} (resp.\,\SI{7.6}{\TeV}).
 }
\footnotesize
\centering
 \begin{threeparttable}
 \begin{tabular}{|L{10cm}||c|c|*2{c|}}\hline
~& \multicolumn{2}{c|}{Common} & \multicolumn{2}{c|}{Separate}\\
~& \multicolumn{2}{c|}{tunnel} & \multicolumn{2}{c|}{tunnels}\\ \hline\hline

Circumference/length collider tunnel~[\SI{}{\km}] & 11 & 11 & 4.8 & 8.7 \\\hline
Number of experiments (\acrshort{ip}s) & 2 & 2 & 2 & 2\\\hline

Injected beam power per beam~[\SI{}{\MW}] & 2.8 & 5.5 & 2.8 & 5.5 \\\hline
\acrshort{com}~energy [\SI{}{\TeV}] & 3.2 & 7.6 & 3.2 & 7.6 \\ \hline
Arc dipole peak field~[\SI{}{\tesla}] & 4.8 & 11 & 11 & 14\\ \hline
Longitudinal polarisation ($\mu^-$ / $\mu^+$) [\%] & \multicolumn{4}{c|}{N/A} \\ \hline

Number of years of operation (total) & 8 & 9 & 5 & 7 \\\hline
Nominal years of operation (equivalent)\tnotex{tab:Parameters_MC:1} & 6 & 7 & 3 & 5 \\\hline

Instantaneous luminosity per \acrshort{ip}~[\(10^{34}\,\SI{}{\per\square\centi\meter\per\second}\)] & 0.9 & 7.9 & 2.0 & 10.1 \\\hline
Integrated luminosity over all \acrshort{ip}s per year of nominal operation~[\SI{}{\per\atto\barn}/y] & 0.18 &  1.58 &  0.4 & 2.02 \\ \hline

Integrated luminosity over all \acrshort{ip}s over the full programme~[\SI{}{\per\atto\barn}] & 1.1 &  11.1 &  1.2 &  10.1 \\ \hline

Peak power consumption~[\SI{}{\mega\watt}] & 117 & 182 & 117\tnotex{tab:Parameters_MC:2} & 182\tnotex{tab:Parameters_MC:2} \\\hline
Electricity consumption per year of nominal operation~[\SI{}{\tera\watt\hour}/y]\tnotex{tab:Parameters_MC:3} & 0.8 &1.2 & 0.8 & 1.2 \\\hline

  \end{tabular}
  \begin{tablenotes}
    \item[a] \label{tab:Parameters_MC:1} This row lists the equivalent number of years of operation at nominal instantaneous luminosity, hence taking into account the luminosity ramp-up.
    \item[b] \label{tab:Parameters_MC:2} The power has been estimated for the common tunnel scenario and would be minimally smaller in the separate tunnel scenario, due the lower static losses of the magnets in the shorter collider tunnel. 
    \item[c] \label{tab:Parameters_MC:3} Computed from the peak power consumption and the assumptions on the operational year (see Table~\ref{tab:operational_year_FCChhMuC}).
  \end{tablenotes}
  \end{threeparttable}
\label{tab:Parameters_MC}
\end{table}

\begin{table}[h!]
\caption{High-level parameters for the \acrfullpl{rcs} of the \acrshort{mc} for the \acrshort{cern} implementation~\cite{bib:imcc_backup_esppu2026}.
}
\footnotesize
\centering
\begin{tabular}{|l||c|c|c|}\hline
~&\acrshort{rcs}1 in \acrshort{sps}&\acrshort{rcs}2 in \acrshort{lhc}&\acrshort{rcs}3 in \acrshort{lhc}\\ \hline\hline
Circumference~[\SI{}{\km}] &6.912&\multicolumn{2}{c|}{26.659} \\\hline
Beam injection energy [\SI{}{\TeV}] & 0.063& 0.35&1.6  \\ \hline
Beam extraction energy [\SI{}{\TeV}] & 0.35&1.6& 3.8\\ \hline
\end{tabular}
\label{tab:Parameters_MC_RCS}
\end{table}

\subsection{Environmental aspects}~\label{sec:env_MC}

The power consumption has been estimated for the two energy stages at \acrshort{cern} at respectively \SI{113} and \SI{172}{\MW}~\cite{bib:imcc_addendum_esppu2026}. The use of NbTi dipoles for the \SI{3.2}{\TeV} collider stage and of \acrshort{hts} dipoles for the \SI{7.6}{\TeV} is assumed.
The main contributions come from the cryogenics for the \acrshort{sc} magnets and the \acrshort{rf} systems all along the collider, the \acrshort{rf} power to drive the cavities, and the power to drive the fast-ramping magnets. 
Additional heat load induced by muon beam decay to the \acrshort{rf} cryostats in the \acrshort{rcs} was not included into the power budget mentioned above. Mitigation for this load is considered and requires detailed study. In the very worst case the full load would have to be removed at \SI{1.9}{\K} and would require a maximum electric power of the order of \SI{12}{\MW} for the \SI{3.2}{\TeV} stage and up to \SI{29}{\MW} for the \SI{7.6}{\TeV} case. At this stage, as a first estimate, one third of the maximum has been included in the numbers given in Table~\ref{tab:Parameters_MC}, leading to respectively 117 and \SI{182}{\MW} for the first and second stage.
The electricity consumption is then obtained using the operation times and the fraction of the peak power needed for each operational phase from Table~\ref{tab:operational_year_FCChhMuC}.

Regarding the impact on land and on \acrshort{ghg} emissions, at \acrshort{cern} the muon collider can potentially benefit from existing infrastructure.
An initial \acrshort{ce} exploration at \acrshort{cern} indicates that the surface installations of the accelerator facility could be built fully on land attributed to \acrshort{cern} provided that the \acrshortpl{rcs} can be integrated in the \acrshort{sps} and \acrshort{lhc} tunnels. Under this assumption, the \acrshort{ghg} impact from the \acrshort{ce} of the underground structures was estimated at \SI{150}{\kilo\tonne} CO$_2$ eq.~\cite{bib:imcc_addendum_esppu2026}, using the evaluation per km of tunnel obtained for \acrshort{clic} (\SI{6.38}{\kilo\tonne} CO$_2$ eq.\,per km---A1--A3 components~\cite{clic_ilc_lca_arup}), for a similar tunnel diameter, and additional scaling factors of 1.3 to account for access shafts and additional galleries, and 1.25 for the A4--A5 components.

Muon decays occurring in the~\acrshort{lss} of the collider produce neutrinos that will exit the earth's surface far from the accelerator facility. Two of these areas located on the Jura mountains will have to be fenced and these could host neutrino detectors on the surface where the neutrino beam emerges. The other exit points are located in the Mediterranean Sea.
The neutrinos arising from the rest of the facility must be diluted so that there is negligible radiation outside the \acrshort{cern} site. A system of movers are required in collider arcs to displace the magnets vertically to achieve this. Tentative studies indicate that it is possible to comply with the \acrshort{rp} regulations. Further optimisation and expansion of the studies to the full complex are required.

\begin{table}[h!]
\caption{Estimated area of land consumed by the~\acrshort{mc} project, i.e., lying outside the fenced \acrshort{cern} site. All surface sites are on \acrshort{cern}-attributed land with the exception of the two sites where the neutrino beam emerges in the Jura, potentially hosting two experiments. These two sites have not been estimated at this moment.}
\footnotesize
\centering
\begin{threeparttable}

\begin{tabular}{|l||c|}\hline
~ & \acrshort{mc} \\\hline\hline
Number of new access shafts & 2 \\\hline
Number of new surface sites & 4\tnotex{tab:MClandoccupation:1} \\\hline
Area of land permanently consumed (fraction outside CERN-attributed land) ~[\SI{}{\square\km}] & 0.2151 (0\%)\\\hline
Area of surface constructions (fraction outside CERN-attributed land)~[\SI{}{\square\km}] & 0.1035 (0\%) \\ \hline
\end{tabular}

  \begin{tablenotes}
    \item[a] \label{tab:MClandoccupation:1} Including two surface sites where the neutrinos emerge in the Jura.
  \end{tablenotes}
    
\end{threeparttable}

\label{tab:MClandoccupation}
\end{table}

\subsection{Technical readiness and \acrshort{rd} requirements}~\label{sec:TRL_MC}

The development of the \acrshort{mc} technologies faces several significant challenges that must be overcome in order to reach a level of maturity comparable to that of other colliders proposed for operation after 2050.

Following the European Accelerator \acrshort{rd} Roadmap, the \acrshort{mc} development has focused on the most critical parts of the complex that provide specific design challenges or require beyond start-of-the-art technologies.
At the time of writing, start-to-end simulations of the muon collider have not been completed and a tool capable of modelling the entire complex is not yet available. This will be essential to consolidate performance predictions, 
optimize key parameters, perform sensitivity analyses and refine background and radiation estimates. A number of the \acrshort{mc} technologies are at a readiness level of 3 or lower~\cite{bib:LDG_RD_review_ESPP2026} and require an extensive experimental development programme. The resources needed are estimated to be around 300 M\acrshort{chf} and 1800 \acrshort{fte}-years for the accelerator and 20 M\acrshort{chf} and 900 \acrshort{ftey} for the detector. The technically-limited timeline for the implementation of the \acrshort{rd} programme is 10~years~\cite{bib:imcc_backup_esppu2026}.

Key milestones include construction and testing of essential magnets and demonstrating cooling technology, both of which drive the schedule.
A detailed description of the \acrshort{rd} requirements is provided in Ref.~\cite{bib:imcc_backup_esppu2026}. Selected elements and key deliverables for the accelerator design are listed in Table~\ref{tab:deliverables}; in general, one can distinguish the following \acrshort{rd} (see also Ref.~\cite{bib:LDG_RD_review_ESPP2026}):
\begin{itemize}
\item 
\textbf{Magnet technology developments:} \acrshort{hts} solenoids for muon production and cooling, large-aperture collider ring dipoles and fast-ramping magnet systems. 
\item
\textbf{\acrshort{rf} technologies:}
klystrons, cavities working in high magnetic field and with high beam loading and test infrastructure.
\item \textbf{Muon cooling technology and demonstration programme:} technologies for muon cooling and their integration into the 6D cooling, the final cooling system and the demonstrator. Performance verification and development of key components like \acrshort{hts} solenoids and \acrshort{rf} systems.
\item 
\textbf{Design and technologies:} further study of key design challenges, including collider modeling lattice optimization, advanced simulations, site impact studies including neutrino flux, proton complex and technical developments as target, \acrshort{rf} and \acrshort{mdi}.
\item \textbf{Detector \acrshort{rd} priorities:} simulation, technology and software to enhance physics output while reducing beam-induced backgrounds.
\end{itemize}

\begin{table}[h!]
  \caption{Selected key deliverables of the proposed \acrshort{rd} programme of the \acrshort{mc}.}
\centering
\footnotesize
\begin{tabular}{|L{4cm}|L{5cm}|L{5cm}|}
 \hline
\textbf{Technologies} & \textbf{Deliverables} & \textbf{Key parameters and goals } \\ \hline\hline

\multicolumn{3}{|c|}{\textcolor{darkgray}{\textbf{Magnets}}} \\ \hline\hline
{Target solenoid }&{Develop conductor, winding and magnet technology}&{\SI{1}{\meter} inner / \SI{2.3}{\meter} outer diameters, \SI{1.4}{\meter} length}, \SI{20}{\tesla} at \SI{20}{\kelvin} \\\hline

{Final cooling solenoid }&{Build and test \acrshort{hts} prototype}&{\SI{50}{\milli\meter} bore, \SI{15}{\centi\meter} length, \SI{40}{\tesla} at \SI{4}{\kelvin}} \\\hline

{Fast-ramping magnet system}&{Prototype magnet string and power converter}&{\SI{30}{\milli\meter} $\times$ \SI{100}{\milli\meter}, \SI{1.8}{\tesla}, \SI{3.3}{\tesla\per\second}} \\\hline\hline

\multicolumn{3}{|c|}{\textcolor{darkgray}{\textbf{Radio frequency}}}\\\hline\hline

{\acrshort{rf} test stands}&{Assess cavity breakdown rate in magnetic field}&{\num{20}--\SI{32}{\mega\volt\per\meter}, \SI{704}{\mega\hertz}--\SI{3}{\giga\hertz} cavities in \num{7}--\SI{10}{\tesla}} \\\hline\hline

\multicolumn{3}{|c|}{\textcolor{darkgray}{\textbf{Muon Cooling}}} \\\hline\hline

{5-cell module }&{Build and test first 5-cell cooling module}&{} \\ \hline\hline

\multicolumn{3}{|c|}{\textcolor{darkgray}{\textbf{Design \& Other Technologies}}}
\\\hline\hline

{Neutrino flux mover system}&{Prototype components and tests as needed} &{Range to reach O ($\pm$\SI{1}{\milli\radian})} \\\hline

 \end{tabular}
\label{tab:deliverables}
\end{table}

The Muon Cooling Technology Demonstration programme, aiming at validating with beam the ionization cooling technology in a typical section of the cooling channel, is a necessary condition for the assessment of the feasibility of the \acrshort{mc} and of its performance.

As pointed out in Section~\ref{sec:env_MC} neutrino flux assessment and mitigation remains a critical issue, particularly with regard to its impact on potential host sites. A thorough evaluation of neutrino flux at different energy levels and site locations is needed~\cite{bib:LDG_RD_review_ESPP2026}.

\subsection{Construction and installation costs}

At this stage of the study an estimate of the construction and installation costs is subject to significant uncertainties and only a cost range can be specified. The construction and installation costs of the \acrshort{mc} at \acrshort{cern} is dominated by the magnets (in particular, solenoids) which account for 37 to 45\% of the total costs depending on the energy (3.2 or \SI{7.6}{\TeV}). Then comes the \acrshort{rf}, at 26 to 30\% of the total. \acrshort{ce} and general infrastructures exhibit a limited cost (between 15 and 17\% of the total), provided that the feasibility of installing the \acrshort{rcs}s in the \acrshort{sps} and \acrshort{lhc} is demonstrated. 

\begin{figure}[h!]
\centering
\includegraphics[width=0.8\textwidth]{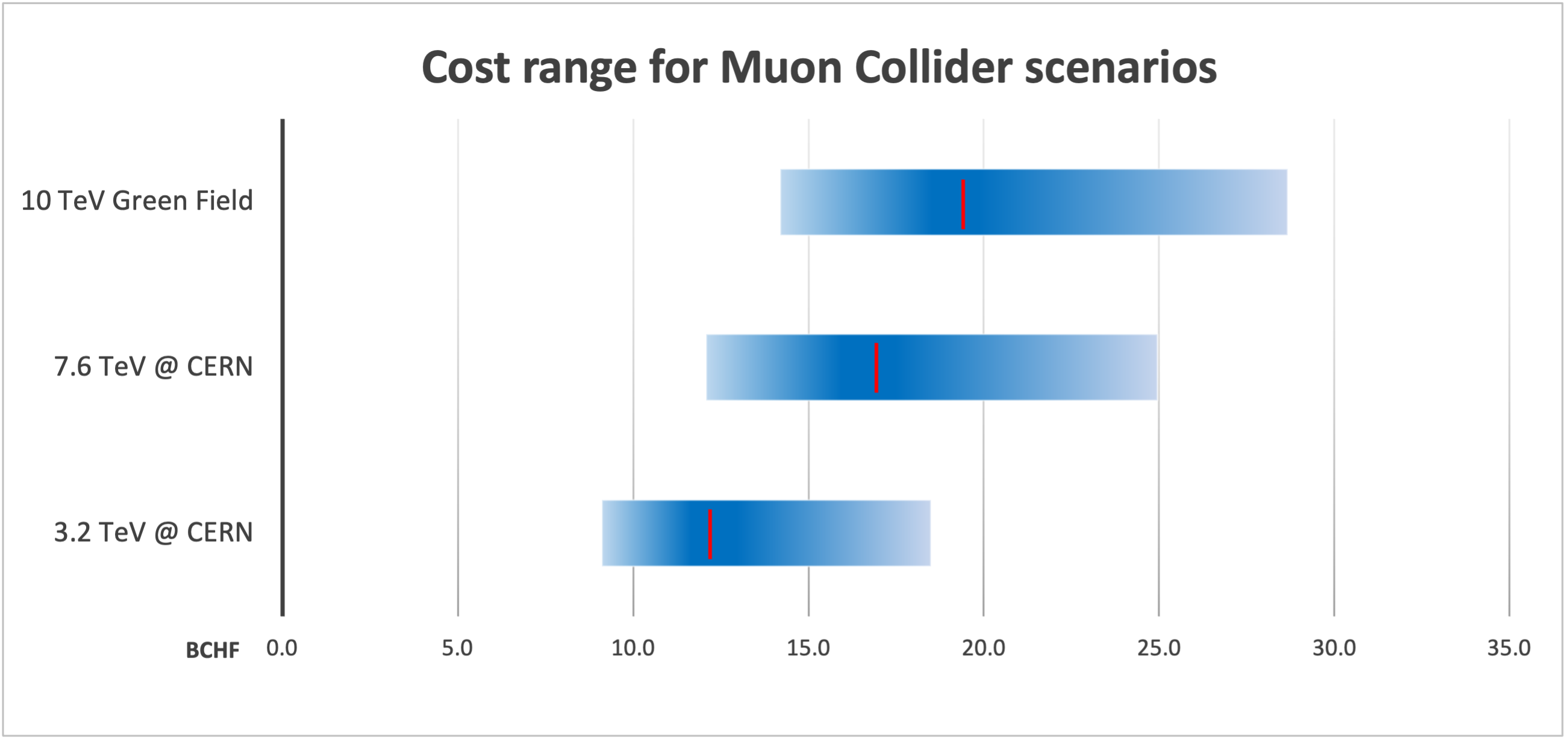}
 \caption{Cost range for a \acrshort{mc} implementation at \acrshort{cern} using existing infrastructure and for a site-independent implementation. The cost of the two experiments is not included in the estimate.}
\label{fig:Cost_MC}
\end{figure}

The estimated capital cost range is provided for the \acrshort{cern} scenario and compared to a green-field realization in Fig.~\ref{fig:Cost_MC}. The cost of the detectors is not included.

\subsection{Project timeline}

To guide the definition of the \acrshort{rd} programme, an ambitious technically-limited timeline has been explored and it is shown in Table~\ref{tab:timeline_MC_revised2}. This scenario assumes a firm commitment to implementing a muon collider in Europe as soon as possible after the \acrshort{hllhc} and the reuse of existing \acrshort{cern} infrastructure to minimize \acrshort{ce} costs and limits the collider’s center-of-mass energy to approximately \SI{3}{\TeV}.

\begin{table}[h!]
\caption{\acrshort{mc} timeline.}
\centering
\footnotesize
\begin{tabular}{|l||c|}\hline
Milestone & Muon Collider \\ \hline\hline
Construction of \acrshort{rf} test stands & 2025--2028 \\
Production of test cavities & 2026--2039\\
  Operation of test stands & 2027--2040\\
\hline
\hline
Demonstration Phase & $\mathrm{T_0}$--($\mathrm{T_0}+7$) \\
\hline
Demonstrator technical design &\\
Construction of initial demonstrator & \\
Construction of muon cooling module (five cells) &\\  
 Definition of the placement scenario for the collider  &  \\
\hline
\hline
Project Preparation Phase & $\mathrm{T_1}$--($\mathrm{T_1}+5$) \\
\hline
Final demonstrator & \\
Implementation studies with the Host states & \\
Environmental evaluation \& project authorisation processes &  \\
Main technologies \acrshort{rd} completion &  \\ 
Industrialisation of key components & \\
Engineering Design completion & \\
\hline
\hline
Construction phase (from ground breaking) & $\mathrm{T_2}$--($\mathrm{T_2}+9 $)\\
\hline
Civil engineering & \\
\acrshort{ti} installation  & \\
Component construction & \\
Accelerator \acrshort{hw} installation
 & \\
\acrshort{hw} commissioning & \\
Beam commissioning & \\ 
\hline
\hline
Physics operation start &  $\mathrm{T_2}+10 $\\
\hline
\end{tabular}
\label{tab:timeline_MC_revised2}
\end{table}

The initial phase of the project focuses on advancing the design and the required technologies to a level that allows the feasibility of a muon collider to be demonstrated with confidence. Key expected deliverables of this phase are listed in Table~\ref{tab:deliverables}. 
Part of the \acrshort{rd} programme (in particular, on the \acrshort{rf} test stands and cavities) has already started.
At present, \acrshort{rd} progress is limited by available resources, both in Europe and particularly in the \acrshort{us}, where organisational structures are still being established. It is therefore important to increase the design effort in order to complete a start-to-end design of the collider and to launch the \acrshort{rd} for technologies that have the strongest impact on the timeline, this could be launched following the outcome of the \acrshort{esppu}.

The construction of the Initial Cooling Demonstrator demands significant resources and will require a specific decision (at time $\mathrm{T_0}$ in Table~\ref{tab:timeline_MC_revised2}) after a revision of the 
Roadmap by the \acrshort{imcc} planned for 2028, following the results of the \acrshort{esppu} and the subsequent mid-term review panel recommended by the \acrshort{p5} report. 
Once the muon cooling technology has been successfully demonstrated, the Project Preparation Phase can be launched ($\mathrm{T_1}$). Upon successful completion, the ground breaking and construction can start ($\mathrm{T_2}$).

%% file: include/05-OptionsoutsideCERN/OptionsoutsideCERN.tex
\section{Future collider proposals outside \acrshort{cern}}\label{sec:outsideCERN}
In this Section we give an overview of two collider projects proposed for other regions:

\begin{itemize}
   \item the \acrfull{cepc} in China;
   \item the \acrfull{ilc} in Japan;
\end{itemize}
both are green-field projects, therefore requiring each the construction of a laboratory campus in support of the research and technical activities at the accelerator complex and the experiments.

The following aspects are discussed:
\begin{itemize}
    \item Main parameters and performance
    \item Technical readiness
    \item Accelerator and experiments construction and installation: material costs
    \item Accelerator construction and installation: human resources
    \item Project timeline.
\end{itemize}

\subsection{Main parameters and performance}

\paragraph*{\acrshort{cepc}}
The~\acrfull{cepc} features a~\SI{100}{\km} collider with two~\acrshort{ip}s. Its injector chain includes a~\SI{30}{\GeV}~\SI{1.8}{\km}-long linac, a~\SI{1.1}{\GeV} positron damping ring and a Booster synchrotron accelerating the electron and positron beams from~\SI{30}{\GeV} to the beam collision energy. The~\acrshort{cepc} baseline design involves three operating modes (ZH, Z, and WW, in chronological order) with a synchrotron radiation power of~\SI{30}{\MW} per beam for the ZH and WW operating modes, and \SI{10}{\MW} per beam for the Z-mode. 
Possible upgrades include \cite{bib:CEPC_TDR_accelerator}:
\begin{itemize}
    \item increase of the power to \SI{30}{\MW} per beam for the Z-mode of operation;
    \item increase of the power to \SI{50}{\MW} per beam for the ZH, Z and WW modes of operation;
    \item increase the \acrshort{com}\ collision energy to~\SI{360}{\GeV} for t$\mathrm{\bar t}$ operation at \SI{30}{\MW} and \SI{50}{\MW} per beam.
\end{itemize}

The transfer lines connecting the pre-injectors with the booster in the main collider tunnel are approximately~\SI{2}{\km} long~\cite{bib:CEPC_TDR_accelerator}. 
A~\acrshort{sc} proton-proton collider, the \acrfull{sppc}, designed to operate at a \acrshort{com}\ energy of \SI{125}{\TeV} with~\SI{20}{\tesla} \acrshort{sc} dipoles, could be installed at a later stage in the same tunnel. This could operate simultaneously with the~\acrshort{cepc} collider and booster~\cite{bib:CEPC_TDR_accelerator}, though the integration of the three accelerators and their services in the same tunnel (including the cryogenics for~\acrshort{sppc}) appears to be challenging given the size of the tunnel.

The main parameters of \acrshort{cepc} as provided in Ref.~\cite{bib:CEPC_TDR_accelerator} are listed in Table~\ref{tab:Parameters_ILC_CEPC}: the integrated luminosity has been estimated assuming data taking over 150 days  (corresponding to a 60\% machine availability over a scheduled operation period of 250 days) \cite{bib:CEPC_TDR_accelerator} while 139 days have been assumed, in general, in Table~\ref{tab:operational_year}. No luminosity ramp-up has been assumed \cite{bib:CEPC_TDR_accelerator}, differently from what has been done for \acrshort{fccee} and \acrshort{clic} in Table~\ref{tab:Parameters_CLIC_FCC_LCF}. The yearly electricity consumption has been determined assuming \SI{5000}{\hour} (full-power equivalent) per year \cite{bib:CEPC_ESPP2026}.

The main~\acrshort{cepc} parameters, re-scaled for the operational scenario defined in Table~\ref{tab:operational_year}, are listed in Table~\ref{tab:Parameters_ILC_CEPC_rescaled}. 
On-axis injection is considered for the ZH mode of operation; this will require switching off the high-voltage of the detectors during this process \cite{bib:CEPC_TDR_accelerator}, with a corresponding reduction of the fraction of time for data-taking when the machine is available (assumed to be 100\% in Table~\ref{tab:operational_year}). 
Finally, the power consumption does not include the contribution of the Science Campus and of the off-line computing.

\begin{table}[h!]
 \caption{High-level parameters of the \acrshort{ilc} 250 in Japan~\cite{bib:IDT_ESPP2026} and of \acrshort{cepc} for different options~\cite{bib:CEPC_TDR_accelerator,bib:CEPC_ESPP2026} (baseline parameters for \acrshort{cepc} are in \textbf{bold}), re-scaled respectively to the \acrshort{lcf} 250~\acrshort{lp} and \acrshort{fccee} operational year (see Table~\ref{tab:operational_year}). For the \acrshort{ilc} we follow an operational scenario adapted from~\cite{bib:ilc_snowmass2022}, with 10 years of operation including three years of ramp-up at one third the luminosity, and no luminosity upgrade (not costed in Ref.~\cite{bib:IDT_ESPP2026}). The integrated luminosity over the full programme is also adapted from~\cite{bib:ilc_snowmass2022}, following such a scenario. The instantaneous and integrated luminosity in-between parentheses includes also the contribution from energies below 99\% of the \acrshort{com}\ energy $\sqrt{s}$. 
 }
\centering
\scriptsize
 \begin{threeparttable}
 \begin{tabular}{|l||cccc|c|}\hline
~& \multicolumn{4}{c|}{\acrshort{cepc}} & \acrshort{ilc} 250 \\ \hline\hline

Circumference/length collider tunnel~[\SI{}{\km}] & \multicolumn{4}{c|}{99.955} & 20.5 \\ \hline
Number of experiments (\acrshort{ip}s) & \multicolumn{4}{c|} {2} & 2 (1)\tnotex{tab:Parameters_ILC_CEPC_rescaled:0} \\\hline
\acrshort{com}~energy [\SI{}{\GeV}] & \textbf{91} & \textbf{160} & \textbf{240} & 360 & 250 \\ \hline

Longitudinal polarisation (e$^-$ / e$^+$) & \multicolumn{4}{c|}{0~/~0\tnotex{tab:Parameters_ILC_CEPC_rescaled:1}} & 0.8 / 0.3 \\ \hline

Number of years of operation (total) & \textbf{2} & \textbf{1} & \textbf{10} & 5 & 10 \\ \hline
Nominal years of operation (equivalent)\tnotex{tab:Parameters_ILC_CEPC_rescaled:2} & \textbf{2} & \textbf{1} & \textbf{10} & 5 & 8 \\\hline

Synchrotron radiation power per beam~[\SI{}{\MW}] & \textbf{10}~/~30~/~50 & \textbf{30}~/~50 & \textbf{30}~/~50& 30~/~50 & -- \\\hline

Instantaneous luminosity per~\acrshort{ip} & \multirow{2}{*}{\textbf{38}~/~115~/~192} & \multirow{2}{*}{\textbf{16}~/~26.7} & \multirow{2}{*}{\textbf{5}~/~8.3} & \multirow{2}{*}{0.5~/~0.8} & \multirow{2}{*}{1 (1.35)} \\
above $0.99~\sqrt{s}$ (total)~[\(10^{34}\,\SI{}{\per\square\centi\meter\per\second}\)] & & & & & \\\hline

Integrated luminosity above $0.99~\sqrt{s}$ (total) & \multirow{2}{*}{\textbf{9.1}~/~27.6~/~46.1} & \multirow{2}{*}{\textbf{3.8}~/~6.4} & \multirow{2}{*}{\textbf{1.2}~/~2.0} & \multirow{2}{*}{0.12~/~0.19} & \multirow{2}{*}{0.12 (0.16)} \\
over all~\acrshort{ip}s per year of nominal operation~[\SI{}{\per\atto\barn}/y]
& \multicolumn{4}{c|}{} & \\ \hline

Integrated luminosity above $0.99~\sqrt{s}$ (total) & \multirow{2}{*}{\textbf{18.3}~/~55.2~/~92.2} & \multirow{2}{*}{\textbf{3.8}~/~6.4} & \multirow{2}{*}{\textbf{12.0}~/~19.9} & \multirow{2}{*}{0.60~/~0.96} & \multirow{2}{*}{ 1 (1.3)} \\
over all~\acrshort{ip}s over the full programme~[\SI{}{\per\atto\barn}] & \multicolumn{4}{c|}{} & \\ \hline

Peak power consumption [\SI{}{\MW}] & \textbf{100}\tnotex{tab:Parameters_ILC_CEPC_rescaled:3}~/~203~/~287 & \textbf{225}~/~299 & \textbf{262}~/~339 & 358~/~432 & 111 \\ \hline
Electricity consumption per year of nominal operation [\SI{}{\tera\watt\hour}/y]\tnotex{tab:Parameters_ILC_CEPC_rescaled:4} & \textbf{0.47}~/~0.95~/~1.3 & \textbf{1.1}~/~1.4 & \textbf{1.3}~/~1.6 & 1.8~/~2.2 & 0.64 \\ \hline

 \end{tabular}
 \begin{tablenotes}
    \item[a] \label{tab:Parameters_ILC_CEPC_rescaled:0} Two experiments at a single \acrshort{ip} (``push-pull'' mode).
    \item[b] \label{tab:Parameters_ILC_CEPC_rescaled:1} No longitudinal polarisation is considered in the baseline, but such a possibility is being explored, especially at the Z energy.
    \item[c] \label{tab:Parameters_ILC_CEPC_rescaled:2} This row lists the equivalent number of years of operation at nominal instantaneous luminosity, hence taking into account the luminosity ramp-up.
    \item[d] \label{tab:Parameters_ILC_CEPC_rescaled:3} Extrapolated from \cite{bib:CEPC_TDR_accelerator}.
    \item[e] \label{tab:Parameters_ILC_CEPC_rescaled:4} Computed from the peak power consumption and the assumptions on the operational year (see Table~\ref{tab:operational_year}).
 \end{tablenotes}
 \end{threeparttable}
\label{tab:Parameters_ILC_CEPC_rescaled}
\end{table}

\paragraph*{\acrshort{ilc}}\label{sec:ILC_intro}

The \acrfull{ilc}~\cite{bib:ILC_TDR_vol3_II} is a linear electron-positron collider relying on Superconducting radio-frequency (\acrshort{srf}) cavities, foreseen to operate at \acrshort{com}\ energies from \SI{250}{\GeV} up to \SI{1}{\TeV}, thanks to possible tunnel extensions and upgrades. The \acrshort{ilc} would be initially operated as a Higgs factory (\SI{250}{\GeV} \acrshort{com}) for 10 years~\cite{bib:ilc_snowmass2022}. The main accelerator fits into a \SI{20.5}{\km}-long tunnel~\cite{bib:ilc_snowmass2022} (\SI{9.5}{\m}-wide~\cite{bib:ilc_snowmass2022} and \SI{5.5}{\m}-high~\cite{bib:ILC_CEP_2020}), and consists of two arms: one of \SI{9.65}{\km} containing the positron main linac and the low emittance transfer line---\acrfull{rtml}, and one of $\SI{10.85}{\km}$ including the electron main linac, the \SI{1.1}{\km} undulator section for the polarised positron source, and the \acrshort{rtml}. A conceptual layout of the \acrshort{ilc} is shown in Fig.~\ref{fig:ILC_layout}.

The \acrshort{rf} powering equipment is separated by a shielding wall of~\SI{1.5}{\m} thickness from the main linac, and access to it is possible when beam operation is stopped but the \acrshort{rf} cavities are powered, as the shielding protects the personnel from the X-rays generated in the \acrshort{rf} cavities. Access during beam operation would have required a wall thickness of~\SI{3.5}{\m} which has been abandoned for cost considerations~\cite{bib:ilc_snowmass2022}.

The electron and positron sources produce both \SI{5}{\GeV} polarised beams (1312 bunches with \SI{554}{\ns} bunch spacing), which are injected into the damping rings sharing the same \SI{3.2}{\km}-long tunnel, before being extracted towards the main tunnel. Operation at the Z pole (\SI{91.2}{\GeV} \acrshort{com}\ energy) is also envisaged, using the same linac and cavities with a reduced klystron power~\cite{bib:ilc_snowmass2022}, albeit with a more complex cycle, as the positrons production requires \SI{125}{\GeV} electrons from the main linac---the latter would then have to alternate between \SI{45.6}{\GeV} and \SI{125}{\GeV}.
The interaction point features two detectors operating on movable platforms, in a ``push-pull'' scheme (exchange is possible in less than 24 hours)~\cite{bib:ilc_snowmass2022}. The machine may also use beam dumps for fixed-target experiments, specially the main dumps, tune-up dumps, and the photon dump from the positron source undulator~\cite{bib:ilc_snowmass2022}. The baseline features polarization of both the electron and positron beams~\cite{bib:ilc_snowmass2022}.

Main summary parameters are given in Table~\ref{tab:Parameters_ILC_CEPC}, while in Table~\ref{tab:Parameters_ILC_CEPC_rescaled} the integrated luminosity and electricity consumption are re-scaled to the \acrshort{lcf} 250~\acrshort{lp} operational year (see Table~\ref{tab:operational_year}), in particular 139 days of data taking per year are considered, as opposed to 170--180 days foreseen in the \acrshort{ilc} operational scenario~\cite{bib:ilc_snowmass2022}.

On top of the baseline scenario, several options are also envisaged and (at least partly) costed: the increase of the centre-of-mass energy from 250 to \SI{500}{\GeV}, the use of an electron-driven positron source instead of the undulator-driven one (as a backup plan), and/or the addition of a second interaction point.

\subsection{Technical readiness and \acrshort{rd} requirements
}\label{sec:TRL_CEPCILC}

\paragraph*{\acrshort{cepc}}

The main technical requirements of \acrshort{cepc} are similar to those of \acrshort{fccee} and the overall technology readiness level is comparable (see Table~\ref{tab:FCCeeTRL}).

The collider \acrshort{rf} system is based on bulk-Nb \acrshort{srf} \SI{650}{\MHz} 2-cell cavities operated at \SI{2}{\kelvin}. The cavity has been designed to meet the following specifications: a vertical acceptance test with Q$_0~>~5\times10^{10}$ at \SI{30}{\mega\volt\per\m}, a horizontal acceptance test with Q$_0 > 4\times10^{10}$ at \SI{28}{\mega\volt\per\m}, and a normal operation gradient of \SI{28}{\mega\volt\per\m} with Q$_0 > 3\times10^{10}$ when integrated in the cryomodule. Q$_0\approx 7\times10^{9}$ at $\approx$~\SI{28}{\mega\volt\per\m} has been achieved in a vertical test of individual cavities, so far~\cite{bib:CEPC_TDR_accelerator}.

The power consumption of~\acrshort{cepc} is mostly driven by the power consumption of the \acrshort{rf} system. The estimated values of the overall electricity consumption are based on the assumption that the \acrshort{rf} klystrons can reach an efficiency of 80\%. An efficiency of 78.5\% has been reached in the laboratory at~\SI{800}{\kilo\watt} in~\acrshort{cw} mode~\cite{bib:CEPC_ESPP2026}.

While the technical readiness for a number of items is similar to that reached at~\acrshort{cern} for similar~\acrshort{fccee} components, the deadline for the completion of the~\acrshort{rd} phase by 2027 (see Table~\ref{tab:CEPCILCtimeline}) is ambitious.

\paragraph*{\acrshort{ilc}}

Several items have to be developed within the four years foreseen for the pre-lab phase of the \acrshort{ilc}, and  are already the subject of intense \acrshort{rd} by the \acrfull{itn}, in particular the \acrshort{srf} cavities, the electron and positron sources, and the nano beam technology. One of the main challenges regards the $\sim 8000$ \acrshort{rf} cavities, which are 10 times more numerous than the \acrfull{xfel} ones, with 30\% more gradient (\SI{31.5}{\mega\volt\per\m}) and a $\mathrm{Q_0}$ above $10^{10}$. Other critical \acrshort{rd} topics are the undulator-based positron source and in particular its rotating photon target, and the damping rings and final focus, aiming at \SI{7.7}{\nm} vertical beam size at the \acrshort{ip}. The technological challenges and corresponding \acrshort{trl} are essentially the same as those of the \acrshort{lcf}, which can be found in Table~\ref{tab:LCFTRL}.

Regarding civil engineering and tunneling, the Kitakami site in the Tohoku region, offers good geological conditions (good quality granite and no active seismic faults). Geological survey and seismic measurements were performed; the impact of a large earthquakes (e.g. the 2011 one---magnitude 9) was checked and will not be an issue for operation. Up to \SI{50}{\km} of tunnel can be dug, covering \acrshort{ilc} energy upgrades---operation may also continue while excavating a longer tunnel for an energy upgrade~\cite{bib:ilc_snowmass2022}.

\subsection{Construction and installation costs}

\paragraph*{\acrshort{cepc}}

A cost estimate (in~\acrshort{cny}---2023 prices) of the~\acrshort{cepc} project has been presented in Refs.~\cite{bib:CEPC_TDR_accelerator,bib:CEPC_ESPP2026}. The cost includes installation and commissioning of the accelerator hardware and technical infrastructure with a provision of 3\% of the corresponding material cost for each of these activities.
The cost summary table~(in~\acrshort{cny}, \acrshort{chf} and after correction for the~\acrshort{ppp} as discussed in Section~\ref{sec:criteriacost}) is presented in Table~\ref{tab:CEPCcost}.

Compensation for resettlement and land acquisition as well as territorial developments, such as the connection to site-specific existing networks (e.g. utility power lines, raw-water adduction, roads) are not included~\cite{bib:CEPC_TDR_accelerator,bib:indicoCEReviewCEPC}.
The laboratory campus has not been costed.

Material price averages over the 2018--2022 period have been considered in Ref.~\cite{bib:CEPC_TDR_accelerator} and therefore they do not fully account for the inflation to 2023 prices; the average 2023 Chinese~\acrshort{ppi} has increased by 3.4\% as compared to its average value in the period 2018--2022.

The cost estimate includes gamma-ray beam lines, accelerator physics studies and project management in addition to the items indicated in Section~\ref{sec:criteriacost}. A contingency of 8\% of the overall cost of the project is assumed in Ref.~\cite{bib:CEPC_TDR_accelerator}, lower than the present level of uncertainty of the cost estimate. 

\paragraph*{\acrshort{ilc}}

First cost estimates were given in the 2013 \acrshort{tdr}~\cite{bib:ILC_TDR_vol3_II} for the \acrshort{ilc} 500 collider with a very detailed breakdown, using a bottom-up approach. These were then re-evaluated in 2017, when the baseline \acrshort{com}\ energy became \SI{250}{\GeV}, and more recently in 2024, the latter version having been submitted to the \acrshort{esppu}~\cite{bib:IDT_ESPP2026}. Around 75\% of the cost were hence re-evaluated in 2024, in particular the accelerator components (\acrshort{srf} technology), the utilities and conventional facilities, and the civil engineering. The costs for preparatory work such as engineering design, land acquisition, local infrastructure’s access to the site (roads, electricity, and water) were not included, as well as the experiments. All costs, except civil engineering, were expressed in 2024 \acrshort{usd} (also called \acrshort{ilcu}, defined by 1 \acrshort{ilcu} = 1 \acrshort{usd} in 2024), thanks to \acrshort{ppp} conversion rates (using specific values for either machinery \& equipment, or material~\cite{bib:IDT_ESPP2026}, which are different from the values of Table~\ref{tab:ppp_exchrates}). The civil engineering costs were instead expressed in \acrshort{jpy}---this part of the cost update was performed following the strict guidelines of the Japanese Government and the national tunnel costing standards as for other major Japanese projects. The costs were reviewed by an international committee~\cite{bib:IDT_ESPP2026}.

The baseline costs of \acrshort{ilc} 250 are shown in Table~\ref{tab:ILCcost}, converted to \acrshort{chf} using the \acrshort{ppp} from Table~\ref{tab:ppp_exchrates}. No cost classes are available, but the uncertainty was evaluated at 30\%, broadly corresponding to cost class~3. 

Three options were evaluated in addition in Ref.~\cite{bib:IDT_ESPP2026}: 
\begin{itemize}
    \item \acrshort{com}\ energy increase from 250 to \SI{500}{\GeV}, costed at 3.9--4.2~B\acrshort{usd} for hardware (accelerator components plus conventional facilities) and 55~B\acrshort{jpy} for civil engineering, for a total (after \acrshort{ppp} conversion) of 4.6 B\acrshort{chf} equivalent,
    \item electron-driven positron source (backup plan if the undulator-based positron source is not available in time), costed at 0.2~B\acrshort{usd} for hardware and 12.5~B\acrshort{jpy} for \acrshort{ce}, for a total of 320 M\acrshort{chf} equivalent,
    \item an additional interaction point, for 0.5~B\acrshort{usd} (480~M\acrshort{chf} equivalent), including a second \acrshort{bds} and interaction point, but excluding \acrshort{ce} work and the new beam splitting systems needed at both \acrshort{bds} upstream ends.
\end{itemize}

\begin{landscape}
\begin{table}[!h]
    \caption{Cost summary table for the~\acrshort{cepc} project for the baseline scenario with two experiments and its possible upgrades to high beam power and energy. The first three columns are derived from the data presented in Ref.~\cite{bib:CEPC_TDR_accelerator}. The following six columns are derived applying the currency exchange rate, and that rate corrected for the~\acrshort{ppp} as described in Section~\ref{sec:criteriacost}. The sub-total represents the sum of the cost items considered for the other projects described in this document. The cost of installation and commissioning has been included in each of the corresponding cost items to have a similar categorisation as for the other projects. Inflation between 2023 and the 2018--2022 period (average \acrshort{ppi} has increased by approximately 3.4\%) has not been included.}
    \centering
    \scriptsize
    \begin{threeparttable}
    \begin{tabular}{|m{3.4cm}||C{1.8cm}|C{1.2cm}|C{1.2cm}||C{1.8cm}|C{1.2cm}|C{1.2cm}||C{1.8cm}|C{1.2cm}|C{1.2cm}|}
    \hline
        \multirow{2}{*}{}
        &\SI{10}{\MW} Z \SI{30}{\MW} W ZH \textbf{baseline}&\SI{30}{\MW} Z W ZH&\SI{50}{\MW} Z W ZH t$\mathrm{\bar t}$&\SI{10}{\MW} Z \SI{30}{\MW}  W ZH \textbf{baseline} & \SI{30}{\MW} Z W ZH& \SI{50}{\MW} Z W ZH t$\mathrm{\bar t}$& \SI{10}{\MW} Z \SI{30}{\MW}  W ZH  \textbf{baseline} & \SI{30}{\MW} Z W ZH& \SI{50}{\MW} Z W ZH t$\mathrm{\bar t}$ \\ 
        & \multicolumn{3}{c||}{[100 M\acrshort{cny}---2023]} & \multicolumn{3}{c||}{[M\acrshort{chf}---2023]} & \multicolumn{3}{c|}{\acrshort{ppp} [M\acrshort{chf}---2023] } \\ \hline\hline
        Booster and collider & 151 & 164 & 212 & 1856 & 2015 & 2601 & 4016 & 4360 & 5626 \\ \hline
        Pre-injectors and transfer lines & 21 & 21 & 21 & 263 & 263 & 263 & 570 & 570 & 570 \\ \hline
        Civil engineering\tnotex{tab:CEPCcost:1}& 71 & 71 & 71 & 872 & 872 & 872 & 1885 & 1885 & 1885 \\ \hline
        Technical infrastructure\tnotex{tab:CEPCcost:2} & 46 & 48 & 65 & 570 & 583 & 802 & 1234 & 1262 & 1735 \\ \hline
        Experiments\tnotex{tab:CEPCcost:3} & 40 & 40 & 40 & 491 & 491 & 491 & 1062 & 1062 & 1062 \\ \hline
         \textbf{Sub-total }& \textbf{330} & \textbf{344} & \textbf{410} & \textbf{4053}& \textbf{4225} & \textbf{5029} & \textbf{8768} & \textbf{9139} & \textbf{10879} \\ \hline
        Gamma ray beam lines & 3 & 3 & 3 & 37 & 37 & 37 & 80 & 80 & 80 \\ \hline
        Accelerator physics & 1 & 1 & 1 & 10 & 10 & 10 & 21 & 21 & 21 \\ \hline
        Project management & 3 & 3 & 3 & 37 & 37 & 37 & 80 & 80 & 80 \\ \hline
        Contingency & 27 & 28 & 33 & 331 & 345 & 409 & 716 & 746 & 885 \\ \hline
       \textbf{Total }& \textbf{364} & \textbf{379} & \textbf{450} & \textbf{4467 }& \textbf{4653} & \textbf{5521} & \textbf{9664} & \textbf{10066} & \textbf{11944} \\ \hline
    \end{tabular}
    \begin{tablenotes}
     \item[a] \label{tab:CEPCcost:1} Including transport and logistics.
     \item[b] \label{tab:CEPCcost:2} Excluding transport and logistics.
     \item[c] \label{tab:CEPCcost:3} Total cost of two experiments.
    \end{tablenotes}
    \end{threeparttable}
    \label{tab:CEPCcost}
\end{table}
\end{landscape}

 \begin{table}[h!]
   \caption{Cost summary table for \acrshort{ilc} 250 in Japan~\cite{bib:IDT_ESPP2026, bib:ILC_cost_review_Abe_2025}, in various currencies and 2024 prices. The last but one column (in M\acrshort{chf}) is derived applying the currency exchange rate (Table~\ref{tab:exchrates}), while the last column includes the \acrshort{ppp} (Table~\ref{tab:ppp_exchrates}).
  }
 \centering
 \footnotesize
 \begin{tabular}{|m{4cm}|C{1.5cm}|C{2.5cm}|C{1.5cm}|C{1.5cm}|C{2cm}|}\hline
 ~ & \multirow{2}{*}{Uncertainty} & \multicolumn{4}{c|}{Cost}  \\
 ~ & & [M\acrshort{usd}] (or [M\acrshort{ilcu}]) & [B\acrshort{jpy}] & [M\acrshort{chf}] & \acrshort{ppp} [M\acrshort{chf}] \\ \hline\hline
\acrshort{srf} & 30\% & 3690 & & 3250 & 3560  \\  \hline
Other accelerator components & 30\% & 1710 & & 1510 & 1650 \\ \hline
Civil engineering & 30\% &  & 196 & 1150 &  2000  \\  \hline
Technical Infrastructure
& 30\% & 1380 & & 1220 & 1330 \\ \hline
Experiments & \multicolumn{5}{c|}{N/A} \\ \hline
\textbf{Total} 
& \textbf{30\%} & & & \textbf{7120} & \textbf{8550} \\  \hline
 \end{tabular}
 \label{tab:ILCcost}
 \end{table}
 
\subsection{Accelerator construction and installation: human resources}\label{sec:manpower_CEPCILC}

\paragraph*{\acrshort{cepc}} ~\label{sec:CEPC_manpower}
According to Ref.~\cite{bib:CEPC_ESPP2026}, the human resources necessary for the accelerator construction and installation amount to about~\SI{15000}{\acrshort{ftey}} over a period of seven years. According to Eq.~\eqref{eq:FTEvsCAPEXconstruction}, approximately~\SI{11000}{\acrshort{ftey}} would be required. A project the size of~\acrshort{cepc}, if conducted in a short timespan as proposed, might exceed the skilled personnel resources available in national institutes.

\paragraph*{\acrshort{ilc}}

After a first, detailed estimate in the 2013 \acrshort{tdr}, the labor force needed for the construction of the \acrshort{ilc} was re-evaluated in 2017, following the decrease in energy to \SI{250}{\GeV}. A simple scaling by 75\% of the \acrshort{tdr} value was then used, obtaining \SI{10120}{\acrshort{ftey}}~\cite{bib:ilc_snowmass2022,bib:ilc_mext}, over a period of nine years. This number has not been updated since then. Applying Eq.~\eqref{eq:FTEvsCAPEXconstruction} would give instead \SI{11300}{\acrshort{ftey}}.

\subsection{Project timeline}

The timelines for the construction of the colliders and the associated technical infrastructure are presented in this Section. These do not include the effort necessary for the construction of the full laboratory infrastructure in a green field, that might absorb a significant amount of resources and require considerable time.  

\paragraph*{\acrshort{cepc}}
The major milestones for the~\acrshort{cepc} project~\cite{bib:CEPC_TDR_accelerator,bib:CEPC_ESPP2026} are listed in Table~\ref{tab:CEPCILCtimeline}. A final selection of the site has not been made. The selection should be followed by the definition of the placement scenario and the preliminary implementation with the Host region and by an \Gls{eiagl} (\acrshort{eia}). Civil engineering is expected to take less than five years. The schedule is aggressive but considered to be feasible~\cite{bib:Osborne_private}, though with a significant uncertainty, pending more detailed geological investigations that will have to follow the site selection.

As mentioned in Section~\ref{sec:TRL_CEPCILC}, the timescale for the completion of the~\acrshort{rd} phase is ambitious, as well as the overall construction and installation schedule. 
The scale of the requirements on production, transport means as well as storage and logistics are significantly larger as compared to previous projects such as~\acrshort{bepcii}, ~\acrshort{csns} and~\acrshort{heps}.

\paragraph*{\acrshort{ilc}}

The technically-limited timeline for the \acrshort{ilc} project is presented in Table~\ref{tab:CEPCILCtimeline}. After project approval, the preparation should last four years (\acrshort{ilc} pre-lab construction and \acrshort{rd})~\cite{bib:ilc_snowmass2022,bib:IDT_ESPP2026}. Then, construction is assumed to last nine years, followed by one year of commissioning---this schedule has remained unchanged since the 2013 \acrshort{tdr}~\cite{bib:ILC_TDR_vol3_II}. 

For a possible energy upgrade, assuming that all cryomodules would be already available before shutting down the machine, a one-year shutdown is foreseen~\cite{bib:ilc_snowmass2022}.

\begin{table}[h!]
  \caption{Timeline for the~\acrshort{cepc}~~\cite{bib:CEPC_TDR_accelerator,bib:CEPC_ESPP2026} and~\acrshort{ilc} projects~\cite{bib:IDT_ESPP2026,bib:ILC_TDR_vol3_II}.}
\centering
\footnotesize
\begin{threeparttable}
\begin{tabular}{|L{8cm}||C{3cm}|C{3cm}|}\hline
Milestone ~&~\acrshort{cepc} & \acrshort{ilc} \\ \hline\hline

Conceptual Design Study & 2018 & 2007\tnotex{tab:CEPCILCtimeline:7} \\
Definition of the placement scenario
& N/A &  N/A \\
Preliminary implementation with the Host region
& N/A &  N/A \\
Feasibility Report ready & N/A\tnotex{tab:CEPCILCtimeline:3} & 2022\tnotex{tab:CEPCILCtimeline:8}  \\
Earliest Project Approval & 2025  &  $\mathrm{T_0}$ \\
Environmental evaluation \& project authorisation processes & N/A & N/A  \\
Main technologies \acrshort{rd} completion & 2027 & $\mathrm{T_0}+4$ \\ 
Technical Design Report ready\tnotex{tab:CEPCILCtimeline:4}&  2027 & $\mathrm{T_0}+4$ \\
Civil engineering  & 2028--2032\tnotex{tab:CEPCILCtimeline:6}  & $\mathrm{T_0}+4 $--$ \mathrm{T_0}+9$ \\
Installation  & 2033--2035\tnotex{tab:CEPCILCtimeline:5} & $\mathrm{T_0}+9 $--$ \mathrm{T_0}+13$ \\
\acrshort{hw} and beam commissioning & 2036--2037 & $\mathrm{T_0}+13$ \\\hline
\end{tabular}
 \begin{tablenotes}
 \item[a] \label{tab:CEPCILCtimeline:7} \acrfull{rdr}~\cite{bib:ILCReferenceDesignReport}.
 \item[b] \label{tab:CEPCILCtimeline:3} The~\acrshort{cepc}~\acrshort{tdr}~\cite{bib:CEPC_TDR_accelerator} does not correspond to a full Feasibility Report as defined in Section~\ref{sec:timelinedef}.
 Neither the final, nor the preliminary implementations with the host regions are discussed.
 \item[c] \label{tab:CEPCILCtimeline:8} The \acrshort{ilc} feasibility study can be considered done since 2022; it is documented in several reports: the 2013 \acrshort{tdr}~\cite{bib:ILC_TDR_vol3_II} for \acrshort{ilc} 500, the 2022 report to Snowmass~\cite{bib:ilc_snowmass2022} for the changes related to \acrshort{ilc} 250, and the 2020 ``Tohoku \acrshort{ilc} Civil Engineering Plan'' for the geological survey and \acrshort{ce} planning~\cite{bib:ILC_CEP_2020}. The latter was reviewed by a committee within the Japan Society of Civil Engineers.
 \item[d] \label{tab:CEPCILCtimeline:4} For~\acrshort{cepc} and~\acrshort{ilc} this is called~\acrfull{edr}.
 \item[e] \label{tab:CEPCILCtimeline:6} Assuming the drilling and blasting method for the tunnel construction. The~\acrshort{ce} could be completed one year earlier if the~\acrshort{tbm} method is used~\cite{bib:CEPC_TDR_accelerator}.
 \item[f] \label{tab:CEPCILCtimeline:5} The timeline for the installation of the detectors in the corresponding experimental caverns is not presented in Ref.~\cite{bib:CEPC_TDR_accelerator}. According to Ref.~\cite{bib:CEPC_ESPP2026} this might occur one or two years after the completion of the installation of the accelerator.
 \end{tablenotes}      
 \end{threeparttable}
\label{tab:CEPCILCtimeline}
\end{table}

Regarding the detectors, it has been noted~\cite{bib:ilc_snowmass2022} that the originally envisioned timeline for the approval of the first set of experiments to proceed towards \acrshort{tdr}, contemplated at least six to seven years from the launch of the calls for \acrfull{eoi}. So  time might be very limited for the standard process of \acrshort{eoi}, \acrfull{loi}, and the actual proposal for experiments at the \acrshort{ilc}.


%% file: include/06-ReuseLHCTunnel/ReuseLHCTunnel.tex
\section{Options reusing the \acrshort{lhc} tunnel}

\subsection{\acrshort{lep3}}

Motivated by first hints of a Higgs boson at the \acrshort{lhc}, a circular \epem~Higgs factory in the \acrshort{lhc} tunnel with a maximum \acrshort{com}\ energy of \SI{240}{\GeV}, \acrshort{lep3}, was first proposed in 2011 \cite{bib:Blondel:1406007}, together with a new machine at twice the circumference, DLEP~\cite{bib:Blondel:1406007}. A few months later also a circular Higgs factory of three times the circumference, TLEP, was considered~\cite{bib:eucardws,bib:Koratzinos:1558076},
 which eventually became the \acrshort{fccee} \cite{bib:FCC:2018evy}.

For the \acrshort{esppu} 2013, a proposal for \acrshort{lep3} was submitted~\cite{bib:lep3espp,bib:LEP3_2012,bib:Aleksan:1628377} (see Table~\ref{tab:tablep3}) but it did not receive significant support~\cite{bib:Nakada:2690131,bib:2020_ESPPU_Brochure}. A~\SI{1.3}{\GHz} \acrshort{srf} system was initially considered but later discarded given the anticipated beam currents. Therefore, already in 2012, an \acrshort{srf} frequency of~\SI{700}{\MHz} was considered \cite{bib:Koratzinos:1969192}. 

Further studies were performed in 2017~\cite{bib:Shatilov_LEP3_2017,bib:Oide_LEP3_2017} considering an~\SI{800}{\MHz} \acrshort{srf} system, four \acrshort{ip}s and a maximum \acrshort{com}\ energy of \SI{240}{\GeV}. The lattice considered has a rather aggressive design with half the cell length of the \acrshort{fccee} ZH mode. This implies a reduction of the magnet filling factor (unless nested magnets are used) and therefore of the main dipole bending radius as compared to that of \acrshort{lep} (see Table~\ref{tab:tablep3}), and also higher gradient quadrupoles and sextupoles than for \acrshort{fccee}.

More recently an additional proposal has been made \cite{bib:Virdee_LEP3_2025} to reuse the \acrshort{lep} tunnel to host an \epem collider providing collision at \acrshort{com}\ energies ranging from  \SIrange{91.2}{230}{\GeV} in two \acrshort{ip}s. 
All the \acrshort{lep3} studies envisage top-up injection from a cycling high-energy booster to the collider, similarly to what is proposed for \acrshort{fccee}. The high-energy booster, located in the same tunnel of the collider, would regularly accelerate the beams delivered from  a pre-injector chain to the collision energy. The machine parameters are listed in Table~\ref{tab:tablep3} together with those of \acrshort{lep2} and those of the earlier proposals and studies. The main dipole bending radius has been increased with respect to the value mentioned in Refs.~\cite{bib:Shatilov_LEP3_2017,bib:Oide_LEP3_2017} implying likely a further increase of the gradient of the quadrupoles and sextupoles to maintain a compact cell length. Optics and beam dynamics simulation still need to be performed \cite{bib:Virdee_LEP3_2025} to support the performance estimate. Scaling from the parameters listed in Refs.~\cite{bib:Shatilov_LEP3_2017,bib:Oide_LEP3_2017}, considering only two~\acrshort{ip}s and the lower maximum \acrshort{com}\ energy (\SI{230}{\GeV}) indicate a possible luminosity of \SI{1.5e34}{\per\square\centi\meter\per\second} (the lowest value of the range considered in Ref.~\cite{bib:Virdee_LEP3_2025} and lower than the value of \SI{1.8e34}{\per\square\centi\meter\per\second} used to estimate the integrated luminosities) at \SI{230}{\GeV} and a higher energy loss per turn by about 8\% as compared to Ref.~\cite{bib:Virdee_LEP3_2025}. 

The 2025 proposal \cite{bib:Virdee_LEP3_2025} is based on two separate \SI{800}{\MHz} \acrshort{srf} systems for the collider and the booster. Each of them would be distributed over two \acrshort{lhc} \acrshort{lss}. The high currents required for the operation at the Z-pole will likely demand the installation of two high-power \SI{400}{\MHz} \acrshort{srf} systems (one per beam and in separate \acrshort{lss}) for the collider for the operation at that energy, in order to compensate for the synchrotron radiation power loss of \SI{50}{\MW}. The \SI{400}{\MHz} \acrshort{srf} system will have to be dismantled and replaced by the \SI{800}{\MHz} \acrshort{srf} system to operate at the maximum energy therefore it will not be realistically possible to interleave low and high energy runs during the operational lifetime of the collider. 

The integration of the collider and booster rings in the~\acrshort{rf} \acrshort{lss} might be challenging and requires a detailed integration study to assess the need of civil engineering work, not contemplated in the proposal, in those areas.

\begin{table}[h!]
\caption{Tentative parameters of a \acrshort{lep3} Higgs factory as proposed in 2011 \cite{bib:lep3espp,bib:LEP3_2012,bib:Aleksan:1628377}, studied in 2017 ~\cite{bib:Shatilov_LEP3_2017,bib:Oide_LEP3_2017} and proposed in 2025~\cite{bib:Virdee_LEP3_2025} and compared with the corresponding parameters achieved by \acrshort{lep2}~\cite{bib:Aleksan:1628377,bib:Assmann_LEP2,bib:electricity_LEP2_2000} and those expected for \acrshort{fccee}.}
\centering
\scriptsize
\begin{threeparttable}
\begin{tabular}{|l|c||c||c|c||c|c||c|c|}
\hline 
Parameter &  \acrshort{lep2} &  \acrshort{lep3} - 2012 &\multicolumn{2}{c||}{\acrshort{lep3} - 2017}& \multicolumn{2}{c||}{\acrshort{lep3} - 2025} & \multicolumn{2}{c|}{\acrshort{fccee}}\\ \hline\hline
\acrshort{com}~energy~[\SI{}{\GeV}] & 209 & 240 & 91.2 & 240 & 91.2 & 230 & 91.2 & 240\\ \hline
Number of \acrshort{ip}s & 4 & 4  &\multicolumn{2}{c||}{4} &\multicolumn{2}{c||}{2} &\multicolumn{2}{c|}{4}\\ \hline
Main dipole bending radius [m] & 3096 & 2600 & \multicolumn{2}{c||}{2755} &\multicolumn{2}{c||}{2958} &\multicolumn{2}{c|}{10021}  \\ \hline
Beam current [\SI{}{\milli\ampere}] & 4 & 7.2 & 346 & 7.2 & 371 & 9 & 1292 & 26.8\\\hline
Peak luminosity per \acrshort{ip} [\SI{e34}{\per\square\centi\meter\per\second}]& 0.01 & 1.1 & 52 & 1.1 & 44 & 1.8  & 144 & 7.5 \\ \hline 
Int. luminosity per year over all \acrshort{ip}s [\SI{}{\per\atto\barn}/y]\tnotex{tab:tablep3:1}& $9\cdot10^{-4}$ & 0.53 & 25 & 0.53 & 10.6 & 0.43  & 69 & 3.6 \\ \hline 
Energy loss per turn $U_0$~[\SI{}{\GeV}] & 3.5 &  7 & 0.144 & 6.92  & 0.13 & 5.4 & 0.039 & 1.86 \\ \hline
\acrshort{rf} frequency [\SI{}{\MHz}] & 352 & 1300 & \multicolumn{2}{c||}{800} & \multicolumn{2}{c||}{800} & \multicolumn{2}{c|}{400}\\\hline
Total \acrshort{rf} Voltage [\SI{}{\giga\volt}] & 3.65 & 12 & 0.2 & 8 & 0.18 & 6 & 0.085 & 2.09\\\hline
Critical photon energy in the arc~[\SI{}{\MeV}] & 1.1  & 1.7  & 0.08 & 1.4   & 0.07  & 1.14 & 0.021 & 0.38 \\ \hline 
Synchrotron radiation power per beam ~[\SI{}{\MW}] & 11.5 &  50 &\multicolumn{2}{c||}{50}  &\multicolumn{2}{c||}{50} & \multicolumn{2}{c|}{50}\\ \hline
Peak power consumption~[\SI{}{\mega\watt}] & N/A & N/A  & N/A  & N/A  & N/A & 250 & 251 & 297 \\ \hline
Electricity consumption per year~[\SI{}{\tera\watt\hour}/y]\tnotex{tab:tablep3:1} & 0.57 & N/A  & N/A  & N/A  & N/A & 1.22 & 1.21 & 1.43 \\ \hline 
\end{tabular}
\begin{tablenotes}
    \item[a] \label{tab:tablep3:1} Computed from the peak power consumption and the assumptions on the operational year of \acrshort{fccee} (see Table~\ref{tab:operational_year}), except for \acrshort{lep2}.
\end{tablenotes}
\end{threeparttable}
\label{tab:tablep3} 
\end{table}

The \acrshort{lep3} electrical power consumption quoted by the proponents amounts to about \SI{250}{\MW} at top energy, comparable to that of the \acrshort{fccee}, but providing a lower luminosity and at lower maximum energy.

The cost of the project has been estimated extrapolating from the cost of the main systems for \acrshort{fccee} and it amounts to 3850~M\acrshort{chf}, including the total cost of two new experiments (795 M\acrshort{chf}), assuming the same cost per experiment considered for the other \epem collider proposals. \acrshort{cern} contribution to the detector construction plus host laboratory responsibility is 146~M\acrshort{chf}.  The cost estimate assumes that no \acrshort{ce} work will be required in the \acrshort{rf} \acrshort{lss}. This hypothesis would need to be validated by integration studies including the \acrshort{rf} system, the booster and the collider in the relevant sections. The requirements in terms of integration and cost of a possible distributed \acrshort{hts} magnet system that might be required to limit the power consumption to the value above-mentioned (but whose feasibility has not been demonstrated) have not been detailed either.

\acrshort{lep3} could be installed only after the end of the \acrshort{hllhc} programme and after the removal of the \acrshort{lhc} machine, implying that
\acrshort{lep3} physics could begin at the earliest in the second half of the 2040s.  The operation of \acrshort{lep3} will likely require a similar amount of resources (material and personnel) to that required for \acrshort{lhc}/\acrshort{hllhc}.

In conclusion, several technical aspects of this collider necessitate detailed design studies, and the performance goals need to be verified before this project, as proposed, can be considered as feasible.

\subsection{\acrshort{helhc}}

In 2010 an inter-departmental 
\acrshort{cern} working group explored the possibility of a \acrfull{helhc}  in the existing~\SI{27}{\km} tunnel, based on assumed~\SI{20}{\tesla} magnets with a \acrshort{com}\ energy of~\SI{33}{\TeV}~\cite{bib:Assmann:1284326}. 
This early \acrshort{helhc} 
effort was terminated at a 2010 EuCARD workshop~\cite{bib:Todesco:1344820}, where---in view of the \acrshort{helhc} technical difficulties 
combined with the limited time available for preparing a post-\acrshort{lhc} machine, 
the unfavourable cost scaling laws when pushing dipole magnetic fields 
to~\SI{20}{\tesla} or beyond, 
and the much higher collision energy required by physics 
(an argument further reinforced by the \acrshort{lhc} results obtained since then)---a future hadron collider of larger circumference, >\SI{80}{\km}, was proposed, for the first time, 
as a much preferred option compared to the \acrshort{helhc}.

The \acrshort{helhc} proposal was nevertheless submitted as input to the 2012/13 \acrshort{esppu}~\cite{bib:Brüning:1471002} and, 
in response to the generic recommendations from the 2013 strategy update~\cite{bib:Nakada:2690131}, also the \acrshort{helhc} design effort was revived as part of the \acrshort{fcc} Conceptual Design Study~\cite{bib:FCC:2018bvk}, as input to the \acrshort{esppu} 2020, but this option was not recommended.

In the~\acrshort{cdr} version, the \acrshort{helhc} dipole magnetic field was reduced to~\SI{16}{\tesla}, the same as for the (then) \acrshort{fcchh}, 
and a collision energy around~\SI{27}{\TeV} targeted~\cite{bib:FCC:2018bvk}. Recent studies indicate that \SI{14}{\tesla} should be the maximum operational field realistically to be assumed for large series of Nb\textsubscript{3}Sn accelerator magnets, further reducing the \acrshort{com}\ energy to \SI{24}{\TeV}. 
This \acrshort{helhc} would require a new superconducting \acrshort{sps} (\acrshort{scsps}) as injector. 

The \acrshort{helhc} must be installed in the existing \acrshort{lhc} tunnel, with an inner diameter of only~\SI{3.8}{\m}, compared with~\SI{5.5}{\m} for the \acrshort{fcc}. These space limitations result in significant constraints on the \acrshort{helhc} machine layout, on the magnets which must be compact and curved, and on the \acrshort{helhc} 
cryogenics system, which needs to be more powerful than the one of the \acrshort{lhc}.  
The \acrshort{helhc} high-field magnets, vacuum system, cryogenics, and tunnel 
integration all appear more challenging than those of the \acrshort{fcchh}.
The cost of \acrshort{helhc} reported in Ref.~\cite{bib:FCC:2018bvk} was 7200 M\acrshort{chf} (excluding the cost of the experiments).

A recent study considered various alternative magnet technologies and dipole fields~\cite{bib:Bottura:2875201}. 
The tentative beam parameters and projected performance figures for \acrshort{helhc} are listed in Table~\ref{tab:tabhelhc}.  
The synchrotron radiation power increases by roughly one to two orders of magnitude with respect to the (HL-)\acrshort{lhc}. 
Assuming a new \acrshort{scsps} as injector, the total electrical power consumption 
is expected to stay below~\SI{200}{\MW} (for~\SI{16}{\tesla} magnets)~\cite{bib:FCC:2018bvk}.

\begin{table}[h!]
\caption{Main parameters of \acrshort{helhc} options, based on
different magnet technologies and dipole fields, 
compared with \acrshort{hllhc} (nominal) and \acrshort{lhc} (achieved), for operation with proton beams~\cite{bib:Bottura:2875201}. All values, except for the injection energy itself, refer to the collision energy. The ring circumference and the straight section length are the same of the \acrshort{lhc} tunnel. The scheduled physics time and accelerator availability correspond to those assumed in Ref.~\cite{bib:Bordry_proposedfuturecolliders}. }
\begin{center}
\footnotesize
\begin{tabular}{|l|c|c|c|}
\hline
	
Parameter & \multicolumn{2}{c|}{\acrshort{helhc}} & (HL-)\acrshort{lhc} \\
\hline\hline  
Centre-of-mass energy~[\SI{}{\TeV}] & 27 & 33  & (14) 13.6 \\
\hline
Injection energy~[\SI{}{\TeV}] & 0.8 & 1.0 & 0.45 \\
\hline
Peak arc dipole field~[\SI{}{\tesla}] &  16 & 20  & (8.33) 8.1  \\
\hline
Instantaneous luminosity (levelled) / \acrshort{ip}~[\(10^{34}\,\SI{}{\per\square\centi\meter\per\second}\)] & 
 16 & 16 & (5) 2.2 \\
\hline
Stored energy / beam~[\SI{}{\giga\joule}] & 1.3 & 1.7  & 
(0.7) 0.41 \\
\hline
Synchrotron radiation power / beam~[\SI{}{\kilo\watt}]& 100 & 251 & (7.3) 4.4 \\
\hline
No.~of high-luminosity \acrshort{ip}s & 
\multicolumn{2}{c|}{2} & 2 \\
\hline
Scheduled physics time per year~[days] & \multicolumn{2}{c|}{160} & 160 \\
\hline
Accelerator availability [\%] & 
\multicolumn{2}{c|}{75} & (80) 79
\\ \hline
Integrated luminosity per year of nominal operation per \acrshort{ip}~[\SI{}{\per\femto\barn}/y]   & 500 & 520 & (250) 123
\\ \hline
\end{tabular}
\end{center}
\label{tab:tabhelhc}
\end{table}

The physics perspectives for the \acrshort{helhc} were discussed in Ref.~\cite{bib:Dainese:2703572} as input to the 2020 \acrshort{esppu} process.     
The 2020 \acrshort{esppu}~\cite{bib:2020_ESPPU_Brochure} called for a future hadron collider with collision energies in excess of~\SI{100}{\TeV}, as could be obtained by the \acrshort{fcchh}.

The installation of the \acrshort{helhc} can only take place after the removal of the \acrshort{lhc} components at the end of the \acrshort{hllhc} programme. This, together with the time required for the development of high-field magnets discussed in Section~\ref{sec:fcchhtrl}, would likely imply an earliest potential start date for \acrshort{helhc} at the beginning of the 2070s for the highest \acrshort{com}\ of~\SI{33}{\TeV}.

In conclusion, based on the current physics landscape emerging from the \acrshort{lhc} and other results, the \acrshort{helhc} offers a very limited physics reach.

\subsection{\acrshort{lhec}}

A~\acrshort{cdr} for the~\acrfull{lhec} was published in 2012~\cite{bib:LHeC_CDR_2012} and updated in 2020~\cite{bib:LHeC_2021}. Exploiting energy recovery technology,~\acrshort{lhec} collides an intense electron beam provided by a new 3-turn high-current~\acrfull{erl} with a proton or ion beam from the~\acrfull{hllhc}.
The accelerator and interaction region are designed for concurrent electron-hadron and hadron-hadron operation and \acrshort{lhec} could operate either in parasitic or dedicated mode.

The~\acrshort{erl} accelerator is located in a new tunnel tangential to the \acrshort{lhc} ring, tentatively in \acrshort{ip}2 where a dedicated experiment would be installed. The length of the~\acrshort{erl} is a fraction of the \acrshort{lhc} circumference as required for the electron and proton matching of bunch patterns. The electron accelerator may be built independently, to a considerable extent, of the status of operation of the \acrshort{lhc}.  

The baseline parameters of~\acrshort{lhec} have evolved since the publication of the 2020 updated \acrshort{cdr}~\cite{bib:LHeC_2021}. 
The \acrshort{erl} circumference considered in Ref.~\cite{bib:LHeC_ESPP2026} is 1/3 of the \acrshort{lhc} circumference as compared to 1/5 considered in the \acrshort{cdr}~\cite{bib:LHeC_2021}. The main baseline parameters for electron-proton operation are summarized in Table~\ref{tab:LHeC}.

After a first year of operation at reduced luminosity \acrshort{lhec} would reach nominal luminosity and operate for five more years (with a \acrshort{ls} of one year after the first three years of operation) aiming to a total integrated luminosity of approximately \SI{1}{\per\atto\barn}. Electron-lead ion operation is also considered in Ref.~\cite{bib:LHeC_ESPP2026}. The design parameters have been selected to limit the power consumption to less than~\SI{100}{\MW} (except for the dedicated mode of operation with an \acrshort{erl} circumference corresponding to 1/5 of the \acrshort{lhc}).

\begin{table}[h!]
\caption{Main parameters of~\acrshort{lhec}. It is assumed that the~\acrshort{erl} does not contribute to significant additional downtime for the machine. The dedicated mode of operation could only take place at the end of the \acrshort{hllhc} programme. The scheduled physics time and accelerator availability correspond to those assumed in Ref.~\cite{bib:Bordry_proposedfuturecolliders}.}
\begin{center}
\scriptsize
\begin{tabular}{|l|c|c|c||c|c|}
\hline

 & \multicolumn{3}{c||}{\acrshort{cdr} 2020~\cite{bib:LHeC_2021}}& \multicolumn{2}{c|}{\acrshort{esppu} 2026~\cite{bib:LHeC_ESPP2026}} \\ \hline
Parameter & Initial & Design & Dedicated & Initial & Nominal \\
\hline\hline  
\acrshort{erl} circumference~[\SI{}{\km}] &  \multicolumn{3}{c||}{5.332 (1/5 of \acrshort{lhc} circumference)} & \multicolumn{2}{c|}{
8.886 (1/3 of \acrshort{lhc} circumference} \\
\hline
\acrshort{com}~energy~[\SI{}{\TeV}] & \multicolumn{3}{c||}{1.18}& \multicolumn{2}{c|}{1.18}\\
\hline
Electron beam energy~[\SI{}{\GeV}] &  \multicolumn{3}{c||}{50} & \multicolumn{2}{c|}{50}\\
\hline
Average electron beam current~[\SI{}{\mA}] & 15 & 25 & 50 & 20 & 50\\ \hline 
Initial luminosity~[\(10^{34}\,\SI{}{\per\square\centi\meter\per\second}\)] & 
0.5 & 1.2 & 2.3 & 0.9 &2.3 \\ \hline
Integrated luminosity per year of nominal operation~[\SI{}{\per\femto\barn}/y]   & 20 & 50 & 180 & 70 &180 \\ \hline
\end{tabular}
\end{center}
\label{tab:LHeC}
\end{table}
The expected maximum power consumption of \acrshort{lhec} with the associated experiment is~\SI{220}{\mega\watt}, when operated in dedicated mode, i.e. approximately~\SI{75}{\mega\watt} more than the nominal \acrshort{lhc}. A yearly electricity consumption of approximately~\SI{1.1}{\tera\watt\hour}/y is expected; this value does not include the electricity consumption of the \acrshort{lhc} injector complex.

Implementation of~\acrshort{lhec} would require the construction of a new tunnel situated at a depth of approximately~\SI{100}{\m} and intersecting the \acrshort{lhc} tunnel in \acrshort{ip}2.

The cost of \acrshort{lhec} (with an \acrshort{erl} circumference 1/3 of the \acrshort{lhc} circumference) was estimated in 2018 to 1600~M\acrshort{chf} (\acrshort{ce} representing approximately 24\% of the overall cost), therefore we can expect a 2024 cost of about 2000~M\acrshort{chf}~\cite{bib:LHeC_ESPP2026}\footnote{This is consistent with an estimate of about 1900 M\acrshort{chf} obtained by considering that the \acrfull{eu} domestic~\acrshort{ppi} has increased from 90 to 127 in the period 2018--2024 while the \acrshort{eur}/\acrshort{chf} exchange rate has decreased from 1.15 to 0.95.}. The cost of the experiment should be added and it has been estimated to 360 M\acrshort{chf} (2024 value) excluding the cost of the magnet, beam pipe, \acrfull{daq}, trigger, mechanical structure and infrastructure. We can therefore estimate an overall cost of approximately 2400~M\acrshort{chf} of which about 2100 M\acrshort{chf} borne by \acrshort{cern}.

In order to achieve high-power efficiency, the~\acrshort{lhec}~\acrshort{erl} design is based on bulk Nb 5-cell \SI{800}{\MHz} \acrshort{srf} cavities operated in \acrshort{cw} mode with an accelerating gradient of~\SI{22}{\mega\volt\per\meter} and $Q_0>3\times10^{10}$ at \SI{2}{\kelvin}. The above requirements coincide with those for the \acrshort{fccee} booster and collider \acrshort{rf}~\SI{800}{\MHz} cavities and they have been achieved in vertical tests conducted at Jefferson Lab in 2018~\cite{bib:LHeC800MHzJLAB}.

The highest energy achieved by an \acrshort{erl} so far is \SI{1}{\GeV} at the \acrfull{cebaf}, while the~\acrfull{jlab} \acrfull{fel} has reached the highest current of all \acrshort{srf} \acrshort{erl} with \SI{10}{\mA}. Larger currents have been achieved in the normal-conducting, lower-frequency~\acrshort{erl} facility at~\acrfull{binp}\cite{bib:AccRDRoadmap}.
High-current multi-turn energy recovery~(i.e. with beam power significantly exceeding the installed \acrshort{rf} power) still needs to be demonstrated for beam powers exceeding~\SI{1}{\MW} (well below the nominal \acrshort{lhec} beam power of \SI{2.5}{\giga\watt}). 4-turn acceleration to~\SI{150}{\MeV} and energy recovery\footnote{Energy recovery efficiency for single-turn acceleration and recovery has been measured to be 99.4\%, when taking into account beam losses, and approximately 100\% for the cavity single particle energy recovery~\cite{bib:CBETA_energyrecovery_2021}.} has been demonstrated recently at the~\acrfull{cbeta} but at significantly lower current (\SI{1}{\nA}) than the design one (\SI{40}{\mA}), limited by beam losses, particularly in the last recovery loop, and radiation protection constraints~\cite{bib:AccRDRoadmap,bib:CBETA_Hofstaetter_2024, bib:CBETA_Bartnik_2020}. \acrshort{cbeta} used a novel and challenging~\acrfull{ffag} transport based on permanent magnets for the arcs; the choice of this technology has imposed a stringent limit on losses and therefore on current to prevent damage to the permanent magnets.

An international collaboration has been established to build a 3-turn~\SI{250}{\MeV}~\acrshort{erl}---the~\acrfull{perle}---at \acrshort{ijclab} Orsay \cite{bib:PERLE_ESPP2026} aiming to deliver a beam current of~\SI{20}{\mA} at top energy in bunches of~\SI{500}{\nC} charge at~\SI{40}{\MHz} as for~\acrshort{lhec}, for a design electron beam power of~\SI{5}{\MW}. Acceleration and energy recovery will take place in one cryomodule equipped with 5-cell~\SI{800}{\MHz} \acrshort{srf} cavities with the same specifications as those required for~\acrshort{lhec}. 

The~\acrshort{perle} facility is supposed to study and address coherent beam instabilities, incoherent effects and to demonstrate the stable operation of the \acrshort{srf} system with the sixfold current (\SI{120}{\mA}) due to the simultaneous acceleration and deceleration of bunches at three different beam energies~\cite{bib:AccRDRoadmap}. 
A~\acrshort{cdr}~\cite{bib:PERLE_CDR_2018} has been written in 2018 and a \acrshort{tdr} is in preparation. An ambitious schedule aiming demonstration of the design performance for a~\SI{250}{\MeV}~\acrshort{erl} at the beginning of the 2030s~\cite{bib:LHeC_ESPP2026,bib:PERLE_ESPP2026} is put forward. 
An intermediate stage with a single-turn operation~(\SI{89}{\MeV}) is contemplated. Future plans include the possibility of reaching \SI{10}{\mega\watt} and a final energy of \SI{500}{\MeV} by the installation of an additional cryomodule.

A 3-turn~\acrshort{erl} configuration had been adopted also for the \acrshort{fcceh}, albeit maintaining the original~\SI{60}{\GeV} energy as default. For \acrshort{fcceh} the preferred position was interaction point L, for geological reasons mainly~\cite{bib:LHeC_2021}, though this straight section is now planned to host the main \acrshort{rf} system of the collider and the injection system of the anti-clockwise beam, but the \acrshort{lhec} \acrshort{hw} could still be used in a different location. 

\acrfull{hom} damping and \acrshort{hom} losses management, emittance preservation of the high-brightness electron beam in the recirculating arcs, beam diagnostics, synchrotron radiation handling at high energy, represent the main technical challenges for the \acrshort{lhec} project, with a~\acrshort{trl} of 4 for the most critical aspects~(see Appendix~\ref{sec:TRLtable})~\cite{bib:SnowMass2021, bib:LDG_RD_review_ESPP2026}. Operation in recirculation mode (no energy recovery) would still be possible, though at significantly lower current (likely by an order of magnitude), and therefore at a correspondingly lower luminosity, with a power consumption smaller than \SI{250}{\MW}.

The time to first physics was estimated to be 13 to 18 years (assuming a technically-limited schedule) in Ref.~\cite{bib:SnowMass2021}, limiting significantly (if not completely) the temporal overlap of~\acrshort{lhec} with the~\acrshort{hllhc} programme expected to be completed by the end of 2041. 
In addition, \acrshort{ip}2 is presently allocated to \acrshort{alice}, the \acrshort{lhc} experiment dedicated to the study of heavy ion collisions. \acrshort{alice} has been upgraded in~\acrshort{ls}2 to operate at higher luminosity during Run 3 and 4 and, following the recommendations of the 2020~\acrshort{esppu}, it is proposing an experimental programme until the end of the \acrshort{hllhc} era. 
Operation of~\acrshort{lhec} after the completion of the \acrshort{hllhc} programme, after a two year-long shutdown for the completion of the \acrshort{ce} and of the installation, is presently contemplated \cite{bib:LHeC_ESPP2026}. This implies the validation of the main concepts and in particular the successful operation of \acrshort{perle} by the beginning of the 2030s. An alternative scenario \cite{bib:PhaseOneLHeC}, considering a staged construction of~\acrshort{lhec} with a first phase completed by the end of \acrshort{ls}4 and allowing collisions of \SI{20}{\GeV} electrons from a single-pass \acrshort{erl} during Run~5, appears not to be realistic for the above-mentioned reasons and it would require an adaptation of the \acrshort{alice3} detector.

The personnel required for the construction of the \acrshort{lhec} accelerator and its infrastructure is expected to be comparable to that required for \acrshort{hllhc}, a project of similar size, and it would amount to \SI{2500}{\acrshort{ftey}}, over seven years \cite{bib:LHeC_ESPP2026}, to be compared with \SI{3400}{\acrshort{ftey}} according to Eq.~\eqref{eq:FTEvsCAPEXconstruction}.

%% file: include/07-NewAcceleratorTechnology/NewAcceleratorTechnology.tex
\section{New acceleration techniques}\label{sec:NewAccTech}

\acrfull{wfa} hold the promise for a path towards more compact, and hence potentially more sustainable, high-energy colliders than those relying on conventional technology. Their distinctive feature is a high peak gradient that can reach one to several orders of magnitude larger values than conventional \acrshort{rf} acceleration schemes. 
The potential of these new techniques is well illustrated by the \SI[per-mode=symbol]{52}{\GeV\per\meter} energy gradient obtained with an electron-driven plasma \acrshort{wfa} in the \acrfull{facet}~\cite{bib:blumenfeld}, or more recently the \SI[per-mode=symbol]{30.7}{\GeV\per\meter} energy gradient achieved at \acrfull{bella} in a laser-driven \acrshort{wfa}~\cite{bib:Picksley_LWFA_PRL}, in both cases with energy gain more than \SI{9}{\GeV}.

In such systems, the wakefields provide the acceleration and are driven by either a pulsed laser, or by particle bunches---typically made of electrons, although protons can also be used, for example in the \acrfull{awake}~\cite{bib:WFA_Muggli_AWAKE_2022}. Wakefields must be sustained either using a plasma, in a \acrfull{pwfa} or \acrfull{lwfa}, or thanks to a dielectric-based structure surrounding the beam, in a \acrfull{swfa}~\cite{bib:SnowMass2021} or \acrfull{dla}. In the case of \acrshort{swfa}, the two beams (drive beam and accelerated beam) may either be traveling in the same structure in a \acrfull{cwa}, or in two different structures in a \acrfull{tba}~\cite{bib:WFA_Jing_2022}. 

Both plasma- and dielectric-based schemes can sustain very large electric fields, thus overcoming the breakdown voltage in conventional \acrshort{rf} structures. \acrshort{swfa} exhibit typically smaller gradients (few hundreds of \SI[per-mode=symbol]{}{\mega\volt\per\m}) compared to plasma-based accelerators (in the \SI[per-mode=symbol]{}{\giga\volt\per\m} range). In principle, compact \epem colliders could be built using such advanced acceleration schemes, and several options were considered in the Snowmass process~\cite{bib:SnowMass2021} (see also update in Ref.~\cite{bib:WFA_Barklow_2023}). 

Such technologies exhibit a disrupting potential, and steady progress has been made in the past decade. For instance, the \acrfull{lux} in \acrshort{desy}, was able to run stable for 24 hours~\cite{bib:WFA_LUX}. A milestone was also reached when \acrfull{fel} amplification was achieved with a \acrshort{lwfa} in China~\cite{bib:WFA_wang2021free} and in Europe~\cite{bib:WFA_labat_natphot2023}, as well as with a \acrshort{pwfa} at the \acrfull{sparclab} facility~\cite{bib:WFA_pompili2022free,bib:WFA_gallettiPRL2022}. Nevertheless, to date no user facility based on wakefield acceleration, aiming at high-energy physics, has been built yet. A \acrshort{fel}, \acrshort{pwfa} project is close to completion, the \acrfull{eupraxia} ~\cite{bib:WFA_EuPRAXIA_CDR2020} (\acrshort{eupraxia}@\acrshort{sparclab}~\cite{bib:WFA_EuPRAXIA_SparcLab_CDR2018}, to be operated as of 2029), but its beam parameters (in particular, beam energy of \SIrange[range-phrase={--}]{1}{5}{\GeV}, less than $0.2\cdot10^9$ e$^-$ per bunch, and no more than \SIrange[range-phrase={--}]{100}{400}{\Hz} repetition rate~\cite{bib:WFA_galletti2024fel}), remain far from those required by a Higgs factory, and no staging of plasmas is foreseen, nor any positron acceleration.

For \acrshort{lwfa} as proposed in Ref.~\cite{bib:WFA_Schroeder_2023}, and for \acrshort{pwfa} such as the electron-driven \acrfull{pwfalc}~\cite{bib:WFA_Delahaye_IPAC2014} or the proton-driven \acrfull{alive}~\cite{bib:WFA_ALiVE_Farmer_2024}, positron acceleration in the plasma remains an outstanding challenge~\cite{bib:WFA_ICFA_Adli}: contrary to electrons, positrons usually experience strong defocusing, as a consequence of the charge asymmetry of the plasma~\cite{bib:WFA_Cao_positronAccel}. Structure-based accelerators such as \acrshort{swfa} and \acrshort{dla} avoid this problem, but suffer from other limitations such as beam break up caused by the strong wakefields generated in the dielectric structure, leading to the need for very small charges per bunch, and hence requiring very high repetition rates (or very small bunch spacing) to maintain an acceptable luminosity. 
In general, structure-based acceleration is currently not considered the preferred option for \acrshort{wfa} technology~\cite{bib:AccRDRoadmap}, and hence \acrshort{rd} is less active in this field than for plasma-based \acrshort{wfa}. $\gamma \gamma$ colliders may also circumvent the positron acceleration problem, since they use only electrons~\cite{bib:WFA_ICFA_Adli}.

In general, the critical challenges common to most wakefield-based accelerators, are related to large-scale staging (with all subsequent issues such as synchronisation, phasing and alignment~\cite{bib:WFA_staging_Lindstrom}), \acrshort{bds} with a final focus aiming at beam sizes of tens of \SI{}{\nano\meter}, emittance preservation along the chain, the relatively small bunch intensity that could be accelerated so far, the very large repetition rates needed to achieve the luminosities required by high-energy physics, and the energy efficiency of the whole process. More details can be found in the \acrfull{icfa} Beam Dynamics Newsletter dedicated to beam dynamics challenges in advanced accelerator concepts~\cite{bib:AdvAccConcepts}, and in the \acrfull{ldg} Accelerator \acrshort{rd} Roadmap~\cite{bib:AccRDRoadmap}. In addition to these, for the specific case of \acrshort{lwfa}, laser technology needs to reach tens of \SI{}{\kilo\watt} of average power with tens of \SI{}{\kHz} repetition rate, while current state of the art is still below \SI{1}{\kilo\watt} average, at the \SI{100}{\Hz} level~\cite{bib:LDG_RD_review_ESPP2026}. Coherent combination of high-efficiency, 
high-repetition rate fibre lasers, to replace the traditional but less efficient 
Titanium:Sapphire technology, might address this issue in the 
future~\cite{bib:WFA_ALEGRO_ESPP2026}. Finally, regarding the more recent concept of proton-driven \acrshort{pwfa} as in \acrshort{alive} or \acrshort{awake}, the challenge of staging is replaced by that of realizing a single plasma of hundreds of meters with a plasma density gradient~\cite{bib:WFA_ALiVE_Farmer_2024}. For such accelerators, the achievable repetition rate is also limited by the high-energy proton complex needed to generate the driver, but rapid cycling synchrotrons or \acrfull{ffag} are being proposed to solve this issue~\cite{bib:WFA_ALiVE_ESPP2026}.

Regarding the wall plug efficiency, three factors~\cite{bib:WFA_limits_Zimmermann} are mainly driving it: the efficiency of the driver generation, that of the drive to wake process, and finally the part of the energy transferred from the wake to the accelerated beam. The last of these requires a trade-off between accelerating gradient (low accelerated bunch charge) and efficiency (high accelerated charge). In addition, the drive-to-wake efficiency may be limited by stability~\cite{bib:WFA_VLebedev_PRAB2017}. 

We summarise in Table~\ref{tab:Parameters_WFA} two examples of wakefield-based collider studies, giving their main parameters and \acrshort{rd} challenges: the \acrfull{halhf}, and a recent concept proposed by the \acrshort{alive} collaboration for proton-driven \acrshort{pwfa}. Other designs proposed earlier for the Snowmass report, or in the aforementioned \acrshort{icfa} newsletter, were not selected here as they are much less developed: the \acrshort{dla} in Ref.~\cite{bib:WFA_England_2022} and the \acrshort{swfa} \acrfull{aflc}~\cite{bib:WFA_Jing_2022} concepts were still very preliminary, the \acrshort{lwfa} presented in Ref.~\cite{bib:WFA_Schroeder_2023} and in the Snowmass report, still lacks a clear design, while the \acrshort{pwfalc} study~\cite{bib:WFA_Adli_2013,bib:WFA_Delahaye_IPAC2014} has not been updated since 2014. Current design studies focus on Higgs factories based on \acrshort{pwfa} such as \acrshort{halhf} and \acrshort{alive}, while \SI{10}{\TeV} \epem collider options including \acrshort{pwfa}, \acrshort{lwfa}, and \acrshort{swfa} components are considered in Ref.~\cite{bib:WFA_10TeV_ESPP2026}. 

For the selected studies, even with the (optimistic) assumption that all the challenges are overcome, the peak luminosity achievable remains a significant factor below what is accessible in major Higgs factory projects (especially circular ones). The wall plug efficiency foreseen lies in the range 3--12\%, comparable to that of linear colliders such as \acrshort{clic}, \acrshort{lcf} and \acrshort{ilc}.

The proton-driven \acrshort{pwfa} proposed by \acrshort{alive} 
misses a conceptual design for the proton driver, and faces the outstanding challenge of the positron acceleration. On the other hand, \acrshort{halhf} circumvents the latter issue thanks to an innovative, asymmetric design in which only the electrons are accelerated to a high energy (\SI{375}{\GeV}) by a \acrshort{pwfa}, while the positrons are accelerated to a lower energy (\SI{41.7}{\GeV}) using an \acrshort{rf} accelerator based on \acrfull{c3} technology~\cite{bib:Vernieri_C3_2023}, which is still in its \acrshort{rd} phase, for a \acrshort{com}\ energy of \SI{250}{\GeV}---note that higher energy options are also envisaged, requiring longer facility length. The size of such a machine would be very competitive, and the critical issue of emittance preservation in the \acrshort{pwfa} e$^-$ linac, is slightly relaxed thanks to the asymmetry. Nevertheless, many \acrshort{pwfa} issues remain to be solved for the electron linac; quoting from Ref.~\cite{bib:HALHF_published}: ``[...] there must be progress toward the use of multiple stages, self-correction mechanisms, higher accelerated charge (by a factor $\sim10$), higher repetition rate (by a factor $\sim 1000$), plasma-cell design required to cope with large power dissipation (by a factor $\sim 1000$) and reduction of beam jitter to an acceptable level (by a factor $\sim 10-100$)''. 
On some of the challenges mentioned above, recent progress were reached, for instance regarding emittance preservation in a single plasma~\cite{bib:WFA_Lindstrom_emittance}, or the issue of plasma heating and its related recovery time, which could limit the collision rate to the tens of \SI{}{\MHz} range when using Argon ions~\cite{bib:WFA_DArcy_recoveryTime} (to achieve a \SI{16}{\nano\second} bunch spacing, as foreseen in \acrshort{halhf}~\cite{bib:HALHF_new_layout2025}, the use of lighter ions may be required~\cite{bib:WFA_pompili2024recovery}). Asymmetric collisions also require \acrshort{rd} on the detector side. A 15 years-roadmap for the \acrshort{rd} needed to validate the technology required by \acrshort{halhf}, was submitted recently to the \acrshort{esppu}~\cite{bib:WFA_HALHF_backup_ESPP2026}; the total resources foreseen amount to 213~M\acrshort{chf} and 341~\acrshort{ftey}.

For any multistage wakefield accelerator, and in particular for all the multi-\SI{}{\TeV} alternatives presented in Ref.~\cite{bib:SnowMass2021}, heavy \acrshort{rd} (and hence the corresponding resources) is needed, and a multi-\SI{}{\GeV} demonstrator required as a first step, before any realistic design could be presented---hence in all these cases a \acrshort{tdr} remains decades away, and no costing is available. Nevertheless, it should be mentioned that so far, no definitive showstopper has been found, and the field is very active, as can be seen from the submissions for the \acrshort{esppu} 2026~\cite{bib:WFA_ALiVE_ESPP2026,bib:WFA_HALHF_ESPP2026,bib:WFA_10TeV_ESPP2026,bib:WFA_AWAKE_ESPP2026} and the number of \acrshort{wfa} projects around the world (see e.g.\ the list in Ref.~\cite{bib:WFA_ALEGRO_backup_ESPP2026}). The \acrshort{lc} Vision is also contemplating the possible use of \acrshort{wfa} for future upgrades of linear colliders~\cite{bib:LCVision_backup_ESPP2026}.

It is finally worth noting that in plasma-based accelerators a number of effects may pose fundamental limits. In the ``blow-out'' or nonlinear regime, where the acceleration gradient is the largest, the strong focusing in the plasma induces energy loss from betatron radiation~\cite{bib:WFA_limits_Zimmermann}, as well as challenges related to the matching of adjacent plasmas~\cite{bib:Shiltsev_RevModPhys} (the matching issue does not apply to a proton-driven \acrshort{pwfa}, which uses a single plasma). 
In general, in \acrshort{wfa} the beam spot size at the interaction point, hence the luminosity, may also be limited by chromatic aberrations~\cite{bib:WFA_Barklow_2023}, or the Oide effect from betatron radiation related to the strong focusing~\cite{bib:OideEffect}. Besides, because of the high energies and ultra short bunches, the beamstrahlung parameter gets well above one, contrary to conventional linear colliders~\cite{bib:WFA_Barklow_2023}, leading to e.g.\ energy spread and background in the detectors.

In summary, in recent years there were many advances towards the possible use of \acrshort{wfa} for the acceleration of higher quality electron beams and the field is very active to try to address the encountered issues through \acrshort{rd}. Nevertheless, the lack of maturity of the technology currently prevents  wakefield-based accelerators to be considered as alternatives to the main projects presented in the above sections, at least in the timescale considered.

\begin{table}[h!]
 \caption{Main parameters for potential wakefield-based \epem colliders. A single \acrshort{ip} is assumed in all cases.}
\begin{center}
\footnotesize
 \begin{threeparttable}
 {\setlength{\tabcolsep}{0.3em}
 \begin{tabular}{|L{4.2cm}||C{2.9cm}|C{2.9cm}|C{4.7cm}|}\hline
Name & \multicolumn{2}{c|}{\acrshort{halhf} (\SI{250}{\GeV})} & \acrshort{alive} \\ \hline 
Main reference & \multicolumn{2}{c|}{\cite{bib:WFA_HALHF_ESPP2026}} & \cite{bib:WFA_ALiVE_ESPP2026} \\ \hline\hline

Main principle & \SI{375}{\GeV} e$^-$

e$^-$ driven \acrshort{pwfa} & \SI{41.7}{\GeV} e$^+$

\acrshort{c3} \acrshort{rf} system~\cite{bib:Vernieri_C3_2023}  & p$^+$ driven 

\acrshort{pwfa} \\ \hline

\acrshort{com}~energy [\SI{}{\TeV}] & \multicolumn{2}{c|}{0.25} & 0.25 \\ \hline

Full facility length [\SI{}{\km}] & \multicolumn{2}{c|}{4.9

} & $\sim$5.5 ~\cite{bib:Farmer_privatecomm}\tnotex{tab:Parameters_WFA:1} 
\\ \hline
Length per linac [\SI{}{\km}] & 1.1 & 1.1 & 0.24~\cite{bib:Farmer_privatecomm} 
\\ \hline

Instantaneous luminosity above $0.99~\sqrt{s}$ [\(10^{34}\,\SI{}{\per\square\centi\meter\per\second}\)] & \multicolumn{2}{c|}{0.76} & 0.25~\cite{bib:Farmer_privatecomm} \\ \hline

Integrated luminosity above $0.99~\sqrt{s}$ per year [\SI{}{\per\atto\barn}/y]\tnotex{tab:Parameters_WFA:3} & \multicolumn{2}{c|}{0.09} & 0.03 \\\hline

Wall~plug~power~[\SI{}{\mega\watt}] & \multicolumn{2}{c|}{106} & $\sim$200~\cite{bib:Farmer_privatecomm} \\ \hline

Acceleration gradient [\SI[per-mode=symbol]{}{\giga\volt\per\meter}] & 1 & 0.04 & 0.5~\cite{bib:Farmer_privatecomm} \\ \hline

Particles per bunch & $10^{10}$ & $3\cdot10^{10}$ & $2\cdot10^{10}$ \\ \hline

Bunches per train & \multicolumn{2}{c|}{160} & 1 \\ \hline

Bunch spacing in train & \multicolumn{2}{c|}{\SI{16}{\nano\second}} & - \\ \hline

Train repetition rate & \multicolumn{2}{c|}{\SI{100}{\Hz}} & \SI{7.2}{\kilo\Hz} \\ \hline

Normalised emittance

$\varepsilon_x/\varepsilon_y$ [\SI{}{\micro\meter}] & 90~/~0.32 & 10~/~0.035 & 0.1~/~0.1 (e$^-$)

0.4~/~0.4 (e$^+$)\\ \hline

Size $\sigma_x/\sigma_y$ at \acrshort{ip} [\SI{}{\nano\meter}] & \multicolumn{2}{c|}{636 / 6.6} & 73 / 13 (e$^-$)

146 /26 (e$^+$) \\ \hline

RMS bunch length [\SI{}{\micro\meter}] & 40 (150 at \acrshort{ip}) & 150 & 105 (e$^-$)

75 (e$^+$) \\ \hline

RMS energy spread at \acrshort{ip} & \multicolumn{2}{c|}{0.0015} & 0.001 \\ \hline

Single beam power [\SI{}{\MW}] & 9.6 & 3.2 & 2.9 \\ \hline

Beam vs wall power ratio & \multicolumn{2}{c|}{0.12} & 0.03 \\ \hline
 \end{tabular}
 
 \begin{tabular}{|L{4.2cm}||L{6cm}|L{4.7cm}|}\hline
R\&D challenges & - For the \acrshort{pwfa}: \textbf{staging}, short bunch separation (plasma recovery issue), self-correction mechanisms, progress needed by orders of magnitude for key parameters: accelerated charge, repetition rate, acceptable power dissipation in plasma cell, beam jitter, alignment; emittance preservation

~

- \acrshort{bds} (both sides)

- e$^+$ source

& - \textbf{e$^+$ acceleration}

- Development of long plasma cell, energy efficiency

- \textbf{Proton driver complex}: design of four \acrshort{ffag} up to \SI{500}{\GeV}, short bunches (\SI{0.7}{\mm}), tailored emittance profile, high repetition rate (\SI{14.4}{\kHz}), novel superconducting magnets and cavities, dump system for $\sim$\SI{30}{\mega\watt} beams

- Demonstrator

- \acrshort{bds}, control of energy spread

\\ \hline
Timescale (technically-limited)
& 15 years R\&D

Construction in 15 years & N/A
 \\ \hline
 \end{tabular}}
 \begin{tablenotes}
    \item[a] \label{tab:Parameters_WFA:1} The proton complex is not included, and may be as large as \SI{6.9}{\km} long~\cite{bib:WFA_ALiVE_ESPP2026}.
    \item[b] \label{tab:Parameters_WFA:3} Computed from the instantaneous luminosity above $0.99~\sqrt{s}$, assuming 139~days of data taking per year (see Table~\ref{tab:operational_year}).

 \end{tablenotes}
 \end{threeparttable}
\label{tab:Parameters_WFA}
\end{center}
\end{table}


%% file: include/08-Annex/A-Criteria.tex
\section{Criteria and metrics for the comparison}
\subsection{Operational year}
\begin{figure}[h!]
\begin{center}
\tikzset{font=\footnotesize} 
\begin{tikzpicture}
\pie[
  text=pin,
  radius=2.0,
   color={cyan!70, 
   yellow!70,
    orange!70,  
   blue!70,
   red!70,
   green!70
    },
  hide number,
  ]{32.9/Annual shutdown - 120,
    8.2/Commissioning - 30,
    2.7/Technical stops - 10,
    5.5/Machine development - 20,
    12.6/Downtime (faults) - 46,
    38.1/Data taking - 139
}
\end{tikzpicture}
\end{center}
\caption{\footnotesize{Breakdown of the operational phases for~\epem colliders (see Table~\ref{tab:operational_year}).}}
\label{fig:operational_year}
\end{figure}
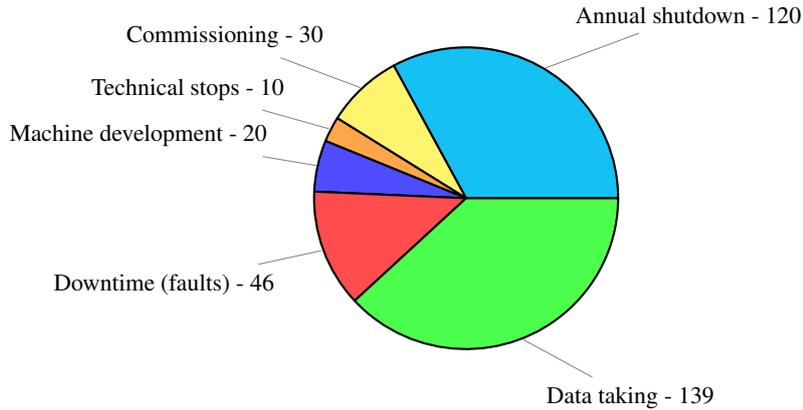

\paragraph*{\epem colliders}~\\
The operational parameters (number of days of operation in each phase, and corresponding fraction of the peak power needed) are presented in Table~\ref{tab:operational_year} and Fig.~\ref{fig:operational_year}.

\begin{table}[h!]
 \caption{\footnotesize {Breakdown of the operational phases~\cite{bib:Bordry_proposedfuturecolliders,bib:FCCee_ESPP2026,Brunner:2022usy} and of the corresponding fractional electricity consumption~\cite{Brunner:2022usy,bib:FCCee_ESPP2026}, for the linear and circular~\epem colliders. For \acrshort{clic} the baseline and energy upgrade are considered, and for \acrshort{lcf}~\cite{LinearCollider:2025lya} the baseline and upgrade options are given---\acrfull{lp} at \SI{250}{\GeV} and \acrfull{fp} at \SI{250}{\GeV} and \SI{550}{\GeV}.
}}
\centering
\footnotesize
\begin{tabular}{|l|c|c|c|c|c|c|c|c|c|c|}\hline
\multirow{3}{*}{Operational phase} & \multirow{3}{*}{Number of days} & \multicolumn{9}{c|}{Fraction of the peak power consumption~[\%]} \\ 
& & \multicolumn{4}{c|}{circular} & \multicolumn{2}{c|}{\acrshort{clic}} & \multicolumn{3}{c|}{\acrshort{lcf}} \\
& & Z & W & ZH & t$\mathrm{\bar t}$ & \SI{380}{\GeV} & \SI{1.5}{\TeV} & 250 \acrshort{lp} & 250 \acrshort{fp} & 550 \acrshort{fp} \\\hline
Annual shutdown & 120 & 12 & 12 & 11 & 11  & 6 & 5 & 28 & 18 & 25 \\
Commissioning & 30 & 58 & 60 & 60 & 62 & 55 & 55 & 77 & 72 & 72 \\
Technical stops & 10 & 27 & 28 & 27 & 28 & 55 & 55 & 54 & 45 & 44 \\
Machine development & 20 & 39 & 45 & 50 & 61 & 55 & 55 & 77 & 72 & 72 \\
Downtime (faults) & 46 & 30 & 36 & 42 & 55 & 55 & 55 & 54 & 45 & 44 \\
\textbf{Data taking} & \textbf{139} & \multicolumn{9}{c|}{100} \\ \hline
\end{tabular}
\label{tab:operational_year}
\end{table}

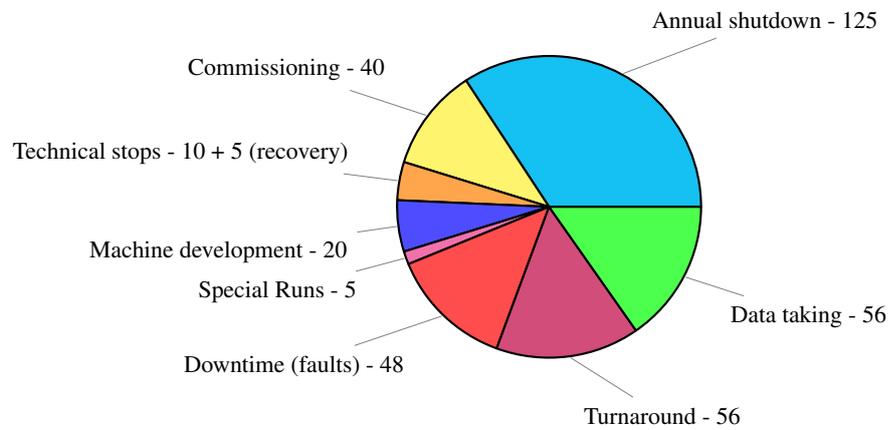
\begin{figure}[h!]
\begin{center}
\tikzset{font=\footnotesize}
\begin{tikzpicture}
\pie[
  text=pin,
  radius=2.0,
   color={cyan!70, 
   yellow!70,
    orange!70,  
   blue!70,
   magenta!70,
   red!70,
   purple!70,
   green!70},
  hide number,
  ]{34.2/Annual shutdown - 125,
    11/Commissioning - 40,
    4.1/Technical stops - 10 + 5 (recovery),
    5.5/Machine development - 20,
    1.4/Special Runs - 5,
    13.2/Downtime (faults) - 48,
    15.3/Turnaround - 56,
    15.3/Data taking - 56
}
\end{tikzpicture}
\end{center}
\caption{\footnotesize{Breakdown of the operational year for hadron colliders (see Table~\ref{tab:operational_year_FCChhMuC}).}}
\label{fig:operational_year_FCChh}
\end{figure}

\paragraph*{Hadron colliders}~\\
The operational parameters are presented in Table~\ref{tab:operational_year_FCChhMuC} and Fig.~\ref{fig:operational_year_FCChh}.

\begin{table}[h!]
 \caption{ \footnotesize{Breakdown of the operational phase and the corresponding fractional power consumption for hadron colliders~\cite{bib:Bordry_proposedfuturecolliders} and the \acrlong{mc}~\cite{bib:imcc_addendum_esppu2026}.  For comparison, the breakdown of the operational phases during~\acrshort{lhc} Run~2 can be found in Refs.~\cite{bib:Salvachua_Evian2019, bib:Todd_Evian2019}.}
 }
\centering
\footnotesize
\begin{tabular}{|L{3cm}|C{2.5cm}|C{2.5cm}|C{1.7cm}|C{1.7cm}|C{1.7cm}|}\hline
\multirow{3}{*}{Operational phase} & \multicolumn{2}{c|}{Number of days} & \multicolumn{3}{c|}{Fraction of the peak power consumption [\%]} \\ 
& \multirow{2}{*}{\acrshort{fcchh}} & \multirow{2}{*}{\acrshort{mc}} & \multirow{2}{*}{\acrshort{fcchh}} & \multicolumn{2}{c|}{\acrshort{mc}} \\
& & & & \SI{3.2}{\TeV} & \SI{7.6}{\TeV} \\ \hline
Annual shutdown & 125 & 125 & 34 & 50 & 38\\
Commissioning & 40 & 40 & 91 & 100 & 100\\
Technical stops & 10 +5 (recovery) &  10 +5 (recovery) &34 & 62 & 61\\
Machine development & 20 & 20 &91 & 100 & 100\\
Special runs & 5 & -- & 91 & -- & -- \\
Downtime (faults) & 48 & 49 & 34 & 72 & 71 \\
Turnaround & 56 & -- & 91 & -- & -- \\
\textbf{Data taking} & \textbf{56} &\textbf{116} & 100 & 100 & 100  \\ \hline
\end{tabular}
\label{tab:operational_year_FCChhMuC}
\end{table}

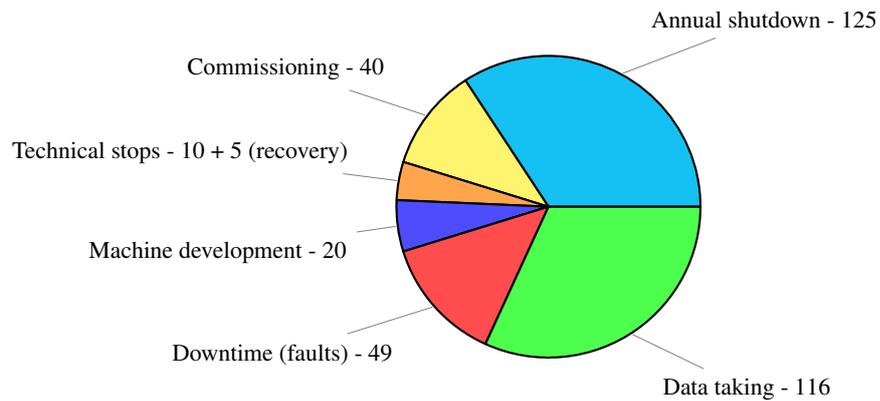
\begin{figure}[h!]
\begin{center}
\tikzset{font=\footnotesize}
\begin{tikzpicture}
\pie[
  text=pin,
  radius=2.0,
   color={cyan!70, 
   yellow!70,
    orange!70,  
   blue!70,
   red!70,
   green!70},
  hide number,
  ]{34.2/Annual shutdown - 125,
    11/Commissioning - 40,
    4.1/Technical stops - 10 + 5 (recovery),
    5.5/Machine development - 20,
    13.4/Downtime (faults) - 49,
    31.8/Data taking - 116
}
\end{tikzpicture}
\end{center}
\caption{\footnotesize{Breakdown of the operational year for the \acrshort{mc} (see Table~\ref{tab:operational_year_FCChhMuC}).}}
\label{fig:operational_year_MuC}
\end{figure}

\paragraph*{Muon colliders}~\\
The main parameters for the~\acrshort{mc} are presented in Table~\ref{tab:operational_year_FCChhMuC} and Fig.~\ref{fig:operational_year_MuC}.

\clearpage
\subsection{Technology readiness level}\label{sec:TRLtable}

The definition of the~\acrshort{trl} is provided in Table~\ref{tab:TRL}.

\begin{table}[h!]
    \caption{Definition of the~\acrfull{trl}~\cite{bib:LDG_RD_review_ESPP2026}. The same \textit{traffic light} colour coding is used as in Ref.~\cite{bib:LDG_RD_review_ESPP2026}.}
    \centering
    \footnotesize
    \begin{tabular}{|c|m{8cm}|}\hline
    Level  & Short description  \\\hline\hline

\cellcolor{red}1 & Basic principles observed and reported. 
\\\hline

\cellcolor{red}2 &  Technology concept formulated. 
\\\hline

\cellcolor{red}3 & Experimental critical proof of concept. 
\\\hline

\cellcolor{yellow}4 & Technology validated in laboratory environment. 
\\\hline

\cellcolor{yellow}5 & Technology (component or breadboard) validation in relevant environment (industrially relevant environment in case of key enabling technologies). 
\\\hline

\cellcolor{yellow}6 & Technology (system/sub-system) demonstration in a relevant environment (industrially relevant environment in case of key enabling technologies). 
\\\hline

\cellcolor{green}7 & System prototype demonstration in an operational environment.
\\\hline

\cellcolor{green}8 & Actual system completed and qualified through test and demonstration.
\\\hline

\cellcolor{green}9 & Actual system proven in an operational environment.
\\\hline

    \end{tabular}
    \label{tab:TRL}
\end{table}

\subsection{Cost classes}\label{sec:costClasses}
The definition of the cost classes used in the report is presented in Table~\ref{tab:AACEcostScheme}.

\begin{table}[h!]
 \caption{Cost estimate classification matrix used in the report. Source:~\acrshort{aace}~\cite{bib:AACE_18R97}.}
 \centering
 \footnotesize
 \begin{tabular}{|C{1.2cm}|C{1.5cm}|L{4.5cm}|C{2.5cm}|C{4cm}|}\hline
 Cost class & Level of definition & Typical estimating technique & Typical purpose of estimate & Expected accuracy ranges, low (L) and high (H) \\ \hline\hline
 Class 5 & 0/2\% & Capacity factored, stochastic, most parametric models, judgement or analogy & Concept screening & L: -20/-50\%; H: +30/+100\% \\ \hline
 Class 4 & 1/15\% & Equipment factored, more parametric models & Study or feasibility & L: -15/-30\%; H: +20/+50\% \\ \hline
 Class 3 & 10/40\% & Semi-detailed unit costs with assembly level line items. Combination of various techniques (detailed, unit-cost, or activity-based; parametric; specific analogy; expert opinion; trend analysis). & Preliminary, budget authorization & L: -10/-20\%; H: +10/+30\%  \\ \hline
 Class 2 & 30/70\% & Detailed unit costs. Combination of various techniques (detailed, unit-cost, or activity-based; expert opinion; learning curve). & Control or bid/tender & L:  -5/-15\%;  H: +5/+20\% \\ \hline
 Class 1 & 50/100\% & Deterministic, most definitive cost estimation. & Check estimate or bid/tender & L:  -3/-10\%;   H: +3/+15\% \\ \hline
 \end{tabular}
 \label{tab:AACEcostScheme}
\end{table}

\clearpage

\subsection{Producer price index (PPI)}\label{sec:PPI}
The data concerning the \acrshort{ppi} have been obtained from the following sources:
\begin{itemize}
    \item China (National Bureau of Statistics of China): \\ \url{https://tradingeconomics.com/china/producer-prices};
    \item European Union (EUROSTAT):\\ \url{https://tradingeconomics.com/european-union/producer-prices};
    \item Japan (Bank of Japan): \url{https://tradingeconomics.com/japan/producer-prices}.
\end{itemize}

\subsection{Exchange rates}\label{sec:exchrates}

\begin{table}[h!]
    \caption{Currency exchange rates used in the report.}
    \centering
    \footnotesize
    \begin{tabular}{|c|c|c|c|}
    \hline
    \acrfull{cny} to \acrshort{chf} & \acrfull{eur} to \acrshort{chf} & \acrfull{jpy} to \acrshort{chf} & \acrfull{usd} to \acrshort{chf} \\ \hline\hline
        0.123 & 0.959 & 0.00587 & 0.881 \\ \hline 
    \end{tabular}
    \label{tab:exchrates}
\end{table}

\subsection{Purchasing power parity (PPP)}\label{sec:ppp_exchrates}

\begin{table}[h!]
    \caption{Currency exchange rates corrected for the~\acrshort{ppp} used in the report (2023 data---Source:~\acrfull{oecd}~\cite{bib:ppp_oecd}).}
    \centering
    \footnotesize
    \begin{tabular}{|c|c|c|c|}
    \hline
    \acrfull{cny} to \acrshort{chf} & \acrfull{eur} to \acrshort{chf} & \acrfull{jpy} to \acrshort{chf} & \acrfull{usd} to \acrshort{chf} \\ \hline\hline
        0.266 & 1.546 & 0.0102 & 0.966 \\ \hline 
    \end{tabular}
    \label{tab:ppp_exchrates}
\end{table}


%% file: include/08-Annex/B-AdditionalTables.tex
\section{Additional tables}

\subsection{Published parameters of \acrshort{cepc} and \acrshort{ilc}}

In Table~\ref{tab:Parameters_ILC_CEPC} we provide the high-level parameters originally given for \acrshort{cepc} and the \acrshort{ilc}, i.e.\ without applying any rescaling as done for Table~\ref{tab:Parameters_ILC_CEPC_rescaled}.

\begin{table}[h!]
 \caption{High-level parameters of the \acrshort{ilc}~\cite{bib:IDT_ESPP2026}, and of \acrshort{cepc} for different options~\cite{bib:CEPC_TDR_accelerator,bib:CEPC_ESPP2026} (baseline parameters for \acrshort{cepc} are in \textbf{bold}). For the \acrshort{ilc} we follow an operational scenario adapted from Ref.~\cite {bib:ilc_snowmass2022}, with 10 years of operation including three years of ramp-up at one third the luminosity, and no luminosity upgrade (not costed in Ref.~\cite{bib:IDT_ESPP2026}). The integrated luminosity over the full programme is also adapted from Ref.~\cite{bib:ilc_snowmass2022}, following such a scenario. The instantaneous and integrated luminosity in-between parentheses includes also the contribution from energies below 99\% of the \acrshort{com} energy $\sqrt{s}$.
 }
\centering
\scriptsize
 \begin{threeparttable}
 \begin{tabular}{|l||cccc|c|}\hline
~& \multicolumn{4}{c|}{\acrshort{cepc}} & \acrshort{ilc} 250 \\ \hline\hline

Length collider tunnel~[\SI{}{\km}] & \multicolumn{4}{c|}{99.955} & 20.5 \\\hline
Number of experiments (\acrshortpl{ip}) & \multicolumn{4}{c|} {2} & 2 (1)\tnotex{tab:Parameters_ILC_CEPC:0} \\\hline
\acrshort{com}~energy [\SI{}{\GeV}] & \textbf{91} & \textbf{160} & \textbf{240} & 360 & 250 \\ \hline

Longitudinal polarisation (e$^-$ / e$^+$) & \multicolumn{4}{c|}{0~/~0\tnotex{tab:Parameters_ILC_CEPC:7}} & 0.8 / 0.3 \\
\hline

Number of years of operation (total) & \textbf{2} & \textbf{1} & \textbf{10} & 5 & 10 \\ \hline
Nominal years of operation (equivalent)\tnotex{tab:Parameters_ILC_CEPC:2} & \textbf{2} & \textbf{1} & \textbf{10} & 5 & 8 \\\hline

Synchrotron radiation power & \multirow{2}{*}{{\textbf{10}~/~30~/~50}} & \multirow{2}{*}{{\textbf{30}~/~50}} & \multirow{2}{*}{{\textbf{30}~/~50}} & \multirow{2}{*}{{30~/~50}} & \multirow{2}{*}{--} \\
per beam~[\SI{}{\MW}] & & & & & \\ \hline

Instantaneous luminosity per \acrshort{ip} & \multirow{2}{*}{\textbf{38}~/~115~/~192} & \multirow{2}{*}{\textbf{16}~/~26.7} & \multirow{2}{*}{\textbf{5}~/~8.3} & \multirow{2}{*}{0.5~/~0.8} & \multirow{2}{*}{1 (1.35)} \\
above $0.99~\sqrt{s}$ (total) of nominal c.o.m. energy~[\(10^{34}\,\SI{}{\per\square\centi\meter\per\second}\)] & & & & & \\\hline

Integrated luminosity above $0.99~\sqrt{s}$ (total) & \multirow{2}{*}{\textbf{10}~/~30~/~50} & \multirow{2}{*}{\textbf{4.2}~/~6.9} & \multirow{2}{*}{\textbf{1.3}~/~2.2} 
& \multirow{2}{*}{0.13~/~0.2} & \multirow{2}{*}{0.15 (0.2)} \\
over all~\acrshort{ip}s per year of nominal operation~[\SI{}{\per\atto\barn}/y] & \multicolumn{4}{c|}{} & \\ \hline

Integrated luminosity above $0.99~\sqrt{s}$ (total) & \multirow{2}{*}{\textbf{20}~/~60~/~100} & \multirow{2}{*}{\textbf{4.2}~/~6.9} & \multirow{2}{*}{\textbf{13}~/~21.6} & \multirow{2}{*}{0.65~/~1} & \multirow{2}{*}{1.2 (1.6)} \\
over all~\acrshort{ip}s over the full programme~[\SI{}{\per\atto\barn}] & & & & & \\ \hline

Peak power consumption [\SI{}{\MW}] & \textbf{100}\tnotex{tab:Parameters_ILC_CEPC:3}~/~\textbf{203}~/~287 & \textbf{225}~/~299 & \textbf{262}~/~339 & 358~/~432 &111 \\ \hline

Electricity consumption per year & \multirow{2}{*}{\textbf{0.5}~/~1.0~/~1.4} & \multirow{2}{*}{\textbf{1.1}~/~1.5} & \multirow{2}{*}{\textbf{1.3}~/~1.7} & \multirow{2}{*}{1.8~/~2.2} & \multirow{2}{*}{0.73} \\
of nominal operation [\SI{}{\tera\watt\hour}/y] & & & & & \\ \hline
 \end{tabular}
 \begin{tablenotes}
    \item[a] \label{tab:Parameters_ILC_CEPC:0} Two experiments at a single \acrshort{ip} (``push-pull'' mode).
    \item[b] \label{tab:Parameters_ILC_CEPC:7} No longitudinal polarisation is considered in the baseline, but such a possibility is being explored, especially at the Z energy.
    \item[c] \label{tab:Parameters_ILC_CEPC:2} This row lists the equivalent number of years of operation at nominal instantaneous luminosity, hence taking into account the luminosity ramp-up.
    \item[d] \label{tab:Parameters_ILC_CEPC:3} Extrapolated from Ref.~\cite{bib:CEPC_TDR_accelerator}.
 \end{tablenotes}
 \end{threeparttable}
\label{tab:Parameters_ILC_CEPC}
\end{table}

%% file: include/08-Annex/ZZ-AcronymsandGlossary.tex
\input{acronyms}

\printglossary[type=\acronymtype,title=List of acronyms, toctitle=List of acronyms]
\clearpage
\input{glossary}

\printglossary[title=Glossary, toctitle=Glossary]
\label{sec:Glossary}

%% file: acronyms.tex
\newacronym{aace}{AACE\textsuperscript{\textregistered}}{Association for the Advancement of Cost Engineering International}
\newacronym{ad}{AD}{Antiproton Decelerator}
\newacronym{aflc}{AFLC}{Argonne Flexible Linear Collider}
\newacronym{ai}{AI}{Artificial Intelligence}
\newacronym{alegro}{ALEGRO}{Advanced LinEar collider study GROup}
\newacronym{alice}{ALICE}{A Large Ion Collider Experiment}
\newacronym{alice3}{ALICE3}{A Large Ion Collider Experiment Upgrade~3}
\newacronym{alive}{ALiVE}{Advanced Linear collider for Very high Energy}
\newacronym{aps}{APS}{Announced Pledges Scenario}
\newacronym{atf}{ATF}{Accelerator Test Facility}
\newacronym{atlas}{ATLAS}{A Toroidal LHC ApparatuS}
\newacronym{awake}{AWAKE}{Advanced Proton Driven Plasma Wakefield Acceleration Experiment}

\newacronym{bcr}{BCR}{Benefit Cost Ratio}
\newacronym{bds}{BDS}{Beam Delivery System}
\newacronym{bella}{BELLA}{BErkeley Lab Laser Accelerator}
\newacronym{bepc}{BEPC}{Beijing Electron–Positron Collider}
\newacronym{bepcii}{BEPCII}{Beijing Electron–Positron Collider II}
\newacronym{binp}{BINP}{Budker Institute of Nuclear Physics}
\newacronym{bnl}{BNL}{Brookhaven National Laboratory}
\newacronym{bscco}{BSCCO}{Bismuth Strontium Calcium Copper Oxide}

\newacronym{capex}{CAPEX}{CApital EXpenditure}
\newacronym{cbeta}{CBETA}{Cornell BNL ERL Test Accelerator}
\newacronym{cdr}{CDR}{Conceptual Design Report}
\newacronym{cds}{CDS}{Conceptual Design Study}
\newacronym{ce}{CE}{Civil Engineering}
\newacronym{cebaf}{CEBAF}{Continuous Electron Beam Accelerator Facility}
\newacronym{cepc}{CEPC}{Circular Electron Positron Collider}
\newacronym{cembureau}{CEMBUREAU}{European Cement Association}
\newacronym{cern}{CERN}{European Organization for Nuclear Research}
\newacronym{cfc}{CFC}{Chlorofluorocarbons}
\newacronym{chf}{CHF}{Swiss Franc}
\newacronym{clear}{CLEAR}{CERN Linear Electron Accelerator for Research}
\newacronym{clic}{CLIC}{Compact Linear Collider}
\newacronym{cms}{CMS}{Compact Muon Solenoid}

\newacronym{com}{c.o.m.}{centre-of-mass}
\newacronym{cv}{CV}{Cooling and Ventilation}
\newacronym{cw}{CW}{Continuous Wave}
\newacronym{cwa}{CWA}{Collinear Wakefield Accelerator}
\newacronym{cny}{CNY}{Chinese Yuan Renminbi}
\newacronym{csns}{CSNS}{Chinese Spallation Neutron Source}
\newacronym{c3}{C$^3$}{Cool Copper Collider}
\newacronym{daphne}{DA$\mathrm{\Phi}$NE}{Double Annular $\mathrm{\Phi}$ factory for Nice Experiments}
\newacronym{daq}{DAQ}{Data Acquisition}
\newacronym{desy}{DESY}{Deutsches Elektronen Synchrotron}
\newacronym{dla}{DLA}{Dielectric Laser Accelerator}
\newacronym{drd}{DRD}{Detector \acrshort{rd}}

\newacronym{eic}{EIC}{Electron--Ion Collider}
\newacronym{edr}{EDR}{Engineering Design Report}
\newacronym{eia}{EIA}{Environmental Impact Assessment}
\newacronym{elena}{ELENA}{Extra Low ENergy Antiproton ring}
\newacronym{en}{EN}{European Norm}
\newacronym{eoi}{EoI}{Expression of Interest}
\newacronym{erl}{ERL}{Energy Recovery Linac}
\newacronym{espp}{ESPP}{European Strategy for Particle Physics}
\newacronym{esppu}{ESPPU}{Update of the European Strategy for Particle Physics}
\newacronym{eu}{EU}{European Union}
\newacronym{eupraxia}{EuPRAXIA}{European Plasma Research Accelerator with eXcellence in Applications}
\newacronym{eur}{EUR}{Euro}

\newacronym{facet}{FACET}{Facility for Advanced Accelerator Experimental Tests}
\newacronym{faser}{FASER}{Forward Search Experiment at the \acrshort{lhc}}
\newacronym{fasernu}{FASER$\nu$}{\acrshort{faser} neutrino detector}
\newacronym{fcc}{FCC}{Future Circular Collider}
\newacronym{fccee}{FCC\nobreakdashes-ee}{Future Circular \epem Collider }
\newacronym{fcchh}{FCC\nobreakdashes-hh}{Future Circular hadron--hadron Collider}
\newacronym{fcceh}{FCC\nobreakdashes-eh}{Future Circular electron--hadron Collider}
\newacronym{fccpa}{PA}{Point A}

\newacronym{fel}{FEL}{Free Electron Laser}
\newacronym{ffag}{FFAG}{Fixed-Field Alternating-Gradient}
\newacronym{fftb}{FFTB}{Final-Focus Test Beam}
\newacronym{fp}{FP}{Full-Power}
\newacronym{fsu}{FSU}{Field Support Units}
\newacronym{fte}{FTE}{Full-Time-Equivalent}
\newacronym{ftey}{FTEy}{Full-Time-Equivalent-years}
\newacronym{ghg}{GHG}{Greenhouse Gas}
\newacronym{gwp}{GWP}{Global Warming Potential}
\newacronym{halhf}{HALHF}{Hybrid Asymmetric Linear Higgs Factory}
\newacronym{he}{HE}{High-Energy}
\newacronym{helhc}{HE\nobreakdashes-LHC}{High-Energy Large Hadron Collider}
\newacronym{heps}{HEPS}{High Energy Photon Source}
\newacronym{hfm}{HFM}{High Field Magnet}

\newacronym{hllhc}{HL\nobreakdashes-LHC}{High-Luminosity LHC}
\newacronym{hom}{HOM}{High-Order Mode}
\newacronym{horizon}{HORIZON}{Horizon Europe Framework Programme}

\newacronym{hts}{HTS}{High Temperature Superconductor}
\newacronym{hw}{HW}{Hardware}
\newacronym{ibs}{IBS}{Iron Based Superconductor}
\newacronym{icfa}{ICFA}{International Committee for Future Accelerators}
\newacronym{ics}{ICS}{Inverse Compton Scattering}
\newacronym{idt}{IDT}{International Development Team}
\newacronym{iea}{IEA}{International Energy Agency}
\newacronym{ihep}{IHEP}{Institute of High Energy Physics}
\newacronym{ijclab}{IJCLab}{Laboratoire de Physique des 2 Infinis Ir{\`e}ne Joliot-Curie}
\newacronym{ilc}{ILC}{International Linear Collider}
\newacronym{ilcu}{ILCU}{\acrshort{ilc} unit}
\newacronym{ild}{ILD}{International Large Detector}
\newacronym{imcc}{IMCC}{International Muon Collider Collaboration}
\newacronym{ip}{IP}{Interaction Point}
\newacronym{ir}{IR}{Interaction Region}
\newacronym{iso}{ISO}{International Organization for Standardization}
\newacronym{isolde}{ISOLDE}{Isotope Separator On Line DEvice}
\newacronym{itf}{ITF}{Implementation Task Force}
\newacronym{itn}{ITN}{\acrshort{ilc} Technology Network}
\newacronym{jlab}{JLab}{Jefferson Lab}
\newacronym{jpy}{JPY}{Japanese Yen}

\newacronym{kek}{KEK}{High Energy Accelerator Research Organization in Japan}
\newacronym{kekb}{KEKB}{KEK B-factory}
\newacronym{lc}{LC}{Linear Collider}
\newacronym{lca}{LCA}{Life Cycle Assessment}
\newacronym{lcf}{LCF}{Linear Collider Facility}
\newacronym{lcls2}{LCLS\nobreakdashes--II}{Linac Coherent Light Source II}
\newacronym{ldg}{LDG}{Laboratory Directors Group}
\newacronym{lep}{LEP}{Large Electron Positron (collider)}
\newacronym{lep2}{LEP2}{Large Electron Positron (collider) 2}
\newacronym{lep3}{LEP3}{Large Electron Positron (collider) 3}
\newacronym{lhc}{LHC}{Large Hadron Collider}
\newacronym{lhec}{LHeC}{Large Hadron-electron Collider}
\newacronym{loi}{LoI}{Letter of Interest}
\newacronym{lp}{LP}{Low-Power}
\newacronym{liu}{LIU}{LHC Injectors Upgrade}
\newacronym{ls}{LS}{Long Shutdown}
\newacronym{lss}{LSS}{Long Straight Section}
\newacronym{lts}{LTS}{Low Temperature Superconductor}
\newacronym{lux}{LUX}{Laser–plasma driven undulator beamline}
\newacronym{lwfa}{LWFA}{Laser Wakefield Accelerator}

\newacronym{map}{MAP}{Muon Accelerator Programme}
\newacronym{mc}{MC}{Muon Collider}
\newacronym{mdi}{MDI}{Machine--Detector Interface}
\newacronym{mext}{MEXT}{Ministry of Education, Culture, Sports, Science and Technology of Japan}
\newacronym{ml}{ML}{Main Linac}
\newacronym{mps}{MPS}{Machine Protection System}
\newacronym{mtbf}{MTBF}{Mean Time Between Failures}

\newacronym{nc}{NC}{Normal Conducting}
\newacronym{neg}{NEG}{Non-Evaporable Getter}
\newacronym{npv}{NPV}{Net Present Value}
\newacronym{ntof}{n\_TOF}{Neutron Time-Of-flight Facility}
\newacronym{oecd}{OECD}{Organisation for Economic Co-operation and Development}
\newacronym{p5}{P5}{Particle Physics Project Prioritization Panel}
\newacronym{pcm}{pCM}{parton center-of-momentum}
\newacronym{pepii}{PEP\nobreakdashes--II}{Positron Electron Project II}
\newacronym{perle}{PERLE}{Powerful Energy Recovery Linac for Experiments}
\newacronym{pets}{PETS}{Power Extraction and Transfer Structures}
\newacronym{ppi}{PPI}{(industrial) Producer Price Index}
\newacronym{ppp}{PPP}{Purchasing Power Parity}
\newacronym{ps}{PS}{Proton Synchrotron}
\newacronym{pwfa}{PWFA}{Plasma Wakefield Accelerator}
\newacronym{pwfalc}{PWFA\nobreakdashes-LC}{Plasma Wakefield Acceleration Linear Collider}

\newacronym{rd}{R\&D}{Research and Development}
\newacronym{r2e}{R2E}{Radiation to Electronics}
\newacronym{rcs}{RCS}{Rapid-Cycling Synchrotron}
\newacronym{rdr}{RDR}{Reference Design Report}
\newacronym{rebco}{ReBCO}{Rare-earth barium copper oxide}
\newacronym{rf}{RF}{radio frequency}
\newacronym{rte}{RTE}{R\'eseau de Transport d'\'Electricit\'e} 
\newacronym{rtml}{RTML}{Ring-to-Main Linac system}
\newacronym{rp}{RP}{Radiation Protection}

\newacronym{sc}{SC}{Super Conducting}
\newacronym{scsps}{scSPS}{Superconducting Super Proton Synchrotron}
\newacronym{sid}{SiD}{Silicon Detector}
\newacronym{slac}{SLAC}{Stanford Linear Accelerator Center}
\newacronym{slc}{SLC}{Stanford Linear Collider}
\newacronym{sm}{SM}{Standard Model}
\newacronym{sparclab}{SPARC\_LAB}{Sources for Plasma Accelerators and Radiation Compton with Laser And Beam}
\newacronym{sppc}{SPPC}{Super Proton Proton Collider}

\newacronym{sps}{SPS}{Super Proton Synchrotron}
\newacronym{sr}{SR}{Synchrotron Radiation}
\newacronym{srf}{SRF}{Superconducting radio frequency}
\newacronym{steps}{STEPS}{STated Energy Policies Scenario}
\newacronym{skekb}{SuperKEKB}{Super KEK B-factory}
\newacronym{swfa}{SWFA}{Structure Wakefield Accelerator}
\newacronym{us}{US}{United States}
\newacronym{usd}{USD}{US Dollar}
\newacronym{tba}{TBA}{Two-Beam Accelerator}
\newacronym{tbm}{TBM}{Tunnel Boring Machine}
\newacronym{tdr}{TDR}{Technical Design Report}
\newacronym{ti}{TI}{Technical Infrastructure}
\newacronym{trl}{TRL}{Technology Readiness Level}
\newacronym{wfa}{WFA}{Wakefield Accelerator}
\newacronym{xfel}{EU-XFEL}{European X-Ray Free-Electron Laser Facility}
\newacronym{yets}{YETS}{Year End Technical Stop}

%% file: glossary.tex

\newglossaryentry{eiagl}
{
    name=Environmental Impact Assessment,
    description= {The process of identifying, predicting, evaluating and mitigating the biophysical, social, and other relevant effects of development proposals prior to major decisions being taken and commitments made~\cite{bib:EIA_def}.}
}

\newglossaryentry{gwpgl}
{
name= Global Warming Potential,
description={(\acrshort{gwp}) is a term used to describe the relative potency, molecule for molecule, of a ~\acrshort{ghg}, taking account of how long it remains active in the atmosphere. The global-warming potentials currently used are those calculated over 100 years. Carbon dioxide is taken as the gas of reference and given a 100-year \acrshort{gwp} of 1~\cite{bib:gwpdef}.}
}

\newglossaryentry{lcagl}
{
    name=Life Cycle Assessment,
    description= {\acrfull{lca} is a process of evaluating the effects that a product has on the environment over the entire period of its life thereby increasing resource-use efficiency and decreasing liabilities. It takes into account a product's full life cycle: from the extraction of resources, through production, use, and recycling, up to the disposal of remaining waste. ~\acrshort{lca} is commonly referred to as a ``cradle-to-grave'' analysis~\cite{bib:LCA_EU,bib:LCA_EEAdef}.  \acrshort{lca} modules A1 to A5 (``cradle-to-gate'' analysis) include the raw material extraction to construction activities on site (including energy consumption). B1--B8 modules are related to the use of the product over the entire life cycle of the project, with the end-of-life stage (C1--C4 modules) for the deconstruction and demolition of the construction project including the impacts of transport to waste processing sites and the disposal of said waste. D module covers the net benefits and loads arising from the reuse of products or the recycling or recovery of energy from waste materials resulting from the construction stage, the use stage and the end-of-life stage~\cite{clic_ilc_lca_arup}.}
}

\newglossaryentry{ppigl}
{
    name=(industrial) Producer Price Index,
    description={The \acrfull{ppi}, also called output price index, is a business-cycle indicator showing the development of transaction prices for the monthly industrial output of economic activities.~\cite{bib:EUROSTAT_PPIdef}.}
}

\newglossaryentry{pppgl}
{name= Purchasing Power Parity,
description={Purchasing power parities (\acrshortpl{ppp}) are the rates of currency conversion that try to equalise the purchasing power of different currencies, by eliminating the differences in price levels between countries. The basket of goods and services priced is a sample of all those that are part of final expenditures: final consumption of households and government, fixed capital formation, and net exports. This indicator is measured in terms of national currency per \acrshort{us} dollar~\cite{bib:OECD_PPPdata}.
}
}